\documentclass[preprint]{aastex}
\usepackage{natbib}
\usepackage{epsfig,lscape,rotating}
\usepackage{color}
\usepackage{amsmath}
\usepackage{amssymb}

\renewcommand{\vec}[1]{\mbox{\boldmath$#1$}}

\shorttitle{Magnetized Accretion and Dead Zones}
\shortauthors{Dzyurkevich et al.}

\begin{document}

\title{ Magnetized Accretion and Dead Zones in Protostellar Disks }

\author{Natalia Dzyurkevich\altaffilmark{1,2}, Neal J. Turner\altaffilmark{3}, 
Thomas Henning\altaffilmark{1}, Wilhelm Kley\altaffilmark{4}}
\affil{ $^1$Max Planck Institute for Astronomy, K\"onigstuhl 17, 69117
  Heidelberg, Germany\\
  $^2$Laboratoire de radioastronomie, UMR 8112 du CNRS, Ecole Normale
  Superieure et Observatoire de Paris, 24 rue de Lhomond, 75231, Paris Cedex
  05, France\\
  $^3$Jet Propulsion Laboratory, California Institute of Technology, Pasadena, California 91109, USA\\
  $^4$University of T\"ubingen, Auf der Morgenstelle 10, T\"ubingen, 72076 Germany}

\begin{abstract}
  The edges of magnetically-dead zones in protostellar disks have been
  proposed as locations where density bumps may arise, trapping
  planetesimals and helping form planets.  Magneto-rotational
  turbulence in magnetically-active zones provides both accretion of
  gas on the star and transport of mass to the dead zone.  We
  investigate the location of the magnetically-active regions in a
  protostellar disk around a solar-type star, varying the disk
  temperature, surface density profile, and dust-to-gas ratio.  We
  also consider stellar masses between 0.4 and 2~$M_\odot$, with
  corresponding adjustments in the disk mass and temperature.  The
  dead zone's size and shape are found using the Elsasser number
  criterion with conductivities including the contributions from ions,
  electrons, and charged fractal dust aggregates.  The charged
  species' abundances are found using the approach proposed by
  Okuzumi.  The dead zone is in most cases defined by the ambipolar
  diffusion.  In our maps, the dead
  zone takes a variety of shapes, including a fish-tail pointing away
  from the star and islands located on and off the midplane.  The
  corresponding accretion rates vary with radius, indicating locations
  where the surface density will increase over time, and others where
  it will decrease.  We show that density bumps do not readily grow
  near the dead zone's outer edge, independently of the disk
  parameters and the dust properties.  Instead, the accretion rate
  peaks at the radius where the gas-phase metals freeze out.  This
  could lead to clearing a valley in the surface density, and to a
   trap for pebbles located just outside the metal freeze-out line.
\end{abstract}

\section{Introduction}

Trapping solid material is a basic requirement for forming planets
enriched in the heavy elements with respect to the central star.  One
way to trap solid particles is through gas pressure gradients.  If the
pressure locally increases with distance from the star, radial force
balance dictates the gas orbit faster than Keplerian.  Embedded
particles then feel a tailwind, receive orbital angular momentum from
the gas, and drift away from the star, collecting near the point where
the general outward pressure decline resumes.  Such local radial
pressure maxima can occur in various ways.  They can form at the
distance where the disk's main heat source switches from accretion
power to stellar illumination \citep{hase10a,hase10b,hase11}, or can
be part of long-lived zonal flows \citep{joh09,uri11,pin12}.  Pressure
bumps can also be made by unsteadiness in the overall inward spiraling
of the disk gas that feeds the growth of the central star.  Material
will pile up where the inflow becomes slower than average.

The best candidate to drive the disk's inflow is turbulence driven by
the magneto-rotational instability or MRI \citep{bal91,haw95,bal98}.
Since the MRI relies on magnetic forces, the disk gas must be ionized
enough to couple to the fields.  The low temperatures rule out thermal
ionization, but the disk surface layers are nevertheless partly
ionized because they absorb X-rays from the young star and cosmic rays
from interstellar space \citep{gam96,gla97}. 
%(Glassgold et al. 1997 ApJ 480,344). 
 The ionization is balanced by recombination on dust grains in
the disk interior \citep{san00}, so that the grains'
%(Sano et al. 2000 ApJ 543, 486)
abundance regulates the flow of material through the disk.

Time-unsteady flow leading to local pressure maxima can occur where
the disk crosses the threshold from magnetically-coupled to decoupled,
at the dead zones' inner and outer radial edges \citep{lyr09,dzy10}.
Away from the edges, pressure bumps could arise from changes in the
rate of flow through the magnetically-active surface layers above and
below the dead zone.  Such interruptions in the layered flow might
come from a jump in the dust recombination cross-section across the
snow line \citep{kre07,bra08b}.

The dead zone itself can participate in the overall radial flow.  Even
where the resistivity is high enough to damp away the MRI, and
magnetic forces cannot directly drive turbulence, the stresses can be
non-zero due to waves propagating from nearby turbulent regions
\citep{fle03,dzy10,oku11} and to hydrodynamical instabilities and
gas-particle interaction \citep{wei93,joh07a,joh11}.  The dead zone
stresses affect the timescales for material to pile up in orbit around
the star, and so must be considered when attempting to use the
time-evolving flow in layered disks to explain episodic accretion
events such as FU~Orionis outbursts \citep{arm01,mar12a,mar12b,zhu10}.
%(Zhu et al. 2010 ApJ 713, 1143).

The size and shape of the dead zone thus govern many aspects of
protostellar disks' evolution.  The dead zone's edges are determined
by the non-ideal terms in the induction equation corresponding to
Ohmic, Hall and ambipolar diffusion.  The Ohmic term is typically
strongest in the dense disk interior, the Hall term at intermediate
densities and the ambipolar term in the disk's low-density atmosphere
and outer reaches 
%(Wardle \& Ng 1999; Wardle 2007; Perez-Becker % Chiang 2011, ApJ 727, 2) 
\citep{war99,war07,san02a,per11b}.

The dead zone determined by the Ohmic diffusivity has been mapped
under ionization by cosmic rays \citep{san00}, X-rays \citep{ilg06}
 and stellar energetic protons \citep{tur09}.
  Recently its extent was estimated with a more precise
magnetic field distribution using a sequence of local shearing-box
chemical-MHD calculations \citep{fla12}.  The impact of the Ohmic
diffusion might be reduced by the associated magnetic dissipation
heating, at least near the star where temperatures approach the
thermal ionization threshold \citep{mur12}.

The Hall term has at most a weak effect on the saturated strength of
the MRI turbulence, at least when the Hall and Ohmic terms are not too
different in magnitude, according to numerical calculations
\citep{san02b}.  \citet{war12} however suggest that the results would
differ at very large ratios of Hall to the other terms.  When the Hall
term is taken into account in this regime, the gas flow through the
magnetically-active layers could be faster.

Ambipolar diffusion becomes important when collisions are rare enough
that the neutral gas drifts with respect to the plasma.  Analysis of
the MRI in the linear regime \citep{bla94,kun04,des04} shows that the
instability is suppressed when the plasma transfers its momentum to
the neutrals slower than the instability grows.  The non-linear
evolution of the MRI was first modeled by \citet{mac95} in two spatial
dimensions, followed by \citet{bra95} for the strong coupling regime.
In 3-D two-fluid local-box simulations the non-linear MRI under
ambipolar diffusion was studied by \citet{haw98} for several different
magnetic field configurations.  \citet{bai11c} show that the turbulent
accretion stress declines smoothly with increasing ambipolar
diffusivity, from around 1\% of the gas pressure at the threshold for
MRI turbulence down to $7\times 10^{-4}$ at diffusivities ten times
greater.  The local shearing-box simulations were unstratified.  In
\citet{sim12}, MRI turbulence on azimuthal background magnetic fields
was modeled for conditions similar to those thought to occur in the
outer reaches of protostellar disks.  The turbulence died out almost
completely below the threshold ambipolar diffusivity.

The ionization degree of the gas is affected by the dust surface area
available to sweep up the ions and electrons.  The recombination
cross-section is often treated as residing on compact spherical grains
\citep{sem04,ilg06,tur08,tur10} so that the area decreases as the
grains grow.  Here we take the alternative approach of \citet{oku09a},
supposing that at least the smaller grains bearing most of the
cross-section remain as loose, fluffy aggregates of smaller particles.
The fractal dimension is close to two and so the area remains
unchanged as the aggregates grow.  Larger particles may be compacted
by collisions or internal forces as the solid material evolves toward
planet formation, but we assume the small fluffy aggregates continue
to dominate the recombination cross-section.  Like Okuzumi we neglect
the electrical polarization that affects small compact grains'
interaction with approaching charges, and compute the abundances of
the ions and electrons, and the charge state distribution of the
aggregates, from his governing equation (his eq.~30).

The situation with the magnetic diffusivity is complicated by the fact
that the grains carry charge.  Small PAH-like particles, due to their
low mass and relatively high cyclotron frequencies, can significantly
change the disk's overall magnetic coupling \citep{bai11c,per11b}
(Mohanty, Ercolano \& Turner, in press).
More-massive compact grains have little direct effect on the
electrical current because of their low cyclotron frequencies, even
when charged to the limit where further electrons are repelled by the
electrostatic force \citet{war07}.  Fluffy aggregates however, due to
their large surface area, can accommodate many more elementary
charges.  We therefore include the aggregates' contributions to the
conductivities.

In this paper we study the factors determining the dead zone's extent
in the presence of fluffy grains.  After presenting a fiducial model
we vary the size of the monomers making up the fluffy aggregates, the
aggregates' abundance relative to the gas, the overall disk mass and
the surface density power-law index, the gas temperature, the stellar
mass and the strength of the cosmic ray ionization.  Across this
parameter space we find the magnetically-dead region has a diversity
of shapes and sizes.

The paper is laid out as follows.  In section~2 we describe how we
calculate the ionization state.  Section~3 has an outline of the
magnetic dissipation mechanisms and the criteria for coupling the gas
to the magnetic field.  In section~4 we present the resulting dead
zone shape as a function of the monomer size, dust-to-gas ratio and
other parameters.  We discuss the implications for building local
pressure maxima in section~5, and summarize our findings in section~6.

%%%%%%%%%%%%%%%%%%%%%%%%%%%%%%%%%%%%%%%%%%%%%%%%%%%%%%%%%%%%%%%%%
\section{Fiducial Disk Model}

\subsection{Gas and magnetic field}

\begin{figure}
\begin{center}
\includegraphics[width=5.6in]{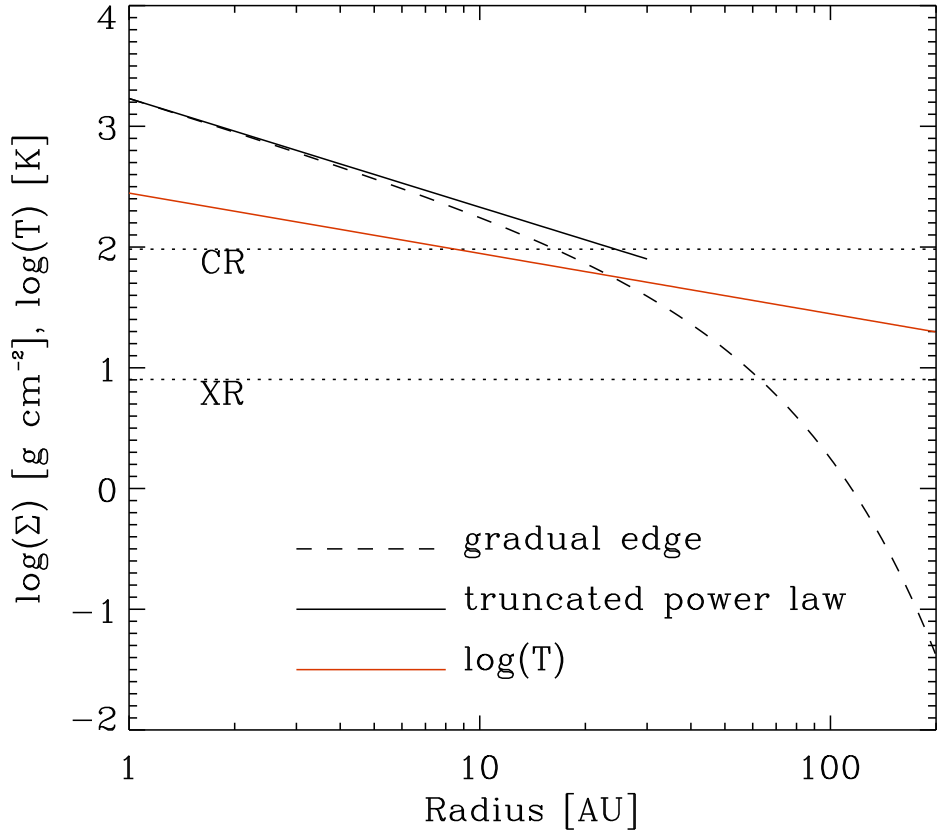}
  \caption{Surface density and temperature profiles in the fiducial
    model disk.  The dashed line shows the surface density power law
    with exponential roll-off.  The solid black line shows the
    power-law alone, while dotted gray lines indicate the penetration
    depths of cosmic rays and X-rays, $96\rm\ g cm^{-2}$ and $8\rm\ g
    cm^{-2}$, respectively.  The red line is the temperature profile.
  }
\label{fidu}
\end{center}
\end{figure}

We begin by mapping the dead zone in a fiducial disk model, taking
stellar mass $M_*=1 M_\odot$ and normalizing the surface density to
$\Sigma=1700\rm\ {g cm^{-2}}$ and the temperature to $T=280$~K, both at
1~AU.  Rather than a power-law surface density profile, we use a
power-law with exponential outer roll-off, the similarity solution
obtained by \citet{har98}.  This profile yields a good match between
the radii of the disk measured in CO emission and in dust continuum
\citet{hug08}.  The surface density falls off with the radius $r$ as
\begin{equation}
  \Sigma = \Sigma_c\left(\frac{r}{r_c}\right)^{-p}
  \exp{\left[-\left(\frac{r}{r_c}\right)^{2-p}\right]},
\end{equation}
where $\Sigma_c$, the surface density at the cut-off radius $r_c$, is
equal to
\begin{equation}
  \Sigma_c=(2-p)\frac{M_{\rm disk}}{2\pi r_c^2}.
\end{equation}
We take the median power-law index $p=0.9$ observed by \citet{and09}
in the Ophiuchus star-forming region, and set the cut-off radius $r_c$
to 40~AU.  The surface density then matches the standard minimum-mass
Solar nebula at 1~AU and again at 100~AU.  The disk's whole mass of
$M_{\rm disk}=0.064 M_\odot$ is stable against self-gravity, as the
Toomre parameter is nowhere lower than~4.  The density is easily
reconstructed as
\begin{equation}
  \rho=\frac{\Sigma}{ \sqrt{2\pi} c_0 r \sin(\Theta) }
  \cdot{} \exp{\left( - \frac{\cos{\Theta}^2}{ 2c_0^2 \sin{\Theta}^2 } \right)},
\end{equation}
where $R=r\sin(\Theta)$ is the cylindrical radius and $c_0=H/(r
\sin{\Theta})$ the isothermal sound speed, with $H$ the pressure scale
height.  Here we use spherical coordinates $(r,\Theta,\phi)$ with the
rotation axis corresponding to $\Theta=0$.  The orbital angular speed
and temperature are
\begin{equation}
  \Omega=\sqrt{ \frac{G M_*}{(r\sin{\Theta})^3}  } {\rm\ \ \ and}\ \ \
  T=T_0\sqrt{\frac{r_{\rm in}}{r\sin\Theta}},
\end{equation}
with $T_0=280$~K at $r_{\rm in}=1$~AU.  The surface density profile
with and without the exponential roll-off is shown in Fig.~\ref{fidu}.
Cosmic rays reach the midplane largely unimpeded at $r>15$~AU.
Scattered X-rays similarly reach the midplane outside $r>60$~AU.
Nevertheless, the thin gas in the outer disk could be
magneto-rotationally stable due to ambipolar diffusion.  To map any
stable regions, we need to obtain the densities of the charged species
in a multi-component plasma.

We calculate the conductivities in the domain extending from 1 to
200~AU in radius and over $\pi/2 \pm{0.4}$ in the polar angle
$\Theta$.  The disk's concave or flaring shape means the model covers
$\pm 12$ pressure scale heights at 1~AU but just $\pm 5$H at 200~AU.

We give the disk a mean vertical magnetic field,
\begin{equation}
B=\sqrt{8 \pi\hat{\rho} \hat{c}_{s}^2/\beta_{p,0},}
\end{equation}
where the midplane plasma beta $\beta_{p,0}=400$ is a constant and
$\hat{\rho}(r)$ and $\hat{c}_{s}(r)$ are the gas density and sound
speed at the midplane.  The RMS vertical magnetic field in the
saturated MRI turbulence varies only slightly with height up to the
location where the plasma beta falls below unity, in stratified local
ideal-MHD shearing-box calculations by \citet{mil00}.  To simplify
matters we therefore make the field strength independent of height.

\subsection{Dust}

We assume a two-component dust population consisting of fractal porous
aggregates and larger, more compact rocks. The fractal aggregates
 are assembled of small, nonaggregated dust (monomers) with radius $a_0$.
 The fluffy aggregates
couple well to the gas through drag forces \citep{mea91,blu04,dom07}
and we assume they remain well-mixed in the gas, with dust-to-gas mass
ratio $f_{dg}$.  Both numerical \citep{kem99} and experimental studies
\citep{wur98,blu98,blu00} show that such aggregates form at early
growth stages when grain-grain collisions are slow.  The outcome is an
ensemble of fractal aggregates with fractal dimension $D\leq 2$, with
a narrow, quasi-monodisperse mass distribution. The projected cross-section
of the aggregate is $\sigma\simeq \sigma_0N$, where $\sigma_0=\pi a_0^2$ is
the geometrical cross-section of the monomer (see \citet{oku09a} for details). 
The fractal growth
continues until the aggregates reach centimeter sizes \citep{suy08}.
The compact bodies in contrast have $D\sim 3$.  They may form in
low-probability collisions between the fluffy particles, with impact
speeds high enough to modify the aggregates' structure but too low for
complete disruption.

The fluffy aggregates dominate the total cross-section for ions and
electrons because of their much greater surface area per unit solid
mass.  The compact particles simply serve as a sink of solid material.
The fluffy grains therefore determine the ionization state, even
though the compact particles may contain most of the mass.

\subsection{Ionization and Recombination}

Thermal ionization is effective only within less than 1~AU of the
central star, where temperatures above 1\,000~K mean alkali metals can
ionize when colliding with hydrogen or helium \citep{ume83}.  At the
lower temperatures found further out, ionization comes from cosmic
rays, stellar X-rays and ultraviolet photons, and radionuclides.  Our
model includes all these except the ultraviolet photons, which
penetrate a mass column less than $0.1 \rm\ g\ cm^{-2}$ (Perez-Becker
\& Chiang 2011) so we do not consider them further.

Most ionizations are of the abundant species, $H_2$ and helium.  In
the total ionization rate
\begin{equation}\label{eq:zeta}
  \zeta=\zeta_{XR}+\zeta_{CR}+\zeta_{RN}
\end{equation}
the cosmic ray contribution is
\begin{equation}\label{eq:zetacr}
\begin{split}
%  \zeta_{CR}=\frac{\zeta_{CR}^0}{2} \times
   \zeta_{CR}= \zeta_{CR}^0 \times 
  &\left[
    \exp{\left(-\frac{D_a}{D_{CR}}\right)}
    \left\{1 + \left(\frac{D_a}{D_{CR}}\right)^{3/4}\right\}^{-4/3}\right.\\
  &\left. +
    \exp{\left(-\frac{D_b}{D_{CR}}\right)}
    \left\{1 + \left(\frac{D_b}{D_{CR}}\right)^{3/4}\right\}^{-4/3}\right].
\end{split}
\end{equation}
where the cosmic ray penetration depth $D_{CR}=96\rm\ g cm^{-2}$,
and the mass column vertically above and below the point of interest
are $D_a$ and $D_b$ respectively \citep{ume09}. 
% (Umebayashi \& Nakano 2009 ApJ 690, 69). 
The cosmic ray ionization rate in the interstellar medium varies from
$10^{-18}\rm\ s^{-1}$ to $10^{-15}\rm\ s^{-1}$ \citep{ume81,mcc03}.
As a fiducial value we choose $\zeta_{CR0}=5\times 10^{-18}\rm\
s^{-1}$.

The X-ray contribution to the ionization rate is
\begin{equation}\label{eq:zetaxr}
\begin{split}
  \zeta_{XR}&=\frac{L_X}{10^{29}\rm\ erg\ s^{-1}}\left(\frac{1\rm\
      AU}{R}\right)^{2.2} \times\\
  &\left(\zeta_1\left[e^{-(N_{H1}/N_1)^{0.4}}
      + e^{-(N_{H2}/N_1)^{0.4}}\right]\right.\\
  &\left.+\zeta_2\left[e^{-(N_{H1}/N_2)^{0.65}}
      + e^{-(N_{H2}/N_2)^{0.65}}\right]\right),
\end{split}
\end{equation}
where the direct X-rays' intensity $\zeta_1=6\times 10^{-12}\rm\
s^{-1}$, the scattered X-rays' intensity $\zeta_2= 10^{-15}\rm\
s^{-1}$ and the corresponding penetration columns $N_1=1.5\times
10^{21} \rm\ cm^{-2}$ or $0.006\rm\ g cm^{-2}$ and $N_2=7\times
10^{23} \rm\ cm^{-2}$ or $3\rm\ g cm^{-2}$ are measured by integrating
the gas density vertically.  This expression for the X-ray ionization
rate was fit by \citet{bai09} to Monte Carlo radiative transfer
results from \citet{ige99}.  We take the young star's X-ray luminosity
to be $10^{29}\rm\ erg\ s^{-1}$.
% based on Chandra observations in the Orion nebula \citep{wol05}.

Finally, we take a radionuclide ionization rate
\begin{equation}\label{eq:zetarn}
  \zeta_{RN}=7\times 10^{-19} (f_{dg}/10^{-2})\rm\ s^{-1},
\end{equation}
with the constant derived from Solar system measurements of the
daughter products of the short-lived nuclide $^{26}$Al \citep{ume09}.
Since the $^{26}$Al is refractory and concentrated in the solids, and
since the only solids well-mixed in the gas are the fluffy dust
aggregates, the ionization rate is proportional to the aggregates'
mass fraction $f_{dg}$.

Eqs.~\ref{eq:zeta}-\ref{eq:zetarn} define the ionization rate
throughout the cold parts of the disk.  To find the ionization state
we need a model for the recombination chemistry.  A complex network of
chemical reactions is involved both in the gas phase and on the
surfaces of dust grains \citep{sem04}.  However in the presence of
grains, the results can be well-approximated with a simplified network
\citep{ilg06}.  We apply the simplified model put forward by
\citet{oku09a}.  His key ansatz is that the grains have an
approximately Gaussian charge state distribution.  This holds if the
grains are fluffy aggregates and (1) capable of holding large numbers
of elementary charges, and (2) only weakly electrically polarized by
approaching electrons and ions.  The grain charge state distribution
is then a function of a single dimensionless parameter $\Gamma$, the
electrostatic energy between a charged grain and an incident electron,
relative to the thermal energy.

Writing the equilibrium rate equations for the electrons and ions,
with the latter averaged over all ion species, and using
electron-grain and ion-grain collision cross-sections averaged over
the Gaussian grain charge state distribution, one can solve
simultaneously for the ion and electron densities in terms of the same
parameter $\Gamma$.  Requiring charge neutrality then yields a single
equation whose root is $\Gamma$.  We solve this equation, Okuzumi's
eq.~30, to obtain the number densities of the ions and electrons and
the mean and width of the grain charge distribution.  Our fractal
aggregates have dust-to-gas mass ratio $f_{dg}$ and are made up of
individual grains or monomers whose radius we choose between 0.005 and
5~$\mu$m.

The most abundant molecular ion is often HCO$^+$, since energetic
particles convert H$_2$ to H$_2^+$ which quickly reacts with CO.
However at temperatures high enough for metal atoms to desorb from the
grains and enter the gas phase, such molecular ions readily transfer
their charge to the metals.  The metal ions recombine more slowly and
hence end up more abundant.  As a representative metal we consider
magnesium following \citet{ilg06}.

The gas quickly reaches a charge equilibrium in the presence of the
fluffy aggregates, with electron and ion densities
\begin{equation}
  n_e=\frac{\zeta(m_d/m_n)}{f_{dg}s_eu_e\sigma}
  \frac{(\sqrt{1+2g}-1)}{g\exp(-\Gamma)}
\end{equation}
and
\begin{equation}
  n_i=\frac{\zeta(m_d/m_n)}{f_{dg}s_iu_i\sigma}
  \frac{(\sqrt{1+2g}-1)}{g(1+\Gamma)},
\end{equation}
where
\begin{equation}\label{eq:gfactor}
  g=\frac{2 c_t\zeta n_{\rm gas}}
  {s_iu_is_eu_e(\sigma\rho_{\rm gas}f_{dg}/m_d)^2}
  \frac{\exp(\Gamma)}{1+\Gamma}.
\end{equation}
Also $c_t$ is the mean gas-phase recombination rate, and the
dimensionless parameter $\Gamma=-\langle Z\rangle e^2/(ak_BT)$ depends
on the mean grain charge $\langle Z\rangle$ in units of the electron
charge $e$, the grain radius $a$ and geometric cross-section $\sigma$,
and the temperature $T$.  Other quantities are the electron and ion
thermal speeds $u_e, u_i$ and sticking probabilities on grains $s_e,
s_i$ for which we use Okuzumi's expressions.

The grain charge state distribution is the Gaussian
\begin{equation}
  n_{\rm dust}(Z) = \frac{n_{\rm dust}}{\sqrt{2\pi\langle\Delta Z^2\rangle}}
  \exp\left[-\frac{(Z-\langle Z\rangle)^2}{2\langle\Delta Z^2\rangle}\right]
\end{equation}
with spread
\begin{equation}
  \langle\Delta Z^2\rangle = \frac{1+\Gamma}{2+\Gamma} \frac{a k_B T}{e^2}
\end{equation}
\citep{oku09a}, where the total dust number density $n_{\rm dust}=\rho
f_{dg}/m_{\rm dust}$.  Approximating the grain charge state
distribution by the Gaussian yields an overall charge-neutral plasma
to high accuracy when the fluffy aggregates are made of at least a few
hundred monomers each.  In Appendix~B we show the net charge that
creeps in when the grains' charge spread is too small to be
represented with a Gaussian.

We compute the conductivity including the contributions from grains
across the charge state distribution, along with the contributions
from the electrons and ions.  The free parameters are the individual
grains' size and mass, the dust-to-gas ratio, the ionization rate and
the gas-phase recombination rate.  Our choices for the parameters
governing the ionization state are listed in Table~\ref{param}.  The
gas-phase recombination is further discussed in the next section.

\subsection{Gas-Phase Recombination Rate\label{sec:gasphaserecomb}}

A key weakness of this ionization recipe when applied over a large
range of disk locations is that it does not capture the transition
from one dominant ion to the next, and in particular between molecular
ions and metals which occurs at the metals' thermal desorption
temperature.  The recipe yields a total ion abundance using a
recombination rate averaged over all ion species $x$:
\begin{equation}
n_i=\sum_x n_x,\ \ \ c_t=\frac{\sum_x c_xn_x}{\sum_xn_x}.
\end{equation}
Here we focus on the transition between our representative metal ion
Mg$^+$, with number density $n_1$, and molecular ion HCO$^+$, with
number density $n_2$.  Under charge equilibrium, the metal ion number
density is given by
\begin{equation}
  \frac{\partial n_1}{\partial t}= n_2n_3\alpha  - c_1n_1n_e -
  s_iu_i\sigma n_{dust}n_1 = 0,
\end{equation}
where $n_3$ is the number density of metal atoms remaining neutral and
$n[{\rm Mg}]=n_1+n_3$ the total number density of metal atoms in the
gas phase.  Also $c_1=3\cdot 10^{-11}/\sqrt{T}$~cm$^3$~s$^{-1}$ is the
metal ions' radiative recombination rate,
$c_2=3\cdot{}10^{-6}/\sqrt{T}$~cm$^3$~s$^{-1}$ the molecular ions'
dissociative recombination rate and $c_3=3\cdot
10^{-9}$~cm$^3$~s$^{-1}$ the molecular-to-metal ion charge transfer
rate coefficient \citep{ilg06}.  Our task is to determine the
effective recombination rate coefficient (eq.~9 in \citet{oku09a}) by
finding the ratio of metal and molecular ion densities $y=n_1/n_2$ in
\begin{equation}
  c_t=\frac{1}{1+y}\left(c_2+yc_1\right).
\label{recrate}
\end{equation} 
First we determine the number density of thermally-desorbed metal
atoms in the gas phase $n[{\rm Mg}]$, which depends on temperature.
As described in Appendix~A we balance adsorption of Mg on grains with
thermal desorption.  The metal-to-molecular ion ratio and gas
ionization state are obtained by applying Okuzumi's recipe first
assuming all the gas-phase metal atoms are ionized, and then
iteratively correcting $c_t$ to get the final $y$.

\begin{figure}
\begin{center}
\includegraphics[width=4.6in]{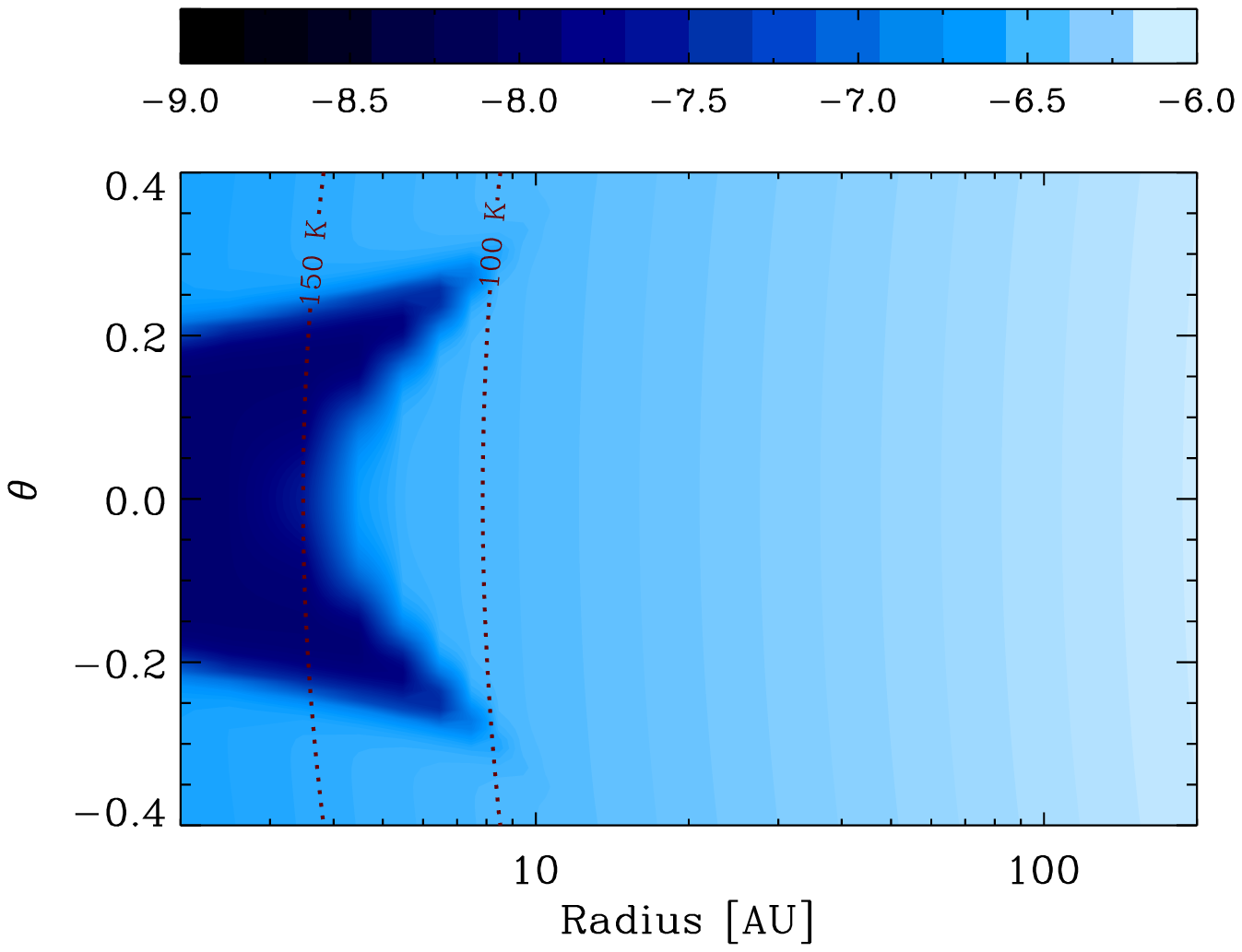}
\includegraphics[width=4.6in]{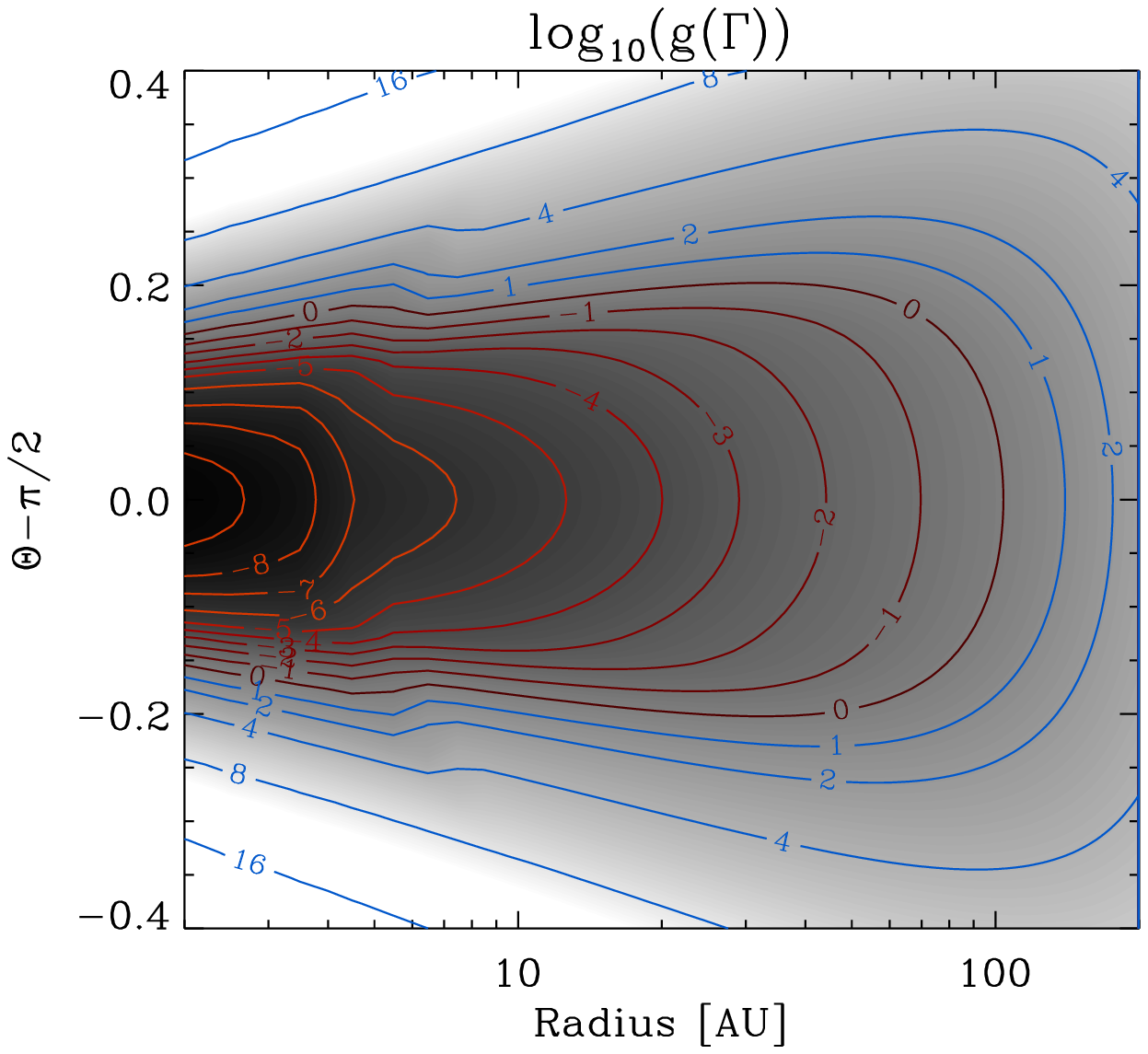}
  \caption{Left panel: logarithm of the total gas-phase recombination
    rate coefficient $c_t$ in the fiducial model after
    eq.~\ref{recrate}, in the $(r,\Theta-\pi/2)$ plane.  Dotted lines
    mark where the temperature is 150~K (left) and 100~K (right).
    Right panel: logarithm of $g(\Gamma)$; gas-phase recombination
    dominates where $g(\Gamma)\gg 1$ (contour lines drawn in blue).}
\label{neal1-0}
\end{center}
\end{figure}

In the fiducial model disk, the ion Mg$^+$ proves to be the main
source of free electrons at the midplane inside 5~AU.  We introduce
the concept of the {\it metal line}, the location at which a metal
such as magnesium freezes out.  In our fiducial model, the metal line
lies entirely inside 8~AU where $T\simeq 100$~K (Fig.~\ref{neal1-0},
top).

In Fig.~\ref{neal1-0} (bottom) we show contours of the dimensionless
factor $g(\Gamma)$ introduced in eq.~\ref{eq:gfactor}.  For
$g(\Gamma)\gg 1$ gas-phase recombination dominates.  The isolines of
$g(\Gamma)$ follow the gas density isolines in the outer disk
($r>50$~AU), but in the inner disk this is no longer the case.  Still
it is safe to say that for gas densities $\rho<3\times 10^{-15}\rm\ g
cm^{-3}$ the recombination occurs mostly in the gas phase rather than
on the grains.  In the outer disk, or in the inner disk above
5~pressure scale heights, the gas ionization degree is therefore
insensitive to the dust properties.

\begin{table}
\caption{Ionization parameters in the fiducial disk model}
\label{param}
\vspace*{3mm}
{\scriptsize
\begin{tabular}{|l|c|}
\hline
  Value  &  Units  \\
\hline
$\zeta_{RN} = 7\cdot 10^{-19}(f_{dg}/10^{-2})$ & $\rm s^{-1}$ \\
$\zeta_{CR0} = 5\cdot 10^{-18}$ & $\rm s^{-1}$ \\
$\zeta_{1(XR)} = 6\cdot 10^{-12}$ & $\rm s^{-1} $ \\
$\zeta_{2(XR)} = 1\cdot 10^{-15}$ & $\rm s^{-1} $ \\
$N = 400$  &  \\
$a = 20  a_0 $ & cm \\
$a_0=10^{-5}$  &  cm \\
$f_{dg}=10^{-4}$ & \\
$\sigma = \pi a^2 $ & $\rm cm^2$  \\% ;! (see after eq. 51)
$\sigma_0=\pi a_0^2$ & $\rm cm^2$ \\
$s_i = 1 $ & \\
$s_e = 0.3 $ & \\
$d_0= 1.4 $ (ice) & $\rm g\ cm^{-3} $ \\ %;1.4  ;
$m_{\rm dust} = \frac{4}{3}\pi a^2 a_0 d_0 $ & g \\%in g cm^-3
$m_n = 2.3 m_{\rm p} $ & g \\%    ;2.34d0*1.00794/6.022d+24
$m_{\rm metal}=24 m_p$, $m_{\rm HCO }=29 m_p$ & g\\
$c_3 = 3\times 10^{-9} $ &  $\rm cm^3\ s^{-1}$    \\
$c_1= 3\times 10^{-11} $ (Mg)  & $\rm cm^3\ s^{-1}$  \\
$c_2= 3\times 10^{-6}  $ (HCO) & $\rm cm^3\ s^{-1}$ \\
\hline
\end{tabular}
}
\end{table}

\section{Magneto-Rotational Turbulence}

The gas couples to the magnetic fields, and gas movements can bend the
field lines, if the fields diffuse slower than they are advected.  For
protostellar disks, various bending length scales and advection time
scales have been suggested.  If we take the length scale to be the gas
scale height $H$, and the time scale to be the differential rotation
or shear time $\Omega^{-1}$, then the induction and diffusion terms in
the field evolution equation, $\nabla\times(\vec{v}\times\vec{B})$ and
$\nabla\times(\eta\nabla\times B)$, are roughly $\Omega B$ and $\eta
B/H^2$, leading to the condition $c_s^2/(\eta\Omega)>1$ (eq.~35 in
\citet{war07}).

However the turbulence involves structures smaller than the gas scale
height.  The magnetic field is sub-equipartition with the gas, so the
Alfv\'en speed $v_A$ is less than the sound speed $c_s$ and the MRI
wavelength $v_{Az}/\Omega$ is less than the gas scale height
$c_s/\Omega$.  Furthermore the poloidal field components $B_\Theta$
and $B_r$ are weaker than the azimuthal component in 3-D MRI
turbulence calculations \citep{mil00}, allowing the poloidal
components to be bent into sharper curves.  This is borne out by the
observation that the turbulent correlation lengths are anisotropic,
with eddies reaching size $H$ only in the azimuthal direction
\citep{gua11,flo12}.

MRI turbulence under stratified conditions is probably sustained by
the cycle in which (a) $B_\phi$ is buoyant and its rise generates
$B_\Theta$, (b) from $B_\Theta$ the MRI produces $B_r$, and (c) the
shear makes azimuthal field again from $B_r$ \citep{tou92,hir11}. 
%(Tout \& Pringle 1992 MNRAS 259, 604; Hirose \& Turner 2011 ApJ 732, L30).  
Step (b) in
particular creates magnetic structures with a short vertical
wavelength and may be the first step to shut down as the diffusivity
is increased.

The appropriate criterion for sustained MRI turbulence is therefore
that the fields diffuse across the MRI vertical wavelength in less
than the MRI growth time.  This corresponds to the dimensionless
Elsasser number
\begin{equation}
  \Lambda\equiv\frac{v_{Az}^2}{\eta\Omega}>1,
\label{els1}
\end{equation}
where the Alfv\'en speed contains only the vertical component of the
magnetic field.  We use the Elsasser number criterion through the rest
of this paper.  Numerical calculations with an Ohmic diffusivity 
show that weak MRI turbulence operates down to
\begin{equation}
  \frac{v_{Az}^2}{\eta\Omega}\simeq 0.1,
\label{els0.1}
\end{equation}
with larger, slower eddies and lower turbulent stresses
\citep{san99,san04,sim09}.  A similar condition holds for the
ambipolar diffusivity \citep{bai11c}.  We calculate the Elsasser
number $\Lambda$ in eqs.~\ref{els1} and~\ref{els0.1} including the
Ohmic and ambipolar magnetic diffusivities, $\eta=\eta_O+\eta_A$.  We
neglect the Hall diffusivity in the MRI turbulence criterion because
numerical results show it has weak effects when not more than two
orders of magnitude greater than the Ohmic diffusivity \citep{san02b}.

\begin{figure}
\begin{center}
\includegraphics[width=5.6in]{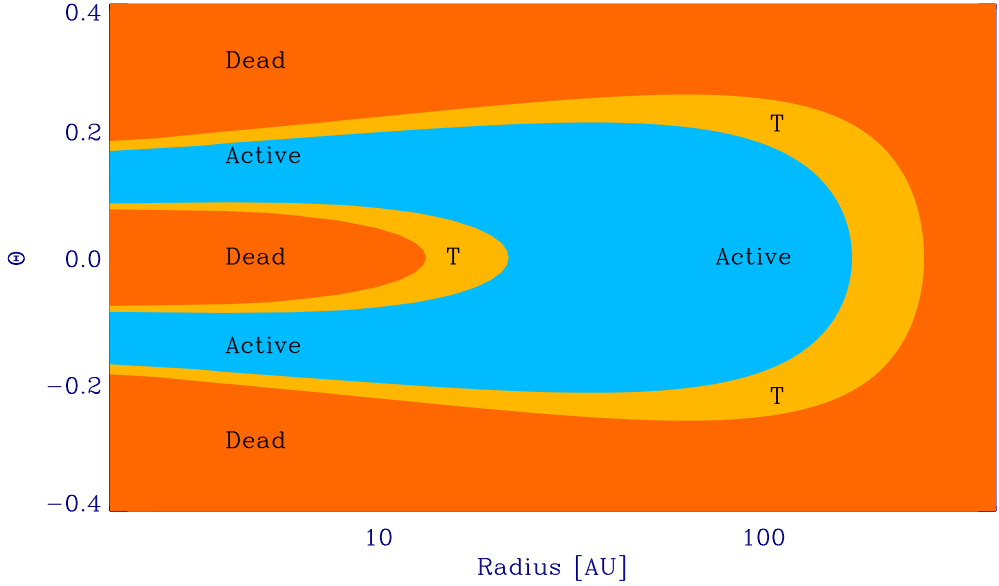}
  \caption{Schematic representation of the MRI-active region (blue),
    the transitional region labeled `T' (yellow) and the magnetically
    dead zone (orange), based on evaluating the Elsasser number in
    eq.~27.  }
\label{AplusO}
\end{center}
\end{figure}

In Fig.~\ref{AplusO} we present schematically the dead zone and
MRI-active region, separated by a transitional zone where
$0.1\leq\Lambda\leq 1$.  In classical layered disk models
(e.g. \citet{arm01}) the disk consists of active and dead zones only.
Here we further divide the dead zone into the transitional and
more-thoroughly dead parts.  The transitional part will prove
important for trapping solids near the dead zone's edge.

An important consequence of using the MRI wavelength rather than the
density scale height is that the resulting Ohmic dead zone varies in
size depending on the magnetic field strength.  Ambipolar diffusion in
contrast yields an Elsasser number independent of the magnetic field
strength, since both the diffusivity and the squared Alfv\'en speed
are proportional to the magnetic pressure.

We impose one more condition for MRI turbulence: since the instability
grows quickly only on sub-thermal magnetic fields \citep{kim00},
%(Kim \& Ostriker 2000 ApJ 540, 372),
we require that the plasma beta $\beta=8\pi P_{\rm gas}/B^2 \geq 1$.

\subsection{Magnetic Diffusivity}

The electrons, ions and charged grains all contribute to the
conductivity, while coupling to the magnetic field in varying degrees.
A charged particle $x$ has cyclotron frequency
\begin{equation}
  \omega_x=|q_x|eB/(m_xc),
\end{equation}
where $q_xe$ is the charge and $m_x$ the mass.  The damping time
of the charged particle's motion relative to the neutrals is
\begin{equation}
  \tau_x=\frac{m_n+m_x}{m_n}\frac{1}{n_n<\sigma v>_{xn}},
\end{equation}
where $\langle\sigma v\rangle_{xn}$ is the momentum-transfer rate
coefficient for colliding with neutrals, averaged over the Maxwellian
velocity distribution.  We adopt $\langle\sigma v\rangle_{en} =
10^{-15}\sqrt{128 k_BT/(9\pi m_e)}$ for the electrons, and
$\langle\sigma v\rangle_{in} = 1.9\times 10^{-9}$ for the ions
\citep{dra83}.  The corresponding rate coefficient for the dust grains
follows eq.~21 in \citet{war99}, where the collision speed is the
neutrals' thermal speed since the grains' drift through the gas is
subsonic.  Putting together the cyclotron frequency and the
collisional damping time, we obtain the parameter $b_x=\tau_x\omega_x$
that describes how well the charged particle couples to the magnetic
field.  For electrons, typically $b_e\gg 1$.  The quality of coupling
to the magnetic fields drops with the mass of the particle.  The ions
are more than 4~orders of magnitude heavier than the electrons, and
the grains are orders of magnitude heavier still.

We treat the magnetic diffusivity as outlined by \citet{war07},
computing the electric currents from the number densities of charged
particles obtained with Okuzumi's recipe.  The current density in the
partially-ionized gas is
\begin{equation}
  \vec{J}=\sigma_O\vec{E}_{\parallel}
  +\sigma_H\hat{\vec{B}}\times{}\vec{E}_\perp
  +\sigma_P\vec{E}_\perp
\end{equation}
where $\sigma_O$, $\sigma_H$, $\sigma_P$ are the Ohmic, Hall and
Pedersen conductivities \citep{cow76,war99},
\begin{equation}\label{eq:sigmaO}
  \sigma_O=\frac{ec}{B}\sum_xn_x|q_x|b_x,
\end{equation}
\begin{equation}\label{eq:sigmaH}
  \sigma_H=-\frac{ec}{B}\sum_x\frac{n_x q_xb_x^2}{1+b_x^2},
\end{equation}
\begin{equation}\label{eq:sigmaP}
  \sigma_P=\frac{ec}{B}\sum_x\frac{n_x|q_x|b_x}{1+b_x^2}.
\end{equation}
The induction equation
\begin{equation}
  \frac{\partial\vec{B}}{\partial t} = \nabla \times{} 
  (\vec{u}\times{} \vec{B})
  -\nabla \times{} \left(
    \eta_O\nabla \times{} \vec{B} + 
    \eta_H(\nabla \times{} \vec{B})\times{} \hat{\vec{B}} +
    \eta_A(\nabla \times{} \vec{B})_\perp 
  \right)
\end{equation}
then has diffusivities
\begin{equation}\label{eq:eta}
  \eta_O=\frac{c^2}{4\pi\sigma_O},
  \ \eta_H=\frac{c^2\sigma_H}{4\pi\sigma_\perp^2}\ \ {\rm and}\
  \ \eta_A=\frac{c^2\sigma_P}{4\pi\sigma_\perp^2}-\eta_O,
\end{equation}
with $\sigma_\perp=\sqrt{(\sigma_H^2+\sigma_P^2)}$.

\subsection{Accretion Rate}

Given the spatial distribution of the Elsasser number through the
model disk, we estimate the mass accretion rate $\dot{M}(R)$ by
summing the effective accretion rates within the magnetically-active,
transitional and dead layers,
\begin{equation}
  \dot{M} = 3\pi c_sH (\Sigma_{\rm ac} \alpha_{\rm ac}+\Sigma_{\rm T}
  \alpha_{\rm T} +\Sigma_{\rm dead} \alpha_{\rm dead}),
\end{equation}
where subscripts $*=$``ac'', ``T'' and ``dead'' indicate the three
respective layers, $\Sigma_*$ are the layers' mass columns and
$\alpha_*$ are the corresponding stress parameters.

There is ample evidence that MRI turbulence yields an accretion stress
$w_{R\phi}$ proportional to the magnetic pressure, that is,
$\alpha_{ac} \equiv w_{R\phi}/P_{\rm gas} \propto 1/\beta$, as long
as $\beta>1$ and the gas dominates the total pressure
\citep{haw95,bai11}.  In vertically-stratified disks, the stress
parameter $\alpha$ increases with height as the local plasma beta
decreases \citep{mil00}, and the turbulent layers closest to the
disk's magnetically-dominated corona transport significant mass
\citep{fle03,tur10,oku11}.  Once the plasma beta falls below unity,
the stress drops off quickly \citep{mil00}.

Since our magnetic pressure is independent of height, we make the
stress uniform in the magnetically-active layer.  We let the stress
fall smoothly with decreasing plasma beta below a floor value
$\beta_f$:
\begin{equation}\label{eq:alphastep}
  \alpha_{ac}=
  \begin{cases}
    \alpha_0 &\text{if $\beta\geq\beta_f$}\\
    \exp\left(-\frac{2}{(\beta+\beta_f-1)}\right) &\text{if
      $\beta<\beta_f$}.
  \end{cases}
\end{equation}
For a fair match to shearing-box results we choose the floor value
$\beta_f=1.85$.  The magnetically-active stress parameter $\alpha_0$
is inversely proportional to the density, $\alpha_0=\frac{\rho_{\rm
    mid}(r)}{400 \rho(r,\Theta)}$, where $\rho_{\rm mid}(r)$ is the
midplane gas density.

Our stress parameter in the transitional layer $\alpha_{\rm T}$
behaves in the same manner with the disk height but is an order of
magnitude weaker, having $\alpha_{\rm T}=\alpha_{\rm ac}/10$.  In the
dead zone, a non-zero Reynolds stress leads to $\alpha_{\rm
  dead}=\alpha_{\rm 0, mid}/10^2$, where $\alpha_{\rm 0, mid}$ is the
magnetically-active value of the stress parameter at the midplane
\citep{dzy10}.

We investigate two approaches to the accretion stresses in disk
regions with poor magnetic coupling.  In the first we simply make the
stress parameter a step function across the Elsasser number threshold
\citep{mar12a,mar12b,arm01}.  However, shearing-box MHD parameter
surveys show a smooth roll-off in the stress at Elsasser numbers below
unity \citep{san04}, suggesting this simple approach might not capture
all the accretion flow.

We therefore also consider a second approach in which the stress
parameter varies smoothly with the Elsasser number.  We use the upper
bound on $\alpha(\Lambda)$ found in unstratified shearing-box
ambipolar-MHD calculations by \citet[][their Fig.~4]{bai11}.  Focusing
on their results with net vertical magnetic flux corresponding to
plasma beta~400, and adding a term for the non-zero dead zone Reynolds
stress, we obtain the approximate fitting formula
\begin{equation}
  \alpha = \alpha_{\rm dead} + \alpha_0\frac{f(\Lambda)}{f(\Lambda=10)},
\end{equation}
where 
\begin{equation}
  \log_{10}f(\Lambda)=-4+\frac{k}{|k|}\sqrt{3k},
\end{equation}
and $k=\log_{10}(\Lambda+1)$.  This results in a smoothly-varying
turbulent stress parameter.  The two approaches yield identical
accretion stresses for Elsasser number $\Lambda=10$, but at other
levels of magnetic coupling, the step-function under- or
over-estimates the accretion rate: $\dot{M}^{\rm step} > \dot{M}^{\rm
  smooth}$ for $1<\Lambda<10$, and $\dot{M}^{\rm step} < \dot{M}^{\rm
  smooth}$ for $10<\Lambda<100$.  Finally we require that the stress
parameter not exceed 0.95, at which point transonic turbulent speeds
would lead to rapid shock dissipation.

%%%%%%%%%%%%%%%%%%%%%%%%%%%%%%%%%%%%%%%%%%%%%%%%%%%%%%%%%%%%%

\section{Results}

In this section we present the dead zone calculated for the fiducial
model disk and then vary the parameters in turn.  The parameter study
is summarized in Table~\ref{sets}.

\begin{table}
\begin{center}
  \caption{Sets of models in the parameter study.  The rows are the
    monomer radius $a_0$, the dust-to-gas mass ratio in the fluffy
    aggregates $f_{dg}$, the disk mass in units of the stellar mass
    $M_d/M_*$, the surface density power-law index $p$, the
    temperature at 1~AU $T_0$, the stellar mass in Solar masses
    $M_*/M_\odot$, the cosmic ray ionization rate outside the disk
    $\zeta_{CR0}$ (eq.~6), and the ionization rate due to radionuclides
    $\zeta_{RN}$ (eq.~9). }
\label{sets}
{\scriptsize
\begin{tabular}{|c|c|c|c|c|c|c|c|c|}\hline
 Models  & fiducial & Set 1 & Set 2 & Set 3 & Set 4 & Set 5 & Set 6 & Set 7 \\ 
\hline
$a_0\rm\ [\mu m]$  & 0.1 & $[5 \to 0.001]$ & 0.1 & 0.1 & 0.1 & 0.1 & 0.1 & 0.1 \\
$f_{dg}$  & $10^{-4}$  & $10^{-4}$ & $[10^{-2} \to 10^{-7}]$ & $10^{-4}$ &
  $10^{-4}$ & $10^{-4}$ & $10^{-4}$  & $10^{-4}$ \\
$M_d/M_*$ & 0.064 & 0.064 & 0.064 &  $[0.2 \to 1.5]\times{}$ & 0.064 & 0.064 &
  0.064& 0.064 \\
& & & & $\times{}0.064$ & & & & \\ 
$p$ & 0.9 & 0.9 & 0.9 & 0.9  & $[0.5 \to 1.5]$ & 0.9 & 0.9 & 0.9 \\
$T_0 \rm\ [K]$& 280 & 280 & 280 & 280 & 280  &  $[150 \to 600]$ &
  $[446 \to 650]$ & 280 \\
$M_*/M_\odot$ & 1 & 1 & 1 & 1 & 1 & 1 & $[0.4 \to 2]$  & 1 \\
$\zeta_{CR0}$  $[s^{-1}]$ & $5\cdot 10^{-18}$ & $5\cdot 10^{-18}$ &
  $5\cdot 10^{-18}$ & $5\cdot 10^{-18}$ & $5\cdot 10^{-18}$ & $5\cdot 10^{-18}$ &
  $5\cdot 10^{-18}$ & $5\cdot [10^{-18}\to 10^{-15}]$ \\
$\zeta_{RN}$  $[s^{-1}]$ & $7\cdot 10^{-21}$  &  $7\cdot 10^{-21}$ & $7\cdot [10^{-19}\to 10^{-24}]$ &  $7\cdot 10^{-21}$ &  $7\cdot 10^{-21}$ &  $7\cdot 10^{-21}$  &  $7\cdot 10^{-21}$  &  $7\cdot 10^{-21}$ \\
\hline
\end{tabular}
}
\end{center}
\end{table}

\begin{figure}
\begin{center}
{\hbox{
\includegraphics[width=3.5in]{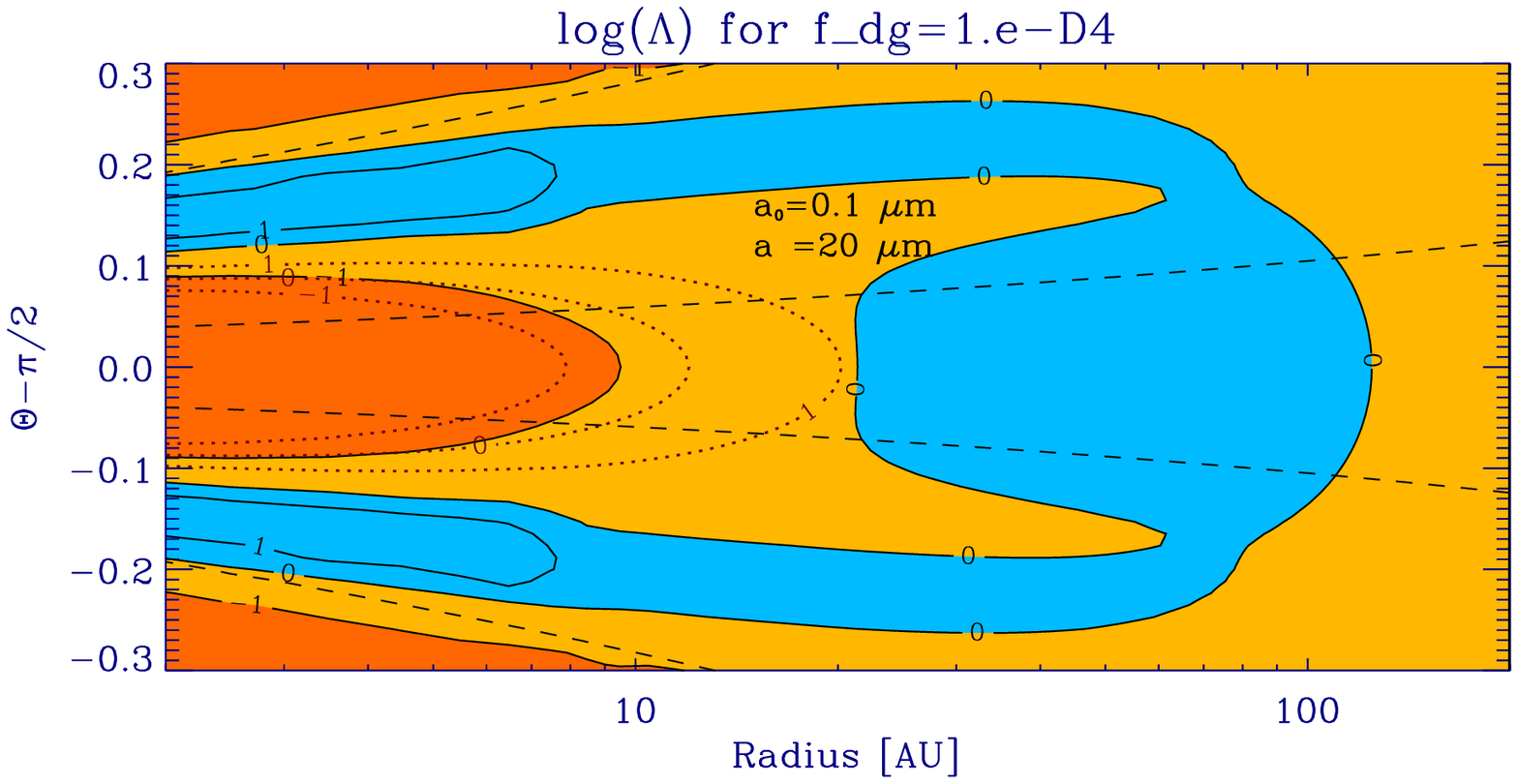}
\includegraphics[width=3.5in]{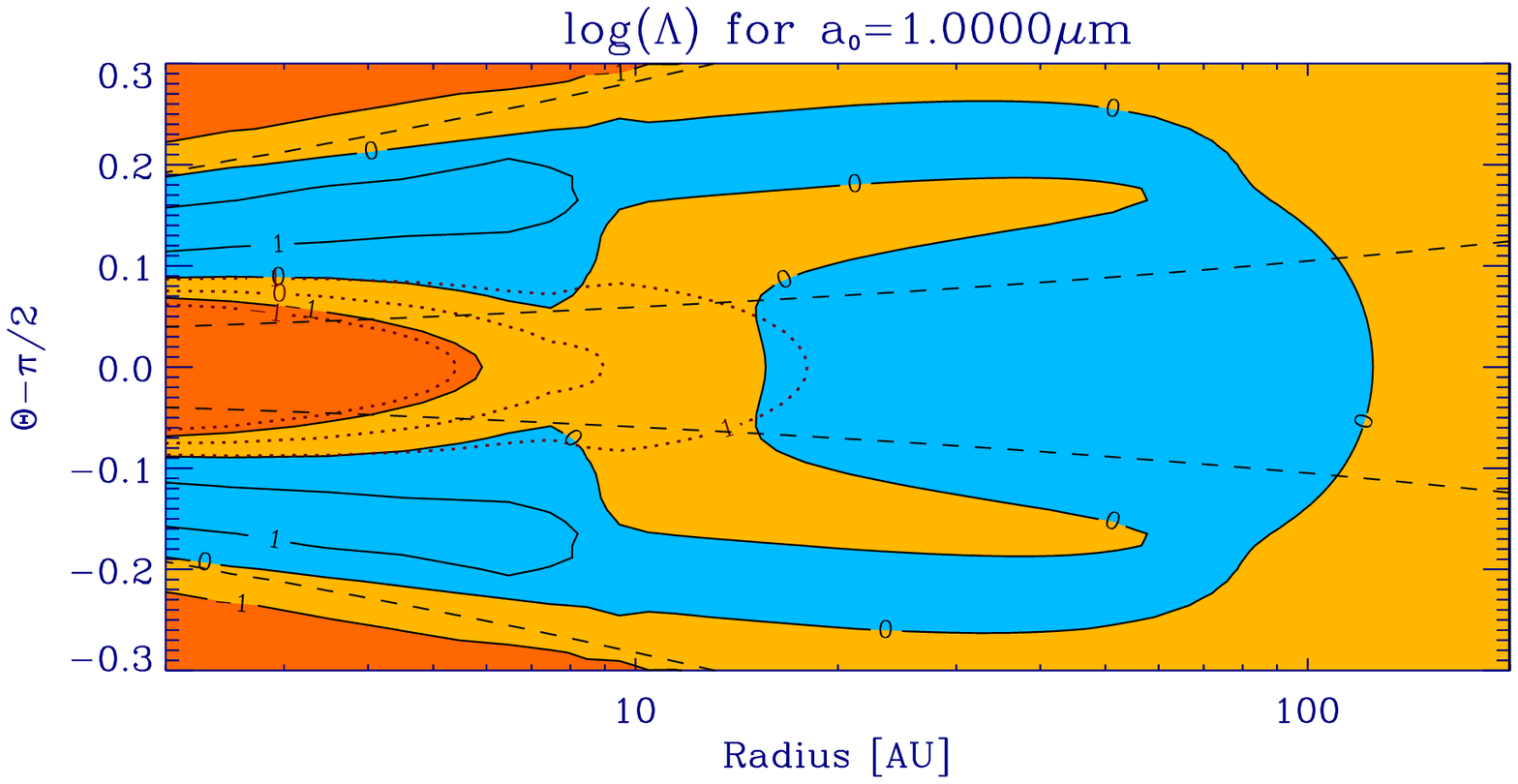}}
}
{\hbox{
\includegraphics[width=3.5in]{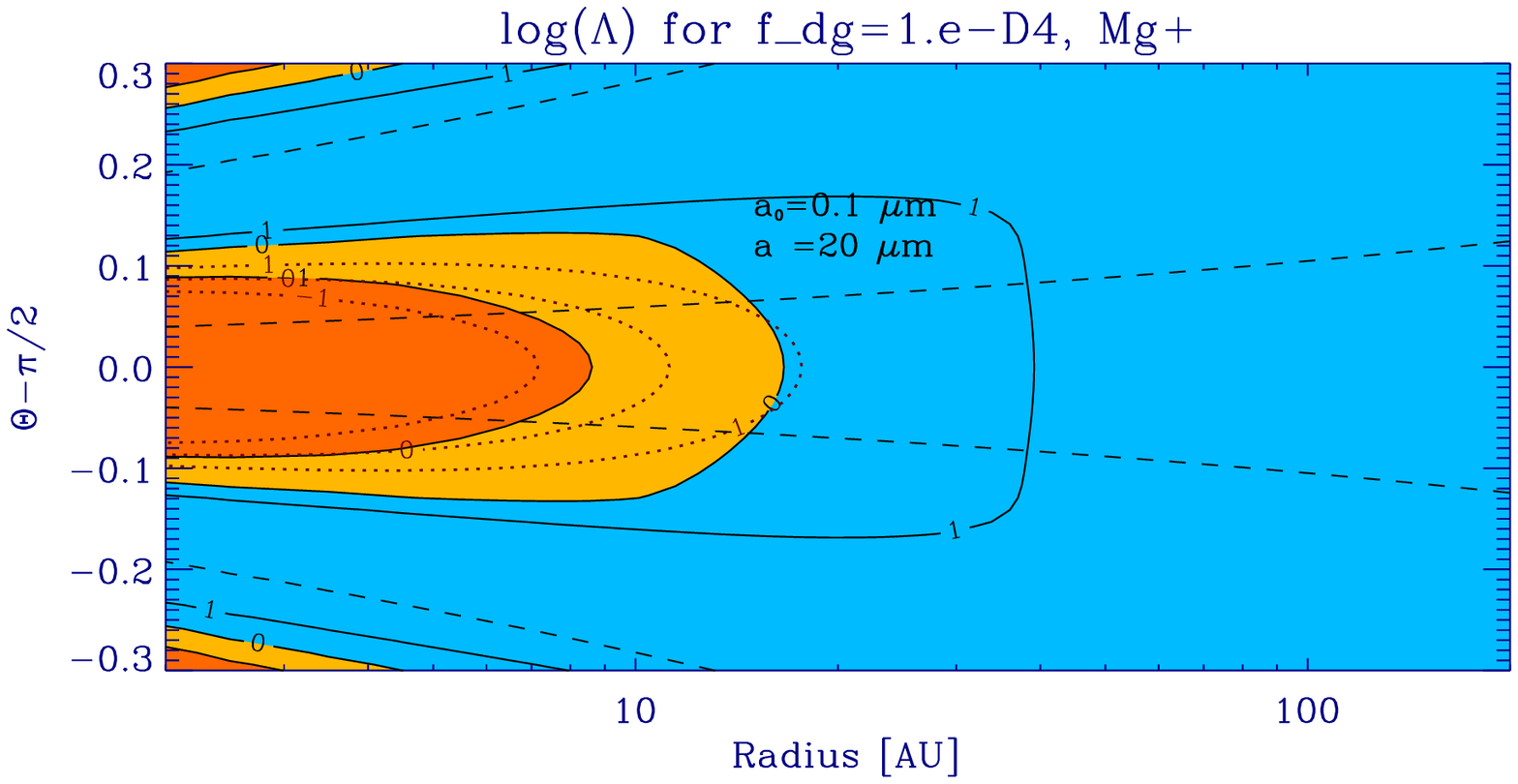}
\includegraphics[width=3.5in]{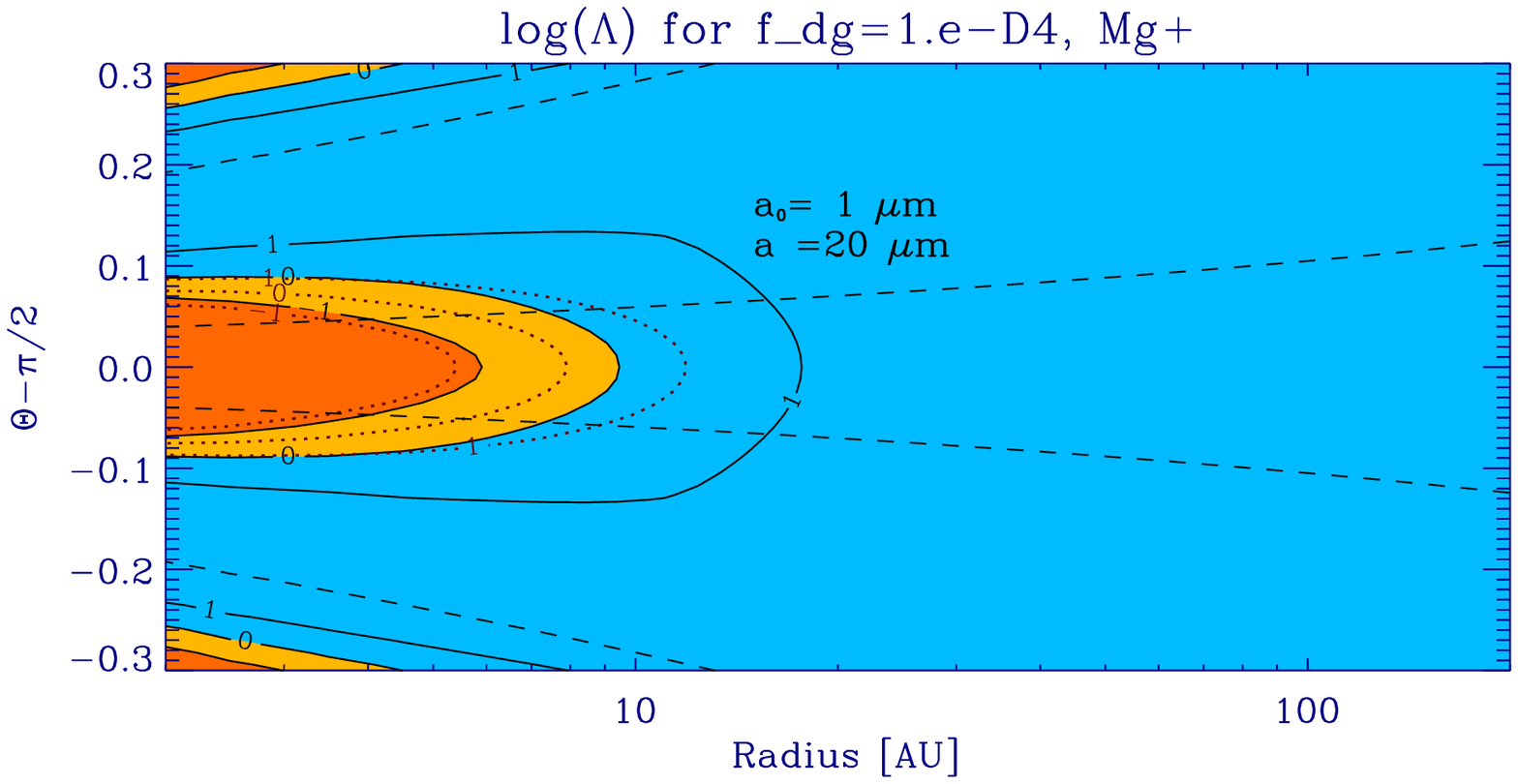}}
}
{\hbox{
\includegraphics[width=3.5in]{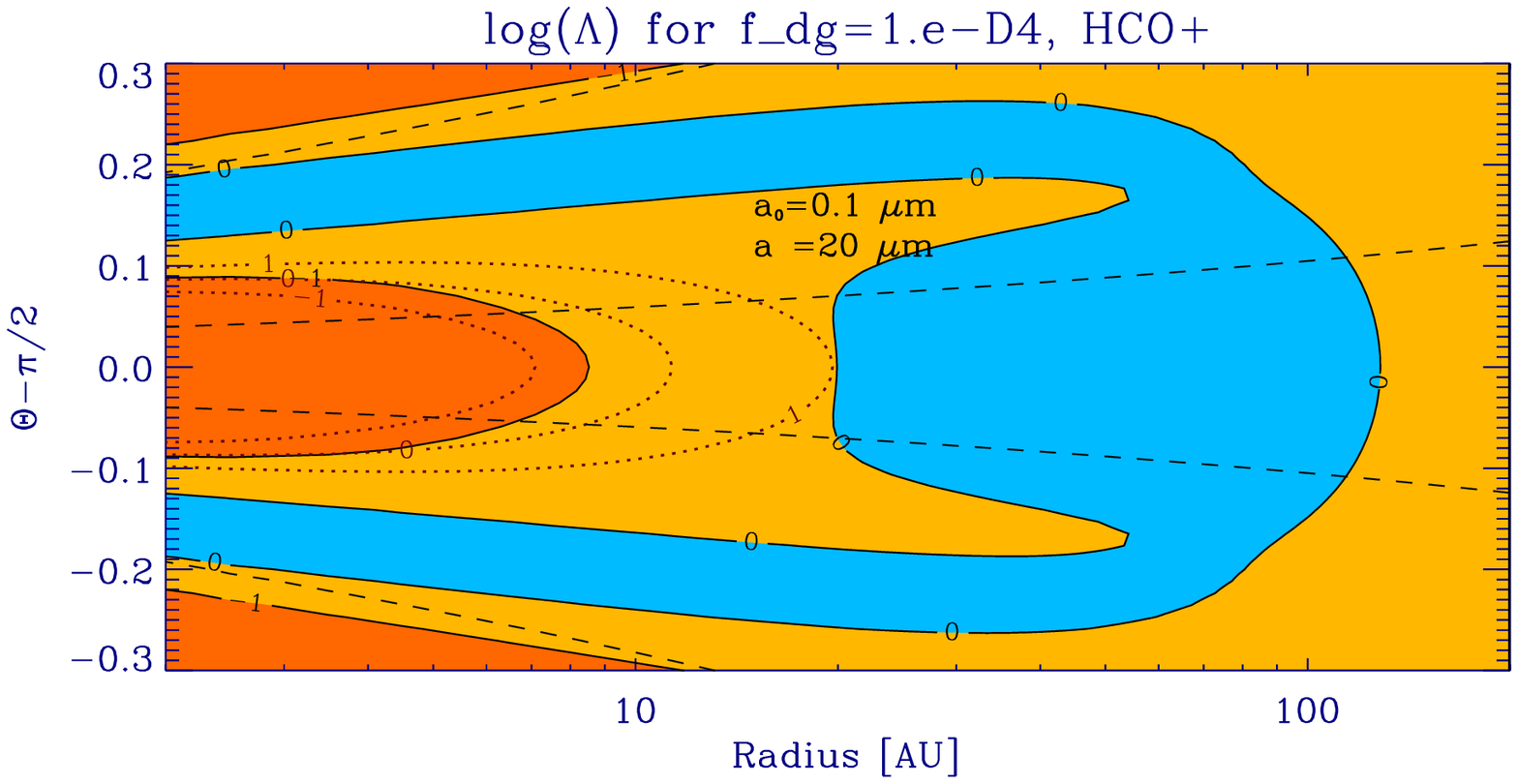}
\includegraphics[width=3.5in]{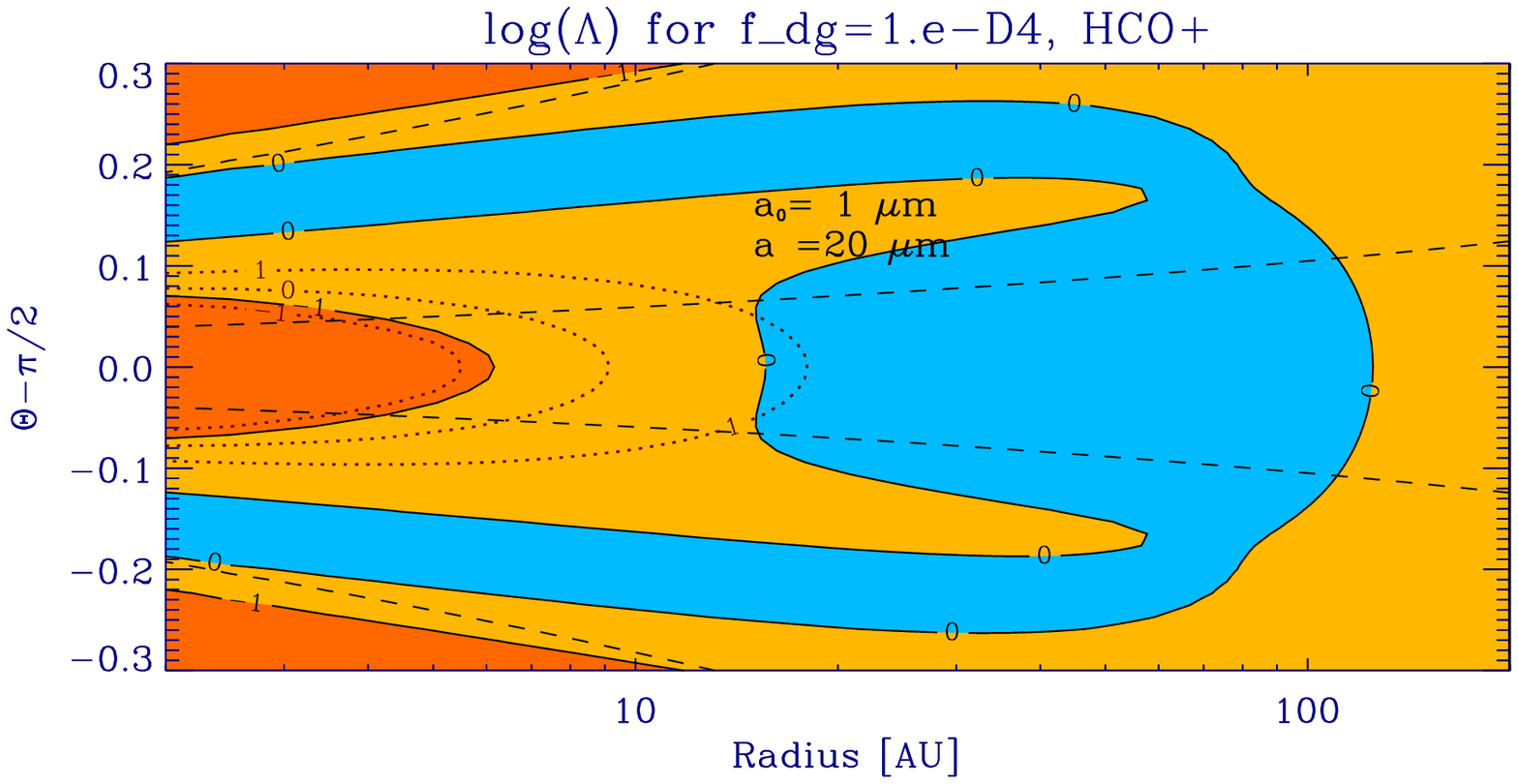}}
}
\caption{Dead, transitional and MRI-active zones for the
    fiducial model (left column) and a version with larger monomers
    $a_0=1\rm\ \mu{}m$ (right column).  Blue regions are MRI-active,
    yellow are transitional and orange are MRI-dead.  The top row
    includes the transition between the Mg$^+$ and HCO$^+$ ions, the
    middle row has Mg$^+$ everywhere and the bottom row has HCO$^+$
    everywhere.  Black solid contours indicate where the overall
    Elsasser number is 0.1, 1 and 10.  Black dotted contours are the
    corresponding surfaces for the Elsasser number computed from Ohmic
    diffusivity alone.  Black dashed lines mark heights $H$ and $5H$. } 
\label{fi-els}
\end{center}
\end{figure}

The fiducial model's dead zone is shown in Fig.~\ref{fi-els} along
with a version in which the monomers are enlarged to 1~$\mu$m, keeping
the dust-to-gas ratio fixed.  The Elsasser number divides the disk
into dead and MRI-active regions.  We further divide the dead zone
into a thoroughly-dead part and the ``transitional'' zone, where the
Elsasser number is less than one order of magnitude below the
threshold for turbulence.  Consider first the top left panel in
Fig.~\ref{fi-els}.  From the midplane to a few scale heights above,
the Elsasser number changes by several orders of magnitude.  By
contrast the radial gradients are gradual, with the transitional
region extending in the midplane from 8 to 20~AU.  The conditions for
MRI turbulence are satisfied at the midplane between 20 and 128~AU.
The dead zone defined by Ohmic diffusion alone, using the Ohmic
Elsasser number $\Lambda_O=v_{Az}^2/(\eta_O\Omega)$, has its customary
almond shape \citep{san00}, indicated in Fig.~\ref{fi-els} by dotted
lines where $\Lambda_O=0.1$, 1 and 10.  When the ambipolar diffusion
is also taken into account, the dead zone is much larger and includes
a transitional region whose outer edge is shaped like a fish's tail.
This shape is sensitive to the disk parameters. The fish's tail fins
are made by the ambipolas diffusion's dependence on the ion density.
There is a gentle dip in the ion density at 2 scale heights.
If  the ion densities were overall just 20\% lower, 
the ambipolar diffusion would decouple the whole interior.
The fish's tail shape is stronger still if the aggregates are made up of bigger
monomers (Fig.~\ref{fi-els}, top right).  The bottom plots in
Fig.~\ref{fi-els} show the importance of treating the transition from
metal to molecular ion.  Using Mg$^+$ throughout leads to dramatically
overestimating the magnetic activity.  On the other hand, HCO$^+$
alone accurately describes the dead zone's outer edge but yields
Elsasser numbers an order of magnitude too low in the MRI-active
layers nearer the star.  The transition between the two ions occurs at
radii $4\leq r \leq 9$~AU, where the temperatures are 150~K to 100~K
(see Fig.~2). 
% ref2201
This transition leads to the jump in the accretion rates. This happens 
because the Elsasser number is roughly one order of magnitude larger when 
the metal ions are present in the MRI-active layers. Thus, the model
with only Mg$^+$ ion produces an accurate estimate for accretion rate for $r<5 AU$, 
and the 'HCO$^+$'-model should be used for the disk outside $r>9$AU (see Fig.~\ref{gasphasem} 
in Appendix~\ref{gasphasemetal}).

\subsection{Characterizing Ambipolar Diffusion \label{sec:Rei}}

\begin{figure}
\begin{center}
\hbox{
\includegraphics[width=2.8in]{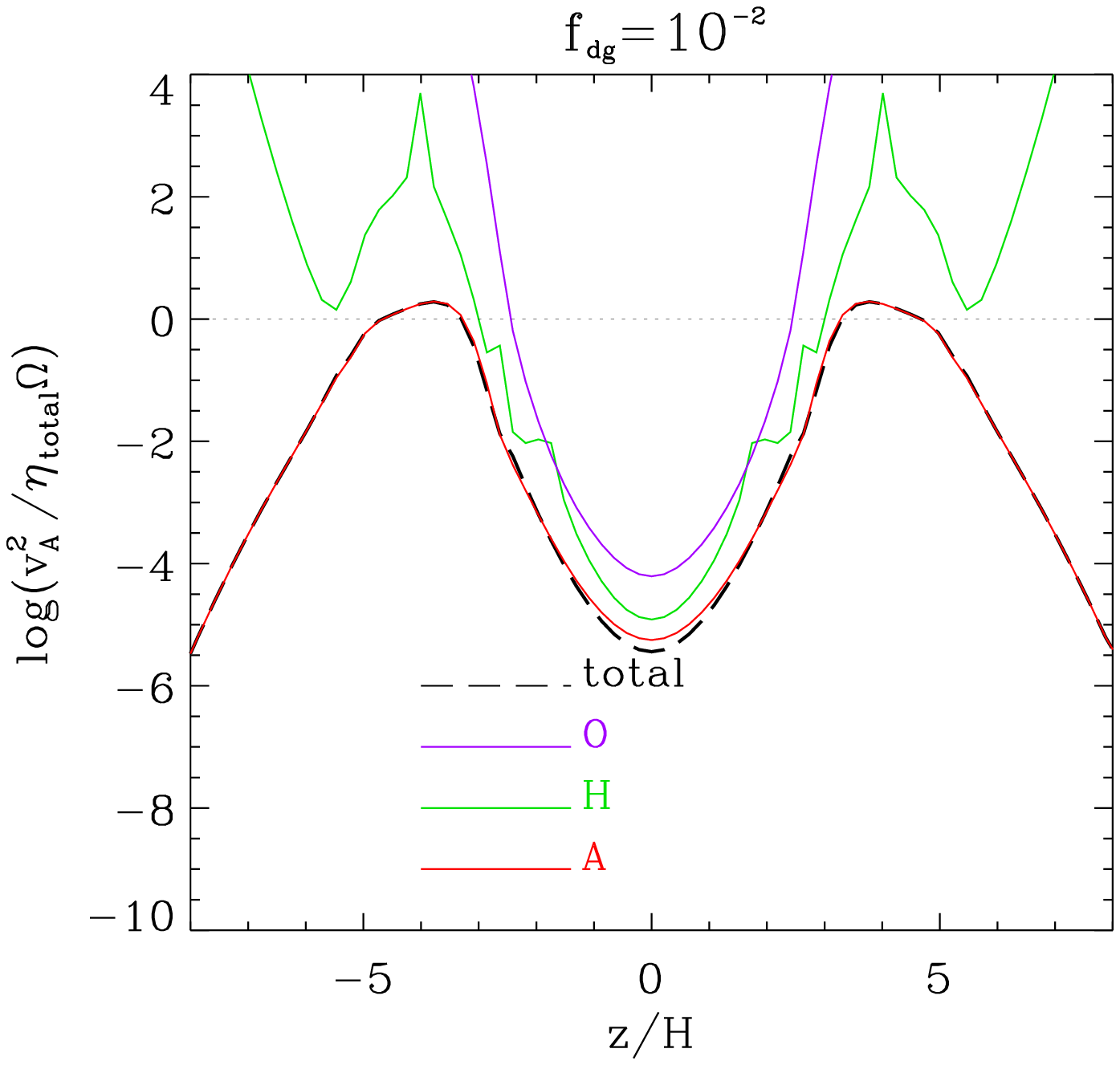}
\includegraphics[width=2.8in]{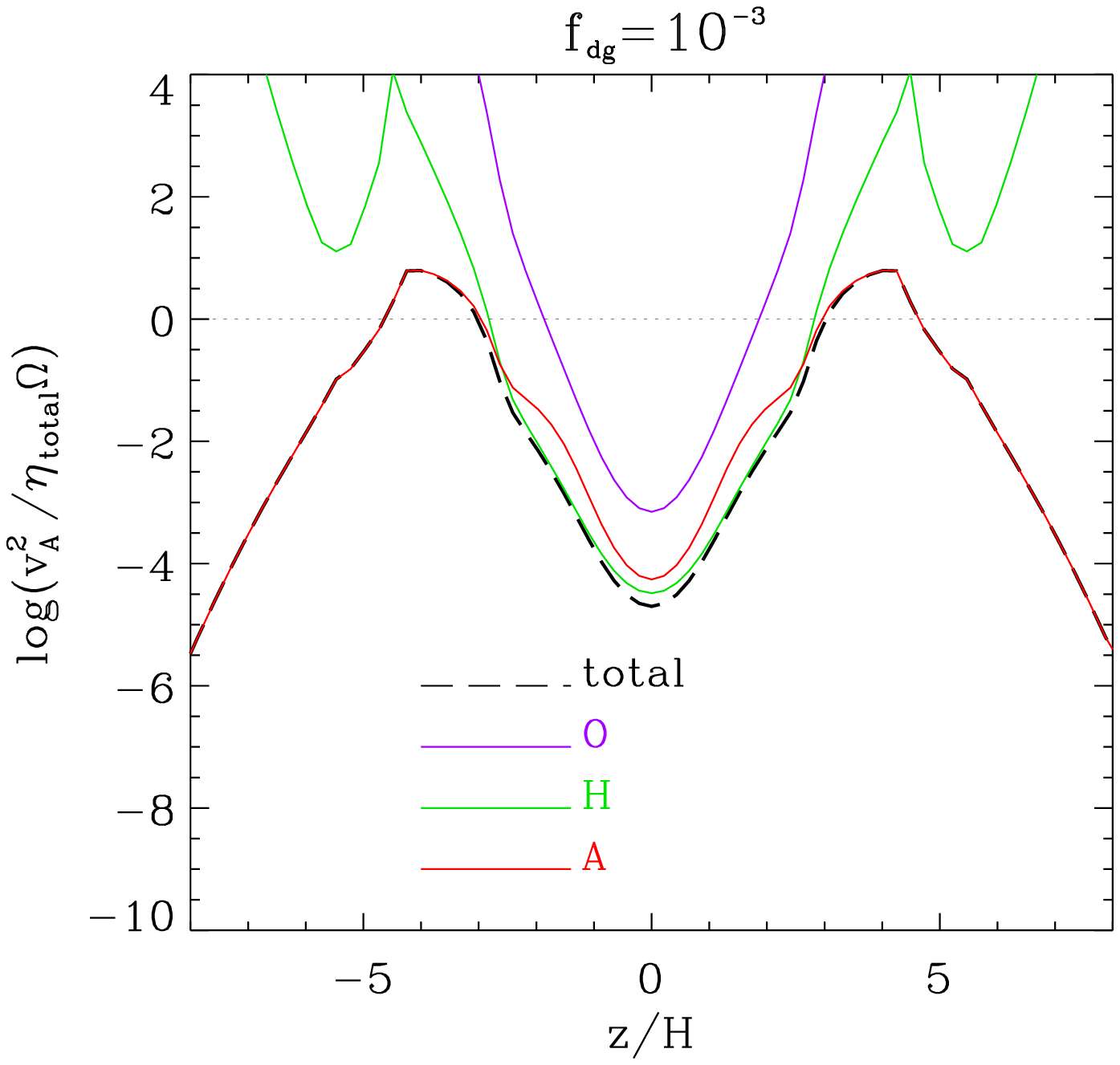}
}
\hbox{
\includegraphics[width=2.8in]{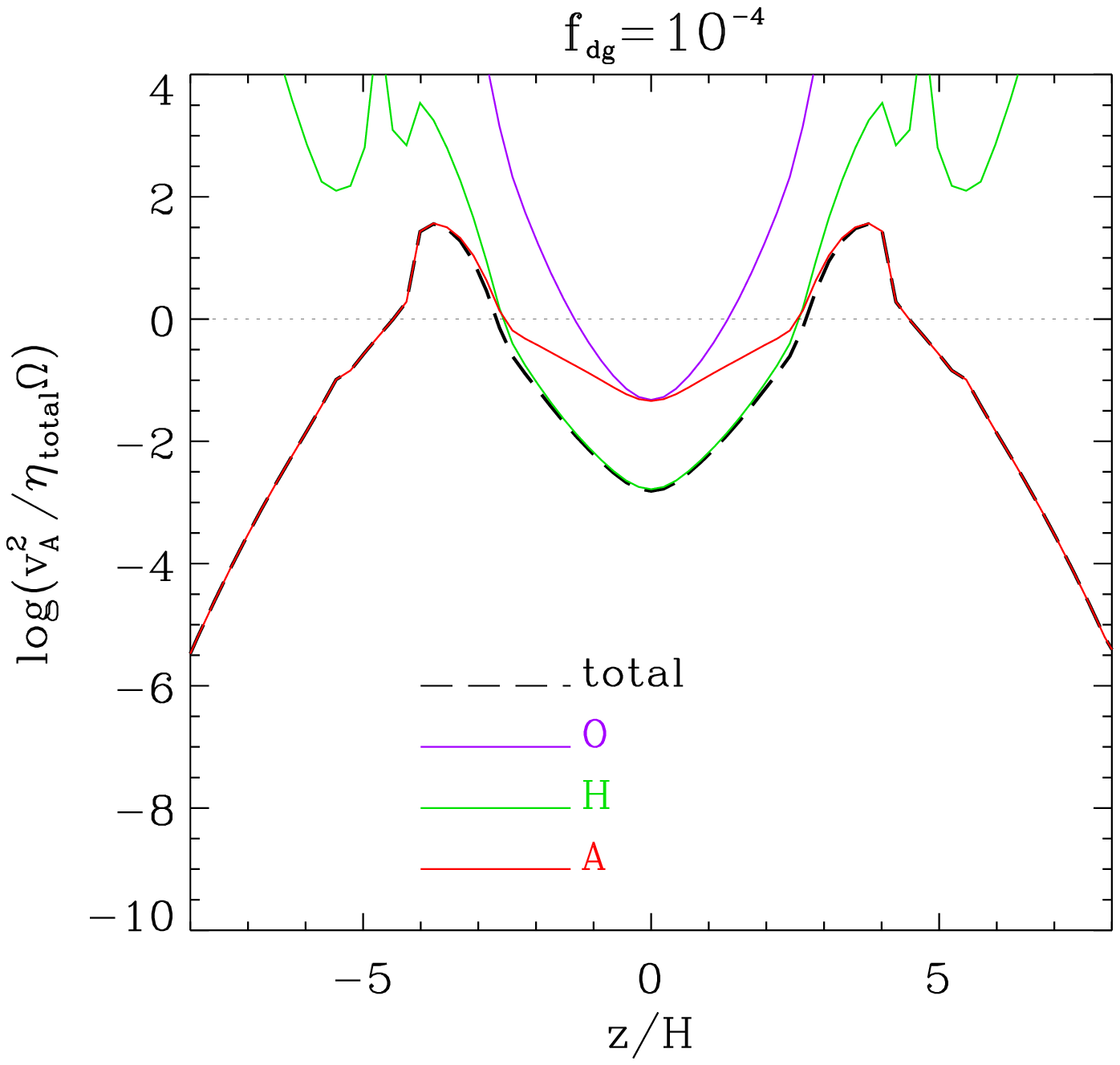}
\includegraphics[width=2.8in]{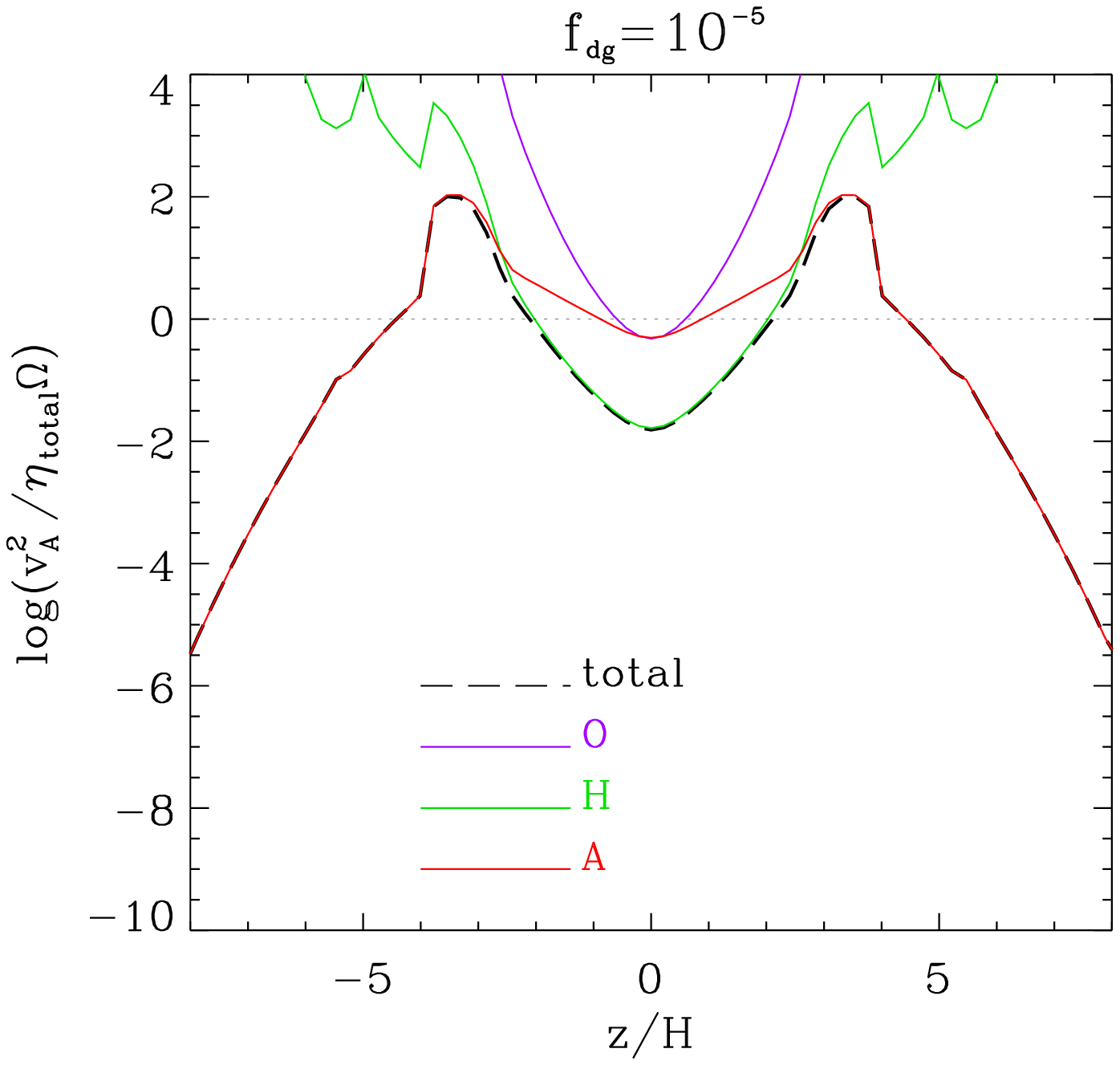}
}
  \caption{ Contributions to the total Elsasser number (black dashed
    curves) from Ohmic diffusion (purple), Hall effect (green) and
    ambipolar diffusion (red) at radius 7.5~AU in the fiducial model
    disk and versions with dust-to-gas ratios from $10^{-2}$ (top
    left) to $10^{-5}$ (bottom right).  The horizontal axis is the
    height in units of the pressure scale height, with $c_0=c_s/u_{\rm
      Kepler}$.  }
  \label{fi-all}
\end{center}
\end{figure}

Figure~\ref{fi-all} shows that ambipolar diffusion is more effective
than Ohmic diffusion in damping MRI in our fiducial model and over a
wide range of dust abundances.  We therefore take a careful look at
the criterion for MRI turbulence in the ambipolar diffusion regime.

Ambipolar diffusion has been characterized by two dimensionless
numbers, the ion Reynolds number $Re_i = \gamma_i\rho_i/\Omega$
\citep{haw98} and the ambipolar Elsasser number $\Lambda_A =
v_{Az}^2/(\eta_A\Omega)$ \citep{bai11}.  The Reynolds number accounts
only for the ions, while the Elsasser number includes the
contributions of all charged species.  The question we ask here is,
where is $Re_i$ good enough, and where must the other plasma
components be considered?

\begin{figure}
\begin{center}
\includegraphics[width=5.2in]{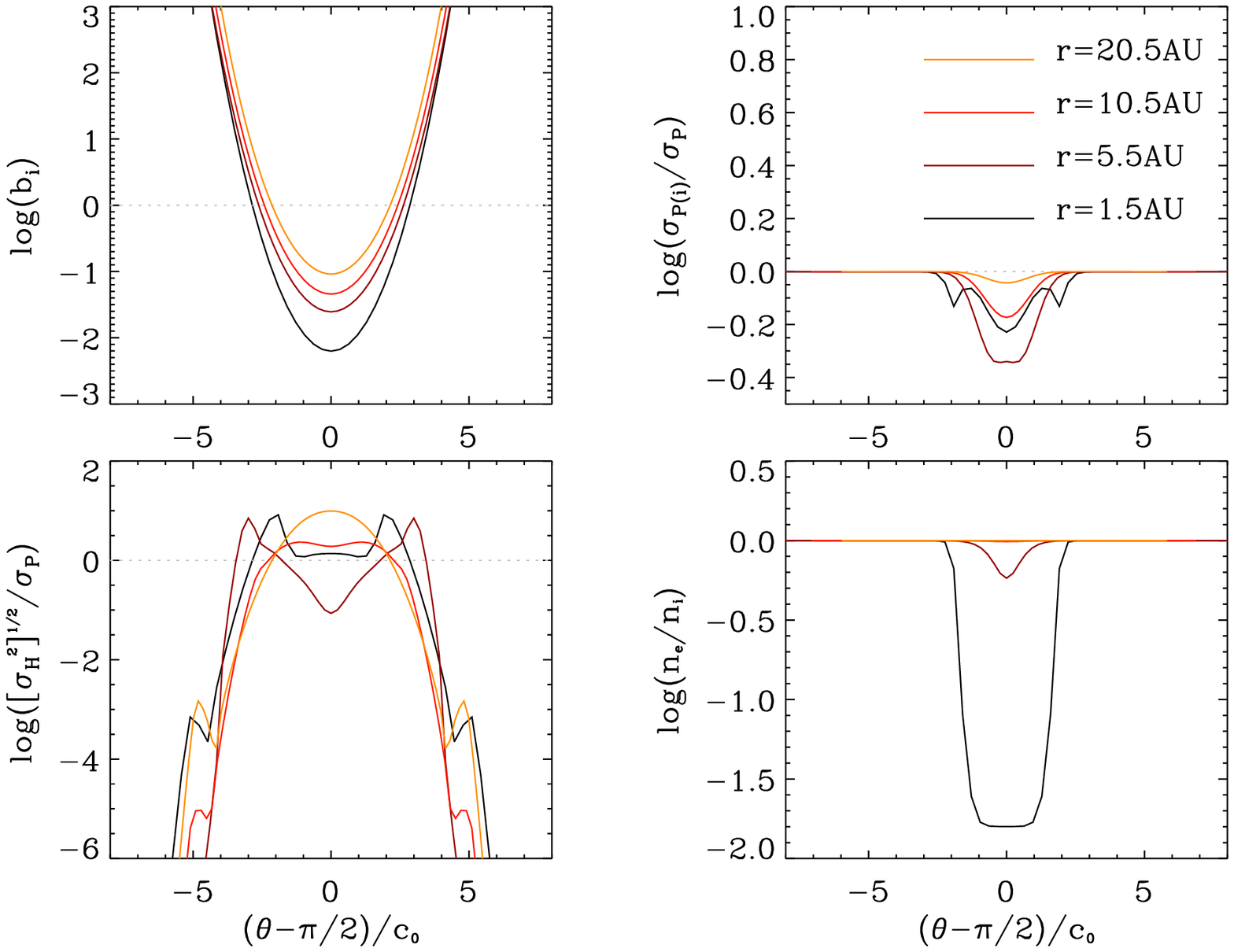}
  \caption{ Top left: ion coupling parameter $b_i$ versus height at
    four radii in the fiducial model disk.  Top right: ratio of ion
    Pedersen conductivity to total Pedersen conductivity.  Bottom
    left: ratio of absolute Hall conductivity to Pedersen
    conductivity.  Bottom right: ratio of electron to ion number
    densities.}
  \label{ambip}
\end{center}
\end{figure}

In Appendix~\ref{mricriterionA} we show that the two numbers are
equivalent if ions couple to the field and dominate the Pedersen
conductivity, which is larger than the Hall conductivity.  In
Fig.~\ref{ambip} we therefore plot the ion coupling parameter $b_i$,
the ions' contribution as a fraction of the total Pedersen
conductivity, the ratio of Hall to Pedersen conductivity, and
the ratio of the electron number density to the ion number density, in the
fiducial model disk.  The ion Reynolds number is unreliable in two
regions.  The first is the dense interior, where collisions thoroughly
decouple the ions from the fields.  The second is the disk atmosphere
above $6H$. 
There, gas densities
   are so low that grains can settle at supersonic speeds.  This would
   make the grains the biggest contribution to the Pedersen
   conductivity.  However since the gas drag and density profile are
   determined by heating and cooling, and potentially also by magnetic
   support, none of which we model in detail, we choose to neglect the
   settling for all our figures.  Fuller treatment may be worthwhile
   in future.
%There, the dust drift velocities may become supersonic due to the 
%settling, what should lead to
% the Pedersen conductivity being largely determined by
%the grains (not shown here). As our sound speed is far from the realistic one  
%in the disk corona, we skip the consideration of the dust vertical drift either.

\begin{figure}
\begin{center}
\includegraphics[width=7.in]{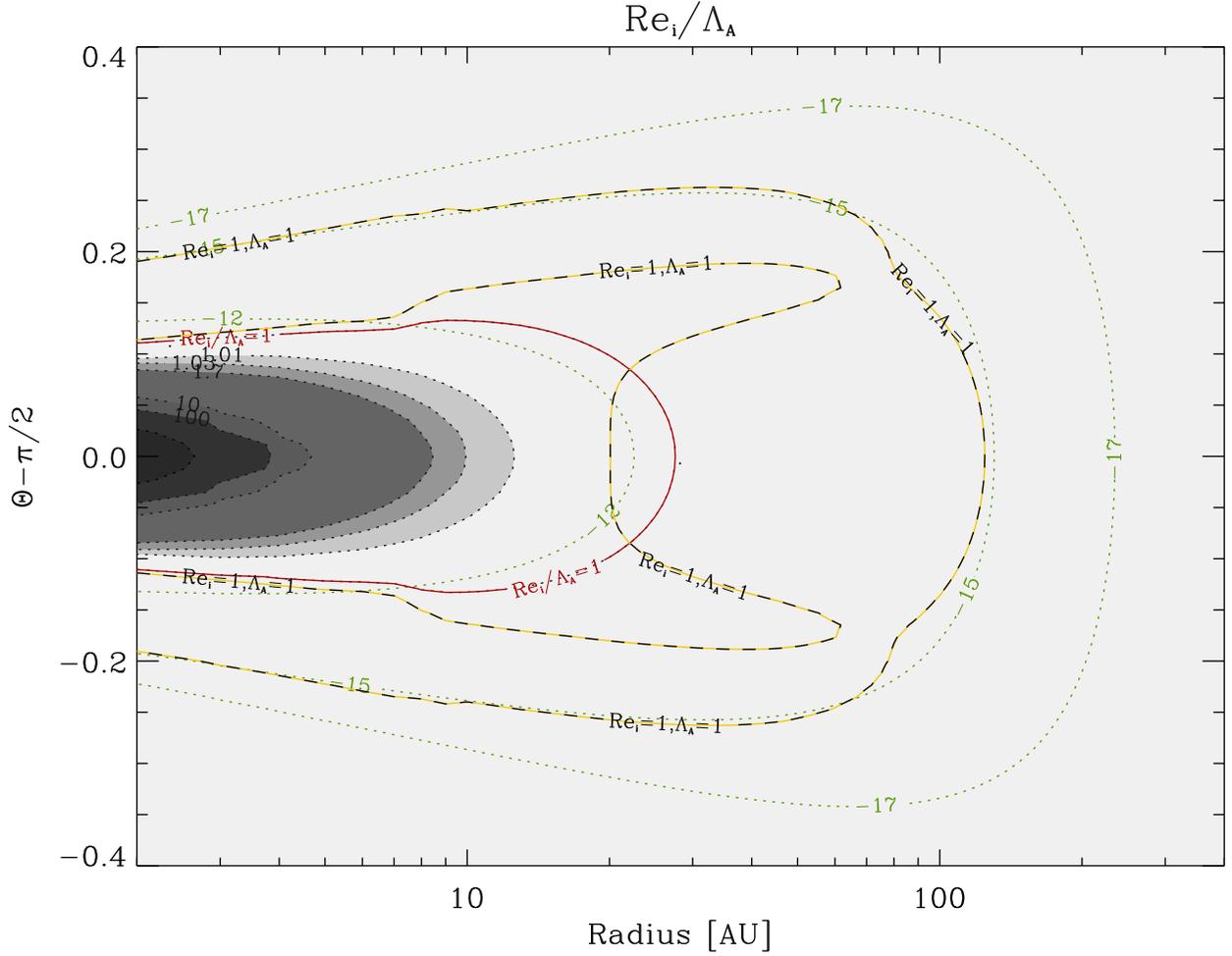}
  \caption{ Gray contours and gray dotted lines show the ratio
    $Re_i/\Lambda_A$ in the fiducial model disk.  The red line marks
    $Re_i/\Lambda_A=1$.  On the black dashed line $Re_i=1$ and on the
    nearly-overlapping yellow solid line $\Lambda_A=1$, two estimates
    of the border of the MRI-active region.  The green dotted lines
    are logarithmically-spaced isodensity contours labelled in $\rm g
    cm^{-3}$.}
\label{ambip2}
\end{center}
\end{figure}

Fig.~\ref{ambip2} is a map of the ratio $Re_i/\Lambda_A$ in the
fiducial model disk.  The two differ by a factor more than 1000 near
1~AU in the midplane.  However the ratio is near unity in the large
region shown in light gray, which includes the boundaries of the
MRI-active region.  The black dashed lines marking where $Re_i=1$
almost overlap the yellow lines indicating where $\Lambda_A=1$.  Yet
the curve where $Re_i/\Lambda_A=1$ is in quite a different location.
This is because even within the light-gray region the two numbers
differ slightly (less than 1\%). 

The map of the MRI-active region looks basically identical whether
drawn using the ion Reynolds or ambipolar Elsasser number.  In our
fiducial model, and in all models from Table~\ref{sets}, the ion
Reynolds number is as good as the ambipolar Elsasser number for
estimating the size of the MRI-active region.  However, far from the
active region's edge, the ion Reynolds number can be unreliable as an
estimator for the turbulent stress. 
% In the following sections we look
%at how these same conclusions hold up over a range of disk properties.

\subsection{Magnetic diffusivities in the ion-dust limit}

Ambipolar is the largest of the three diffusivities near the midplane
at dust-to-gas mass ratio $10^{-2}$ (Fig.~\ref{fi-all}).  This is
unusual, as Ohmic diffusion typically dominates near the midplane for
weak magnetic fields and Hall diffusion for stronger fields
\citep{war07}.  We therefore examine the reasons.  The ionization
balance falls in the ion-dust limit, where the grains sweep up many
electrons and ions, becoming the most abundant plasma component with a
mean negative charge of a fraction of an electron.  Even so, the
grains' large mass means they couple very poorly to the magnetic
fields, leaving the Ohmic and Hall diffusivities dominated by the
well-coupled electrons, and the Pedersen conductivity by the
poorly-coupled ions:
\begin{itemize}
\item The Ohmic conductivity comes from the electrons because they
  couple best to the fields; $\sigma_O\propto n b$ in
  eq.~\ref{eq:sigmaO}, and $b_e\approx 1000 b_i$, while in the
  ion-dust limit $n_e\approx n_i/100$ \citep{oku09a}.
\item The Hall conductivity also comes from the electrons since
  $\sigma_H\propto n b^2/(1+b^2)$ in eq.~\ref{eq:sigmaH}.  The
  coupling $b_e\gg 1\gg b_i$ means the electrons' contribution
  $\approx n_e$ is greater than the ions' $\approx n_ib_i^2$.
\item The Pedersen conductivity is in the ions because they are more
  numerous than the electrons, while the two coupling parameters
  bracket the maximum at $b=1$ in $\sigma_P\propto n b/(1+b^2)$
  (eq.~\ref{eq:sigmaP}).
\end{itemize}
Consequently the Ohmic conductivity is greater than the Pedersen,
which in turn is greater than the Hall, since $(\sigma_O\sim n_eb_e) >
(\sigma_P\sim n_ib_i) > (\sigma_H\sim n_i/100)$.  Thus in our
eq.~\ref{eq:eta}, $\sigma_\perp\approx\sigma_P$ and the ambipolar
diffusivity $\eta_A\propto\sigma_P^{-1}$ is bigger than the Ohmic
diffusivity $\eta_O\propto\sigma_O^{-1}$.  The special circumstance
leading to ambipolar dominance in the ion-dust limit is our chosen
magnetic field strength, which couples the ions too little for them to
contribute to the Hall conductivity, but enough to dominate the
Pedersen conductivity.

For $f_{dg}<10^{-3}$, the reduced grain surface area means the
midplane near 7~AU no longer reaches the ion-dust limit.  The
ionization fraction is larger, the electron density is comparable to
the ion density, and both exceed the density of the charged grains,
which carry many electrons on average.  The Ohmic and Hall
conductivities are both larger in proportion to the increased electron
density, while the Pedersen conductivity scales with the lesser
increase in the ion density.  As a result the Hall overtakes the
Pedersen conductivity, giving the ordering
$\sigma_O>\sigma_H>\sigma_P$.  Since $\eta_H\sim 1/\sigma_H >
\eta_A\sim\sigma_P/\sigma_H^2$, the largest diffusivity is the Hall
one.

The Hall term is special in that it depends on the sign of the
magnetic field, suppressing the MRI for one field direction and
amplifying it for the other.  It is unclear whether the Hall term has
a strong impact on a turbulent eddy penetrating into the transitional
or dead zone, as an eddy typically carries magnetic field of both
signs.  The importance of the Hall effect in the nonlinear evolution
of the MRI remains to be clarified with further numerical
calculations.  Therefore we here confine ourselves to considering the
Ohmic and ambipolar Elsasser numbers.

\subsection{Magnetic Field Strength}

One of the greatest uncertainties in modeling the magnetic coupling in
protostellar disks is the strength and configuration of the magnetic
field.  A lower limit on the field strength is the mean Galactic field
of $\sim 10$~$\mu$Gauss \citep{bec96a,bec96b}.  An upper limit is
equipartition of the vertical magnetic field with the gas, which makes
the shortest unstable MRI mode longer than the disk thickness.  In our
models the midplane plasma beta is independent of radius, while the
magnetic pressure is independent of height (see eq.~5 in section~2.1).

\begin{figure}
\begin{center}
{\hbox{
\includegraphics[width=3.5in]{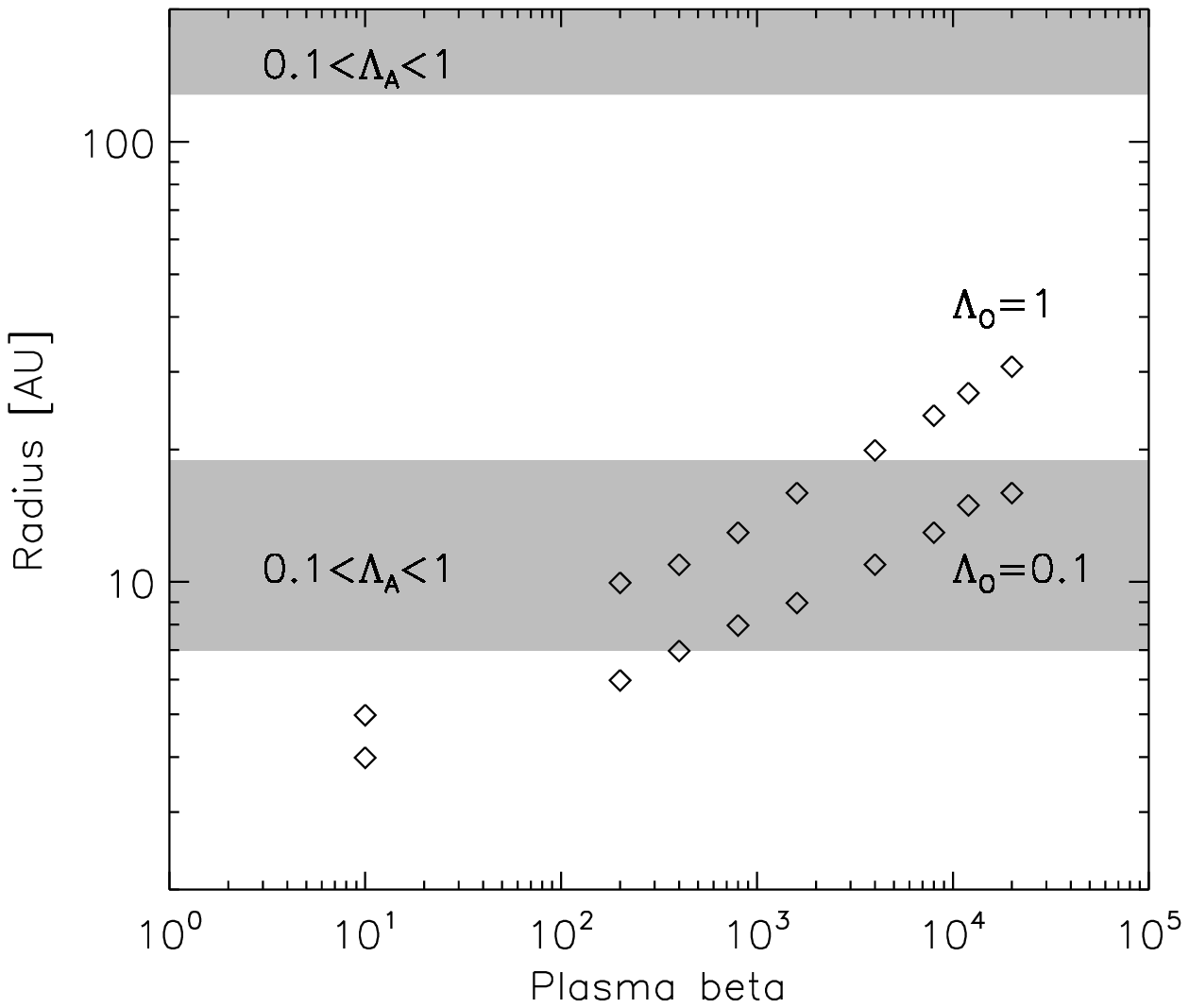}
\includegraphics[width=3.5in]{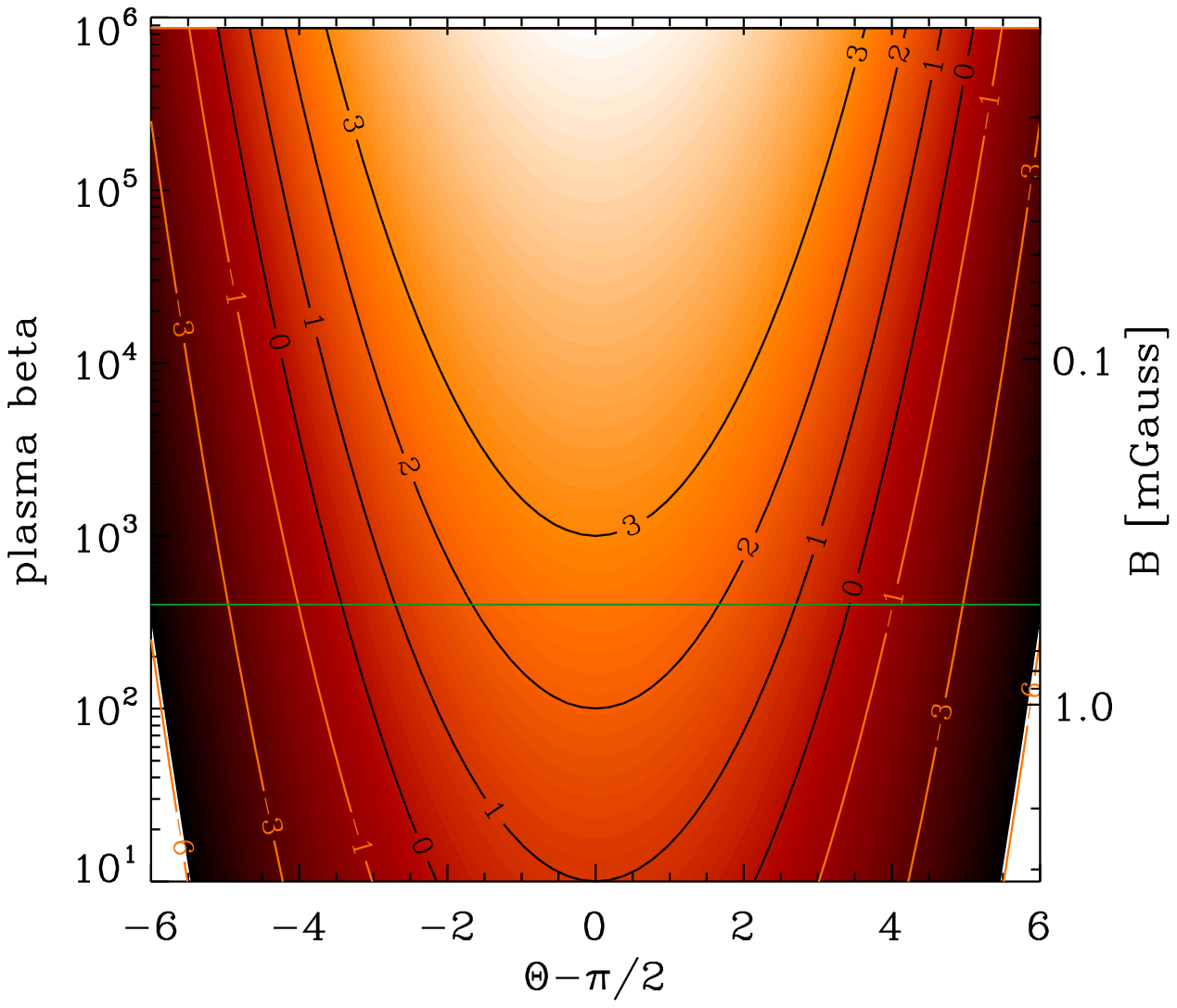}
}}
\caption{ Left: The radial extents of the Ohmic and ambipolar
  transitional layers at the midplane in the fiducial disk model, as
  functions of the plasma beta parameter.  Gray shading indicates
  where the ambipolar Elsasser number is between 0.1 and~1.  Diamonds
  show Ohmic Elsasser numbers of 0.1 (bottom chain) and 1 (top chain).
  Right: Contours of plasma beta as a function of the height in
  pressure scale heights (horizontal axis) and the midplane plasma
  beta (vertical axis).  On the right-hand vertical axis is the
  strength of the corresponding vertical magnetic field in Gauss at
  100~AU.  A green line marks our chosen midplane plasma beta
  parameter $\beta=400$.  }
\label{plasma2}
\end{center}
\end{figure}

The ambipolar Elsasser number is basically independent of the magnetic
field strength, but the Ohmic Elsasser number is proportional to $B^2$
(eqs.~21, 25 and 27).  Fig.~5 shows that the ambipolar Elsasser number
is smaller than the Ohmic Elsasser number only by a small factor near
the midplane.  Thus in a magnetic field just a few times weaker, the
MRI would be damped by Ohmic rather than ambipolar diffusion near the
midplane.  Another concern is above what height the disk is
magnetically-dominated, with plasma beta less than unity.  To clarify
these points, we vary the midplane plasma beta from 10 to $10^6$.  The
corresponding magnetic fields are between 3.25 and $10^{-2}$
milligauss at 100~AU, where the field strength can potentially be
measured using sub-millimeter polarization \citep{krej09}.  In
Fig.~\ref{plasma2} (left) we plot the radial locations where the Ohmic
and ambipolar Elsasser numbers are 0.1 and 1, as functions of plasma
beta.  This range corresponds to the dead zone's outer edge, which we
call the transitional zone.  Its width shows how gradual the
transition is between dead and active zones.  We see that the Ohmic
dead zone grows with plasma beta.  At $\beta>4000$, Ohmic rather
than ambipolar dissipation sets the outer dead zone boundary in the
midplane.  On the other hand, for strong magnetic fields such as
3~milligauss, or midplane plasma beta 10, the Ohmic dead zone lies
deep within the ambipolar dead zone.  

Choosing low midplane plasma
beta also leaves the disk's upper parts strongly magnetized and stable
against MRI.  In our fiducial model, the plasma beta is below unity
above $3.5H$ (Fig.~\ref{plasma2} right, green line).  For a field
strength at 100~AU matching the galactic field of 10~$\mu$G (plasma
beta of $10^6$), the magnetized region is only above $5H$
(Fig.~\ref{plasma2}, right).  Note that ambipolar diffusion defines
the dead zone's {\it minimum radial extent}.  A larger dead zone
indicates that the fields are weak enough for the Ohmic diffusion to
control the evolution of the MRI.

\subsection{Monomer Size}

\begin{figure}
\begin{center}
\hbox{
\includegraphics[width=3.5in]{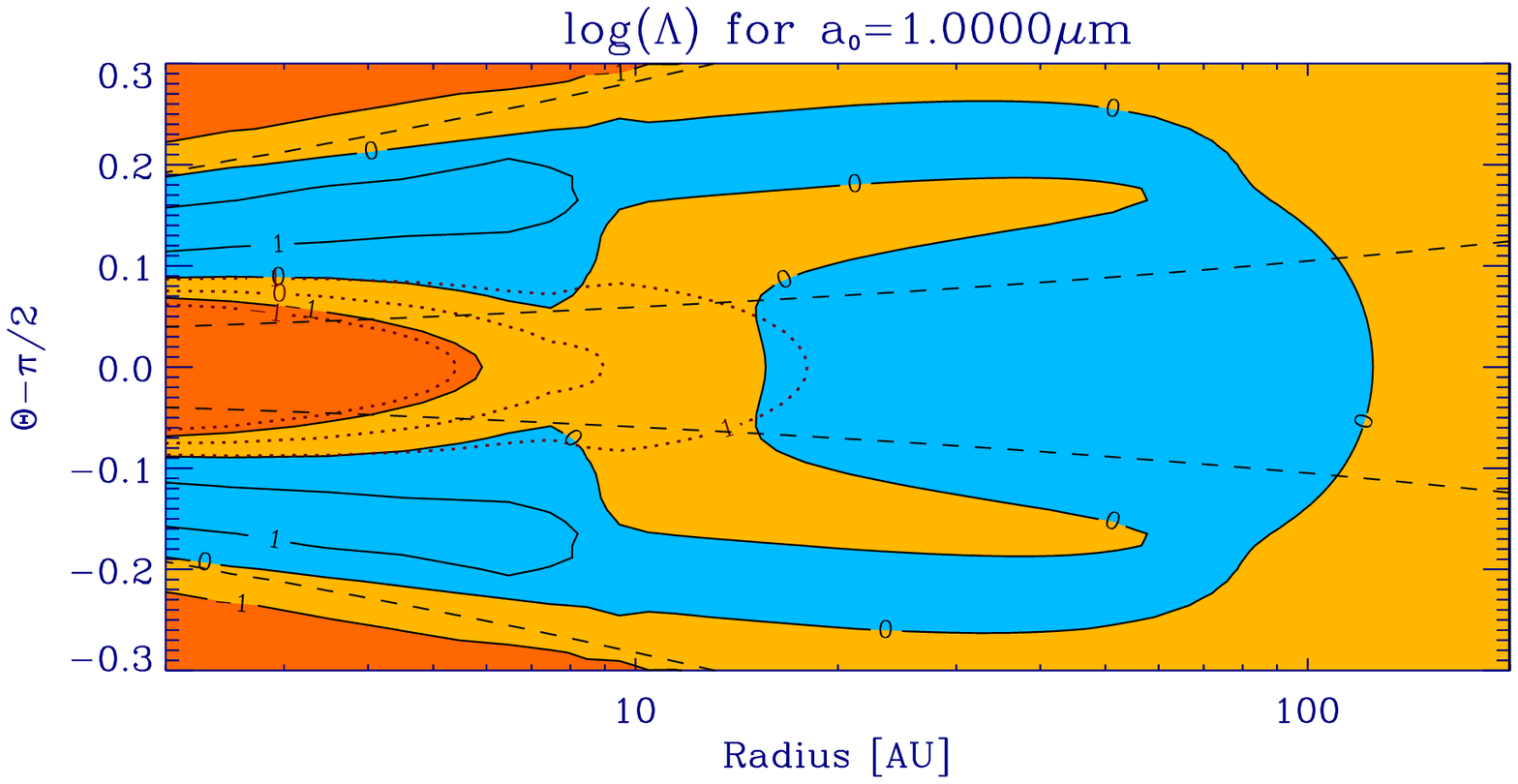}
\includegraphics[width=3.5in]{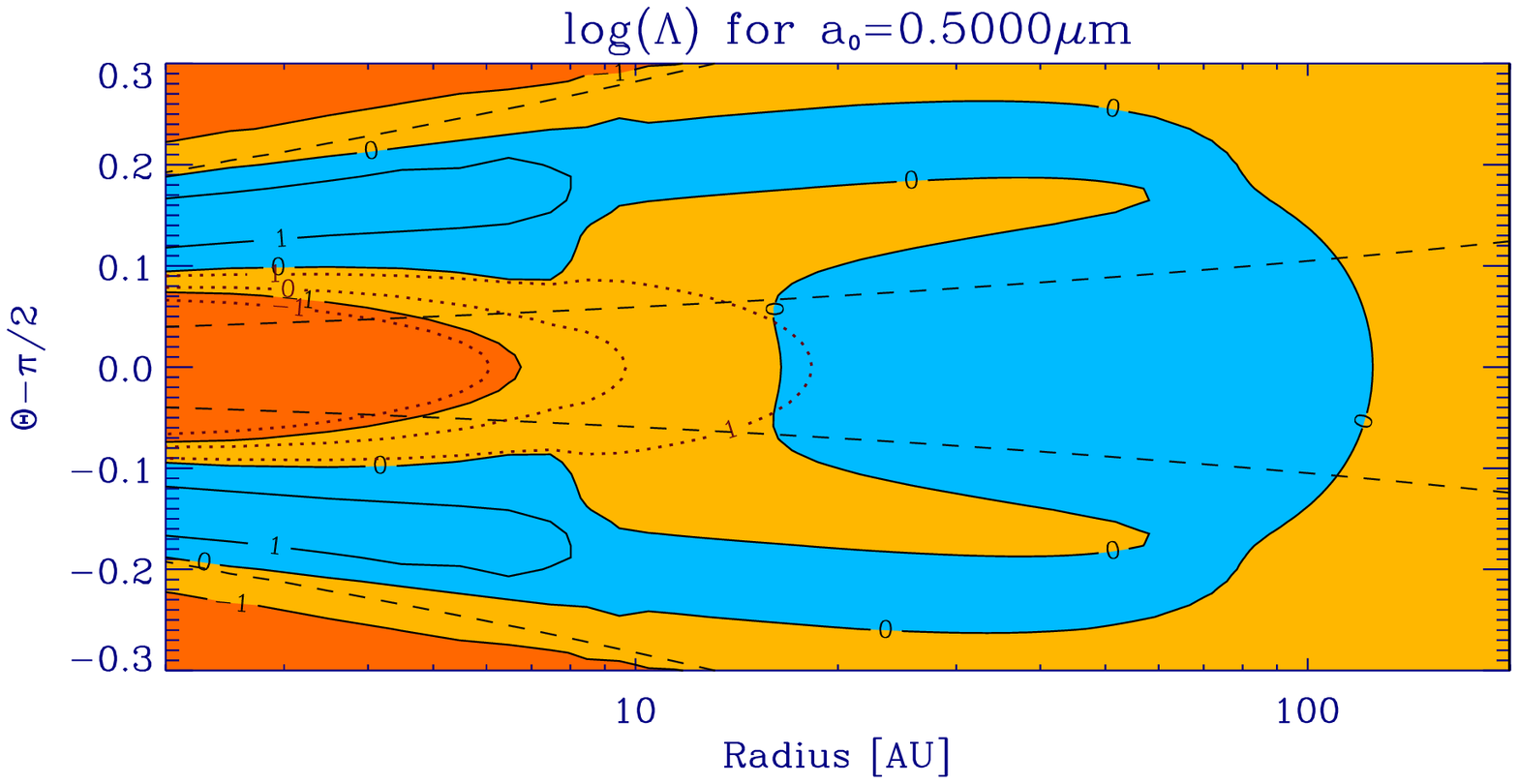}
}
\hbox{
\includegraphics[width=3.5in]{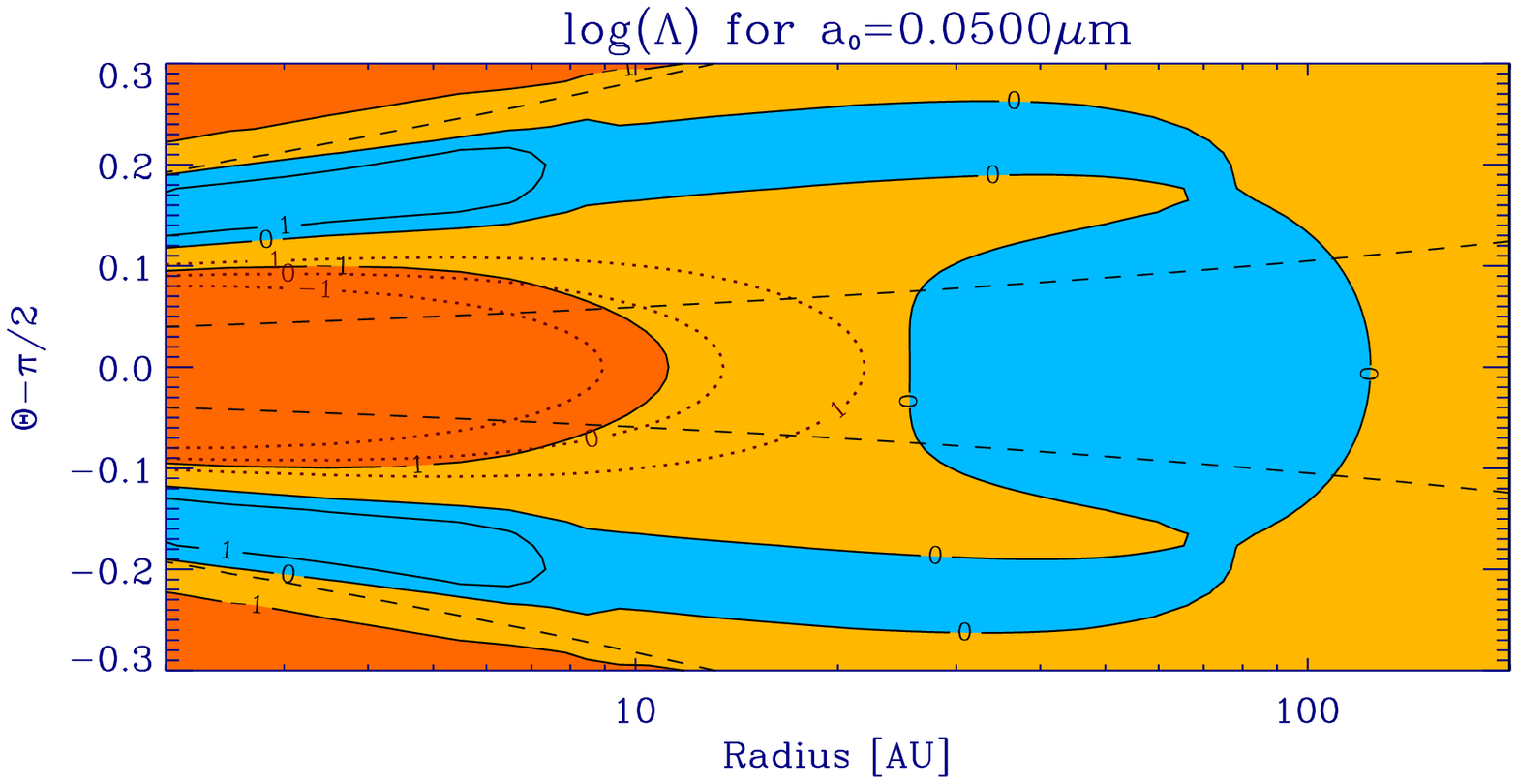}
\includegraphics[width=3.5in]{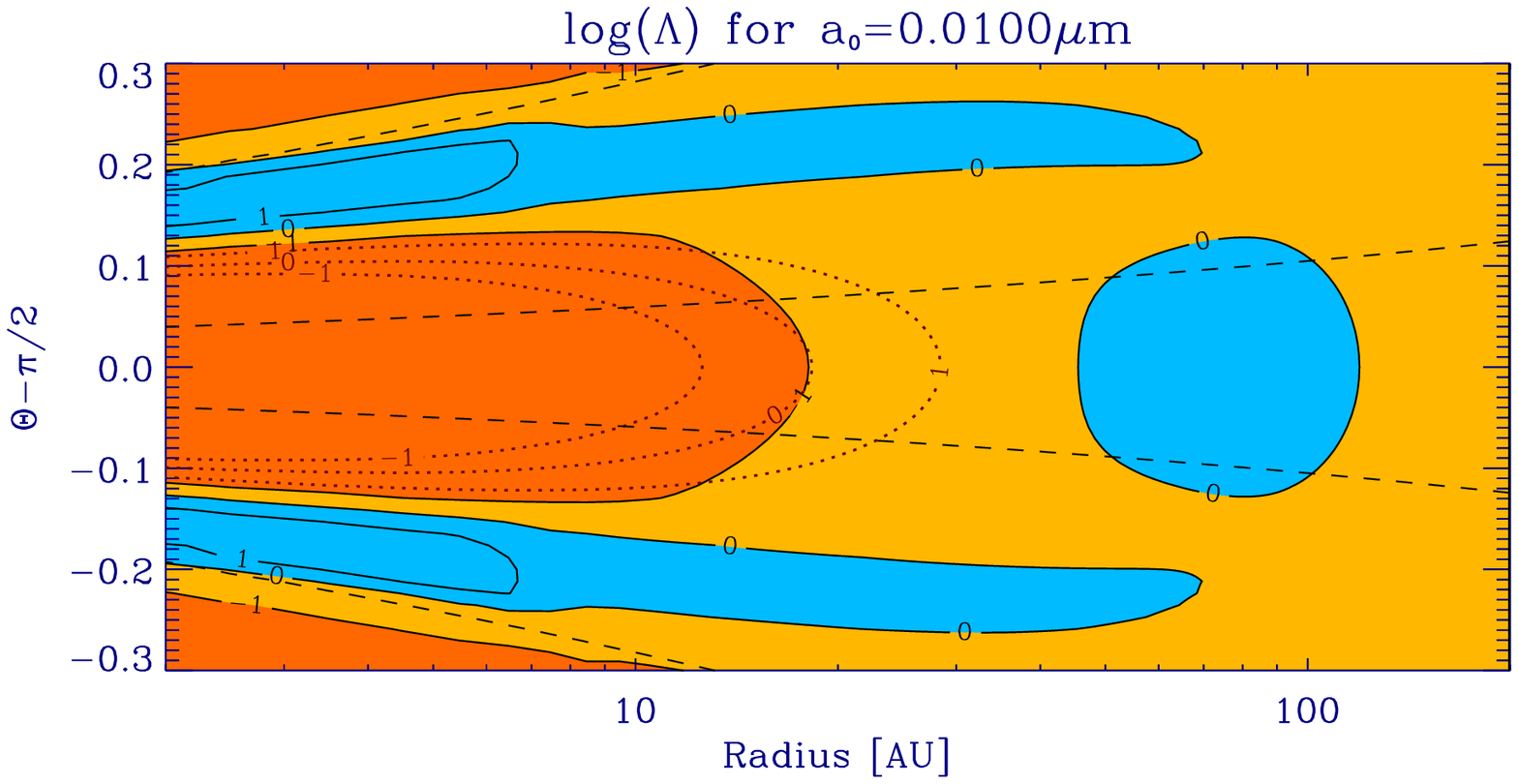}
}
\caption{Contours of the total Elsasser number (black solid lines) and
  ohmic Elsasser number (black dotted lines) in disks containing fluffy
    aggregates composed of monomers of various radii. The MRI-active regions
  are shown in blue, the transitional regions in yellow and the dead
  zone in orange.
%  Right: Contours of relation $Re_{A}/\Lambda_{A}$ (grey), logarithmic gas density 
%  contours (dotted green), contour of  $Re_{A}=1,\Lambda_{A}=1$ in blue
%  and red contour for exact equivalence between the Elsasser and the ambipolar
%  Reynolds numbers.
}
\label{mono1}
\end{center}
\end{figure}

Here we discuss the influence of monomer size on the shape of the
MRI-active region (set~1, Table~\ref{sets}). 
The outer border of
the active zone is well aligned with the gas isodensity surface
$\rho=10^{-15} \rm\ g cm^{-3}$.  The active zone encloses an interior
dead and transitional zone shaped like a fish tail.  Outside 10~AU,
the active zone is defined by ambipolar diffusion.  The Ohmic
diffusion plays hardly any role in the radial extent of the dead zone
for the assumed magnetic field strength.  However Ohmic diffusion is
important for the dead zone's vertical thickness in the inner disk.
The transitional layer widens outside 8~AU, giving the fish-tail
shape.  The widening is caused by the switch from Mg$^+$ to HCO$^+$ as
the main ion, leading to lower ionization fractions since the
molecular ion recombines faster.  For very small monomers, 0.005 (not
shown) and 0.01~$\mu$m in radius, the active layers are thinner and
separated from a midplane island of magnetic activity in the outer
disk (Fig.~\ref{mono1}, bottom).  Also, the ratio $Re_i/\Lambda_A$
deviates strongly from unity.  Small grains couple better to the
magnetic fields and participate in currents in the disk corona.
Meanwhile in the interior dead zone, the small grains readily sweep up
free electrons, leaving $n_i\gg n_e$.  These results suggest that the
effects of very small grains such as PAHs are best characterized using
the Elsasser number criterion, which includes the contributions of all
the plasma components as outlined above in section~\ref{sec:Rei}.

The transitional zone gains in radial extent at the midplane as the
monomer size decreases.  For $a_0=1 \mu$m, the transitional zone runs
from 6~AU where $\Lambda=0.1$ to 15~AU where $\Lambda=1$, giving 9~AU
of transition between $\alpha_{\rm dead}\approx 10^{-5}$ and
$\alpha_{\rm ac}\approx 10^{-3}$ (section~3.2).  For $a_0=0.01\mu$m,
the transitional layer extends from 19 to 45~AU, a width of 26~AU.

\begin{figure}
\begin{center}
\includegraphics[width=4.5in]{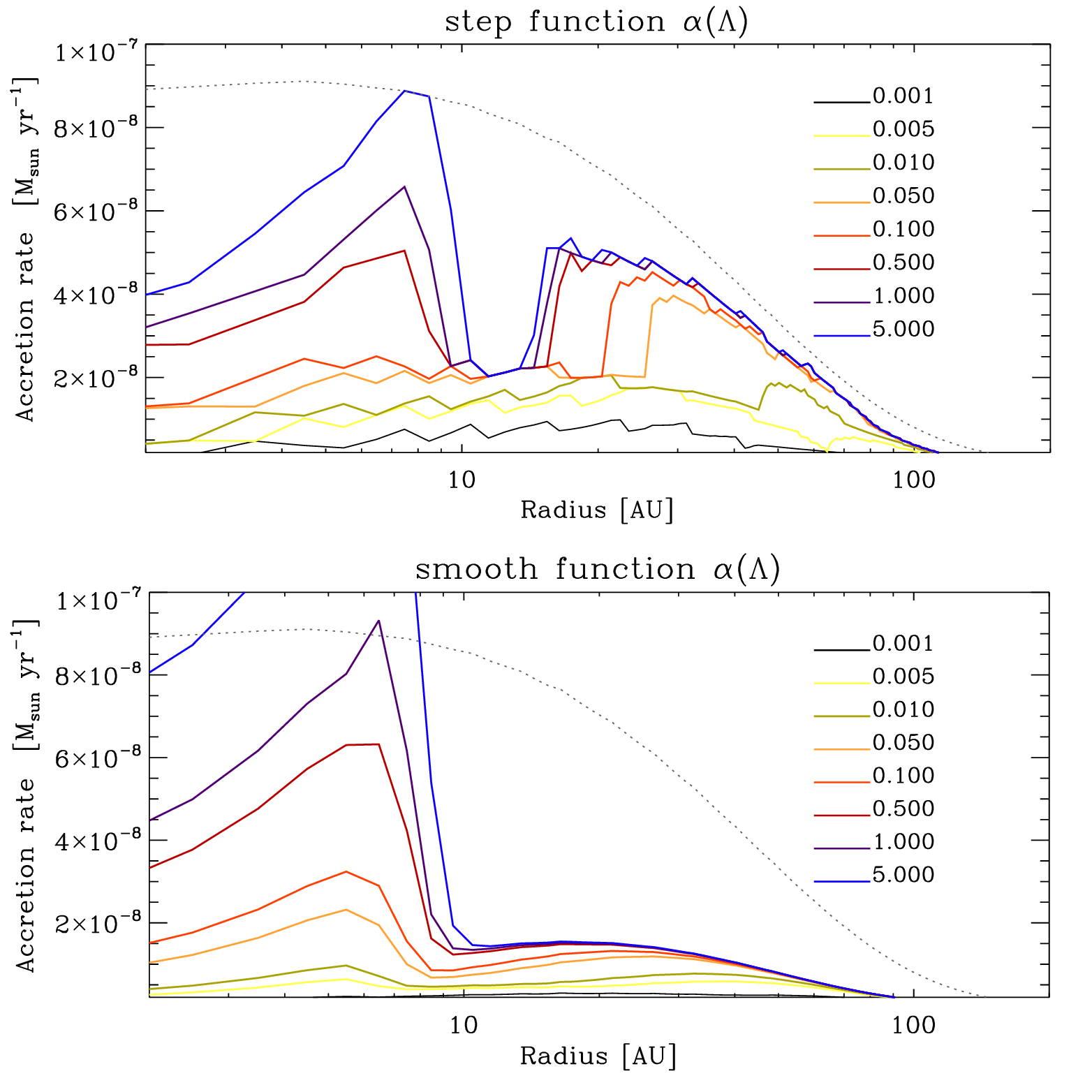}
  \caption{Accretion rate in solar masses per year for various monomer
    sizes.  The dotted line shows the accretion rate for the case when
    the whole disk is active with the turbulent stress from eq.~29.
    Note that the smooth-function turbulent stress can lead to higher
    accretion rates in the $\Lambda>10$ regions (see section
    3.2).  \label{fig:set1mdot}}
\label{mono2}
\end{center}
\end{figure}

The accretion rates are plotted against the distance from the star for
the various monomer sizes in Fig.~\ref{mono2} using the approach set
out in Section~3.2.  Very small monomers with $a_0\le 0.01$~$\mu$m
produce accretion rates below $10^{-8} M_\odot$~yr$^{-1}$ at 1~AU.
The biggest monomers, $a_0=5 \mu$m, give a maximum in the accretion
rate between 7 and 8~AU.  At about 10~AU the accretion rate has a gap,
related to the radial variation in the active layers' thickness shown
in Fig.~\ref{mono1}.  When the stress parameter $\alpha(\Lambda)$ is
given a step-function shape, the accretion rate jumps from $9\times
10^{-8} M_\odot$~yr$^{-1}$ at 7~AU down to $2\times 10^{-8}$ at 12~AU.
Aggregates composed of the intermediate-sized monomers yield jumps
almost as large: from 6.5 to $2\times 10^{-8} \rm M_\odot yr^{-1}$ for
$a_0=1$~$\mu$m, and from 5 to $2\times 10^{-8}$ for $a_0=0.5$~$\mu$m.

However the accretion rate profile changes drastically when the smooth
function is adopted for $\alpha(\Lambda)$.  The accretion rate peaks
near the metal line, and accretion is much slower in the outer disk.
The reason is that the Elsasser number is very near unity along the
whole midplane outside 10~AU, so that the smooth function yields a
turbulent stress that varies slowly across the border between
transitional and active zone.

\subsection{Dust-to-Gas Ratio}

\begin{figure}
\begin{center}
\hbox{
\includegraphics[width=3.5in]{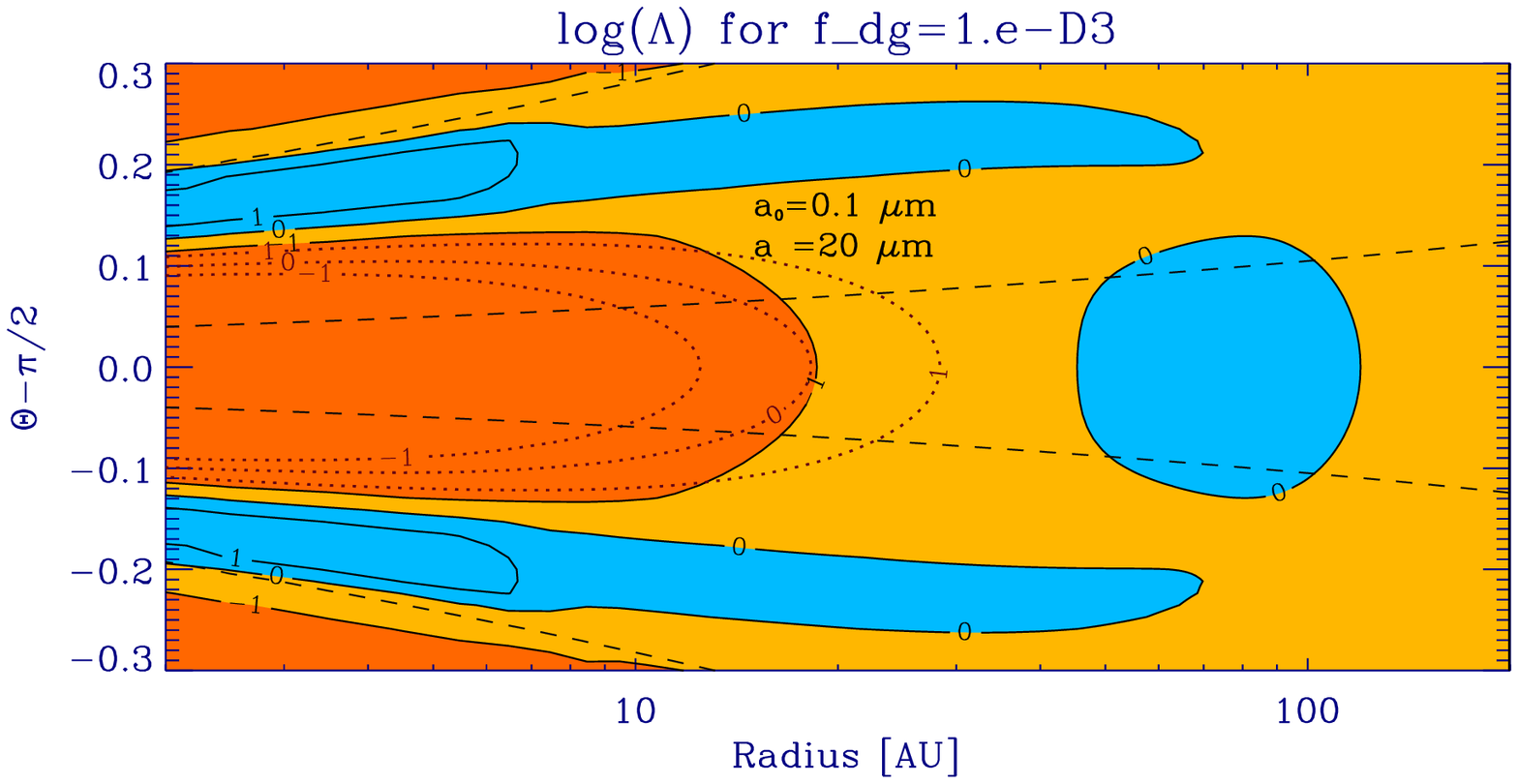}
\includegraphics[width=3.5in]{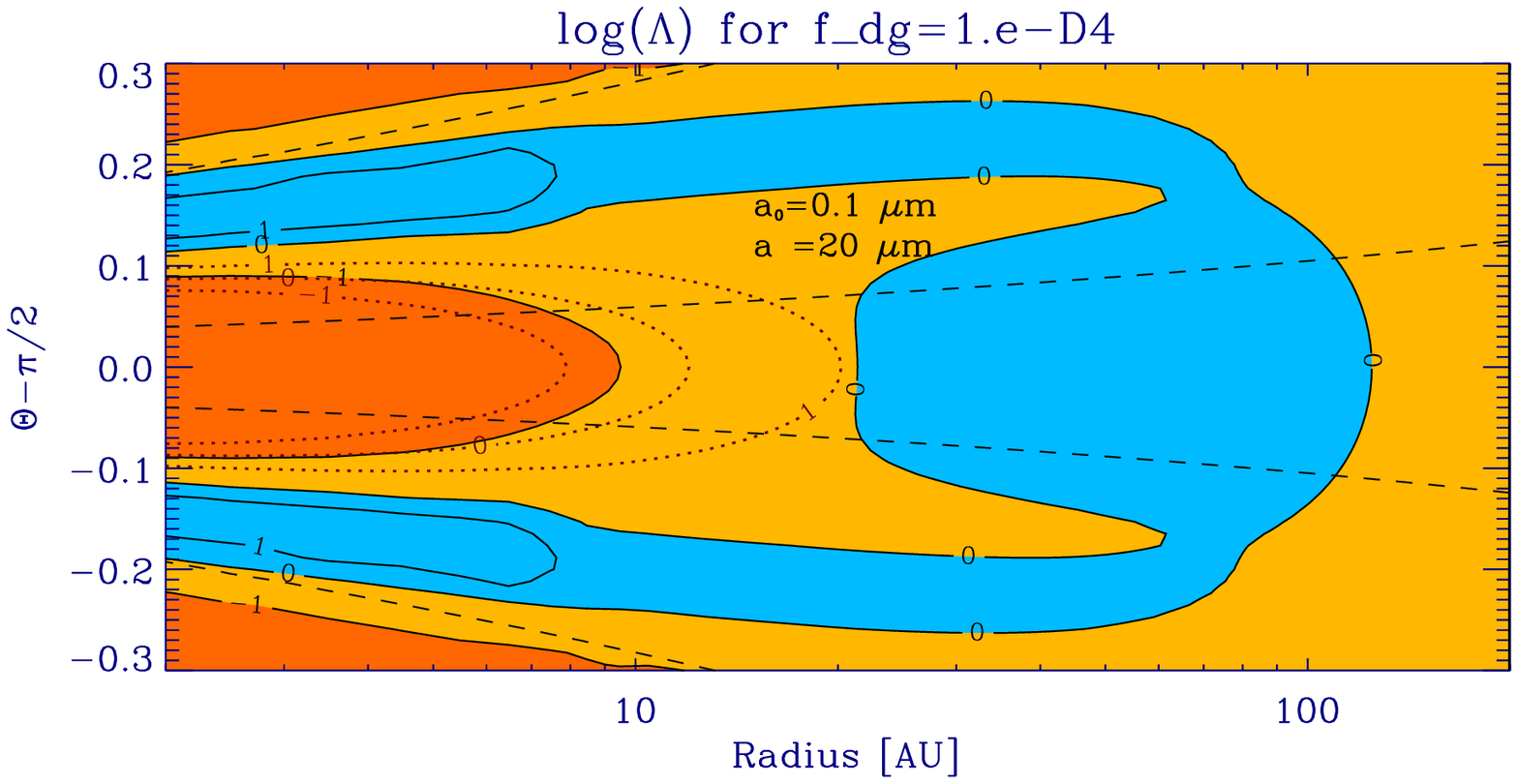}
}
\hbox{
\includegraphics[width=3.5in]{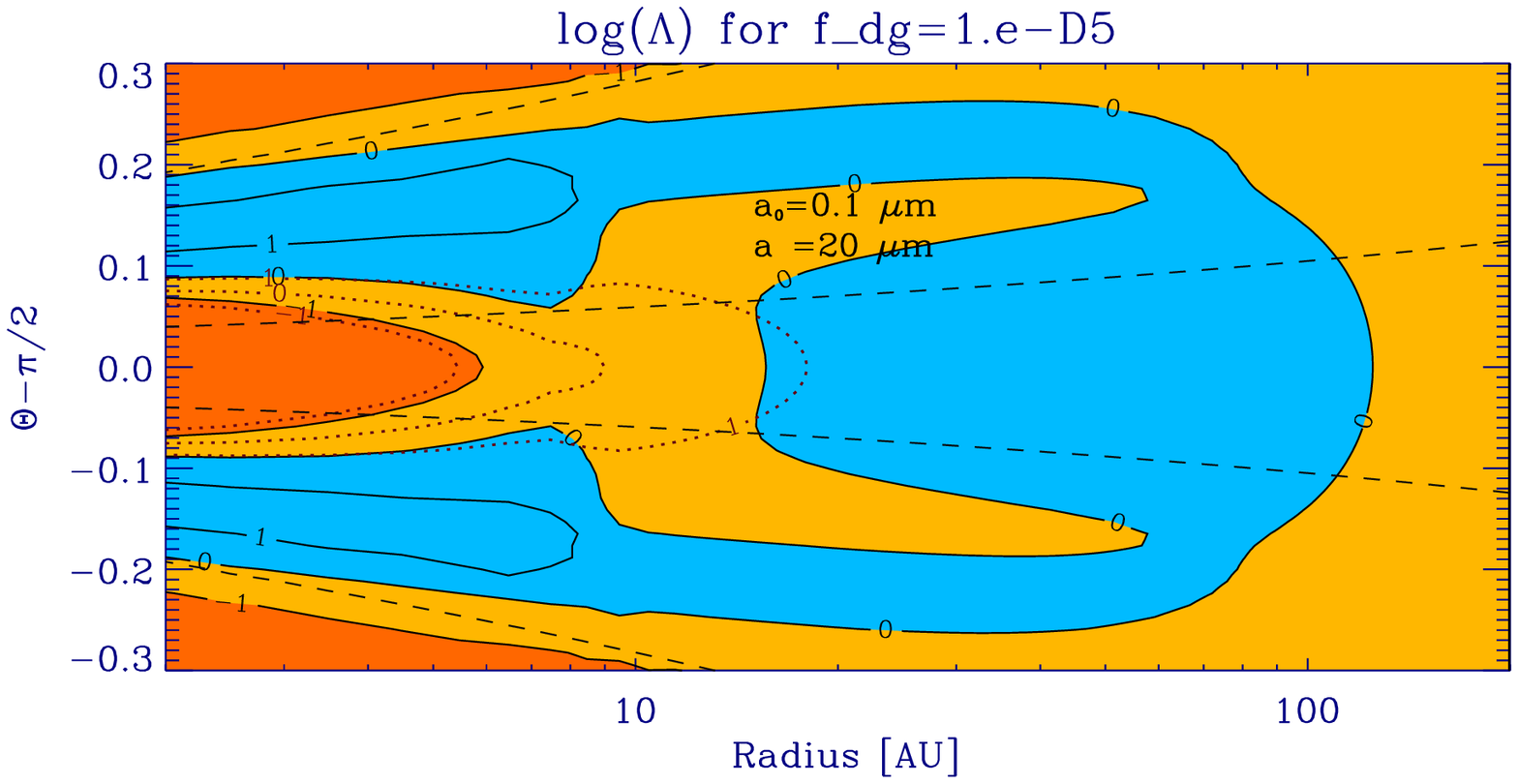}
\includegraphics[width=3.5in]{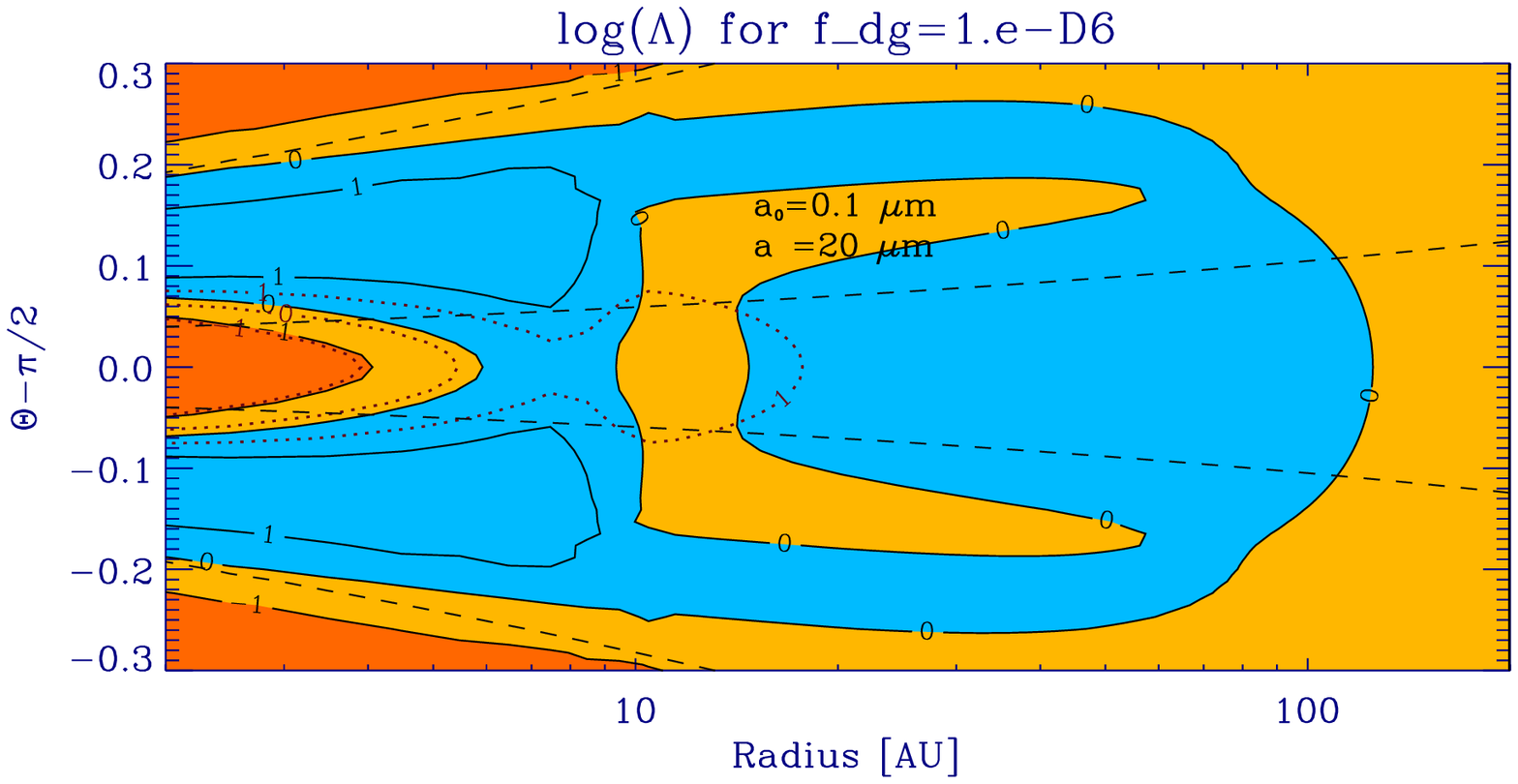}
}
\caption{ MRI-active regions appearance for various dust-to-gas ratio,
  see also Fig.~\ref{mono1}.  Colors and symbols are as in
  figure~\ref{mono1}.  }
\label{dust1}
\end{center}
\end{figure}

In this section we vary the dust-to-gas ratio (set~2,
Table~\ref{sets}), leaving other parameters unchanged from the
fiducial model.  The resulting dead zones and the ambipolar
Reynolds-to-Elsasser number ratio are shown in Fig.~\ref{dust1}.  The
radial extent of the transitional layer again shows striking
variations.  The transitional layer's midplane radial width is 25~AU,
12~AU, 7.5AU and 2~AU with dust-to-gas ratios $10^{-3}$, $10^{-4}$,
$10^{-5}$ and $10^{-6}$.  From the shift in the dead zone shape with
increasing dust depletion, we see that the fish-tail shape appears
when the active surface layers merge with the midplane active zone
(Fig.~\ref{dust1}, top left and  top right).  At the lowest
dust-to-gas ratio, the transitional layer splits into two pieces,
located near 1.5 and 10~AU, and both about 2~AU in radial extent at
the midplane (Fig.~\ref{dust1}, bottom right).  The MRI-active region's
outer boundary again lies at the gas isodensity surface
$\rho=10^{-15}\rm\ g\ cm^{-3}$.

\begin{figure}
\begin{center}
\includegraphics[width=4.5in]{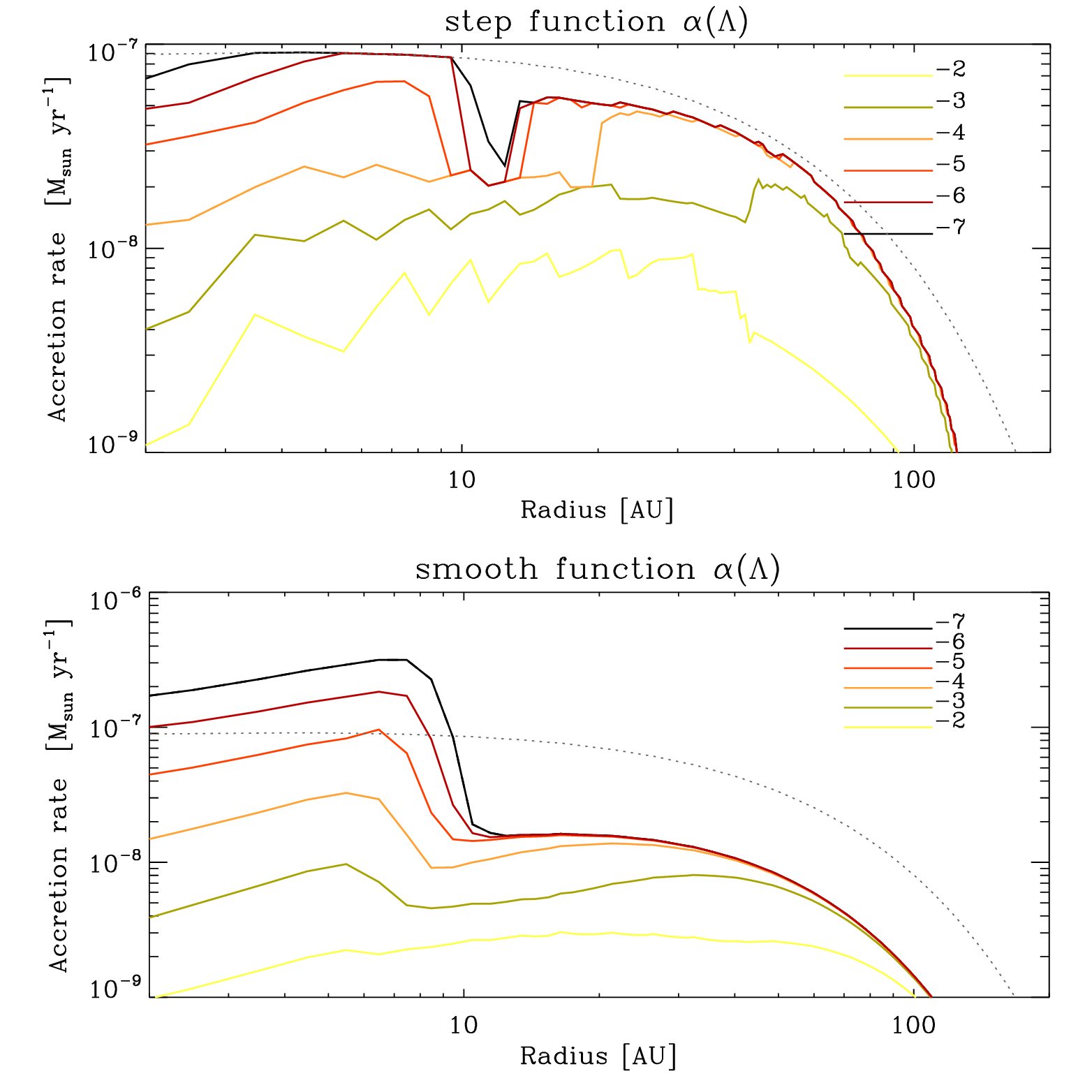}
  \caption{ Accretion rate in solar mass units per year calculated for
    the MRI-active surface density and for $\beta>1$ disk regions, for
    disks with various dust-to-gas ratios, $log(f_{dg})=[-2, -3, -4,
    -5, -6, -7]$. The dotted line shows the accretion rate for the
    case when whole disk is active with the turbulent stress taken
    from eq.~29. Note that the smooth-function turbulent stress can
    lead to higher accretion rates in the $\Lambda>10$ regions (see
    section~3.2).  \label{fig:set2mdot}}
\label{dust2}
\end{center}
\end{figure}

The radial profiles of the accretion rate are shown in
Fig.~\ref{dust2}.  For the step function $\alpha(\Lambda)$ and
dust-to-gas ratios less than $10^{-5}$, the accretion flow shows a gap
from 10 to 12~AU.  The deficit in the gap is as much as a factor of
four for dust-to-gas ratio $f_{dg}=10^{-6}$.  However no such feature
appears with the smooth $\alpha(\Lambda)$ function.  Instead the
accretion rate steps up inside the metal line, located between 5 and
9~AU, and the dead zone edge has no clear signature in the radial
accretion rate profile.

\subsection{Disk Mass}

\begin{figure}
\begin{center}
\hbox{
\includegraphics[width=3.5in]{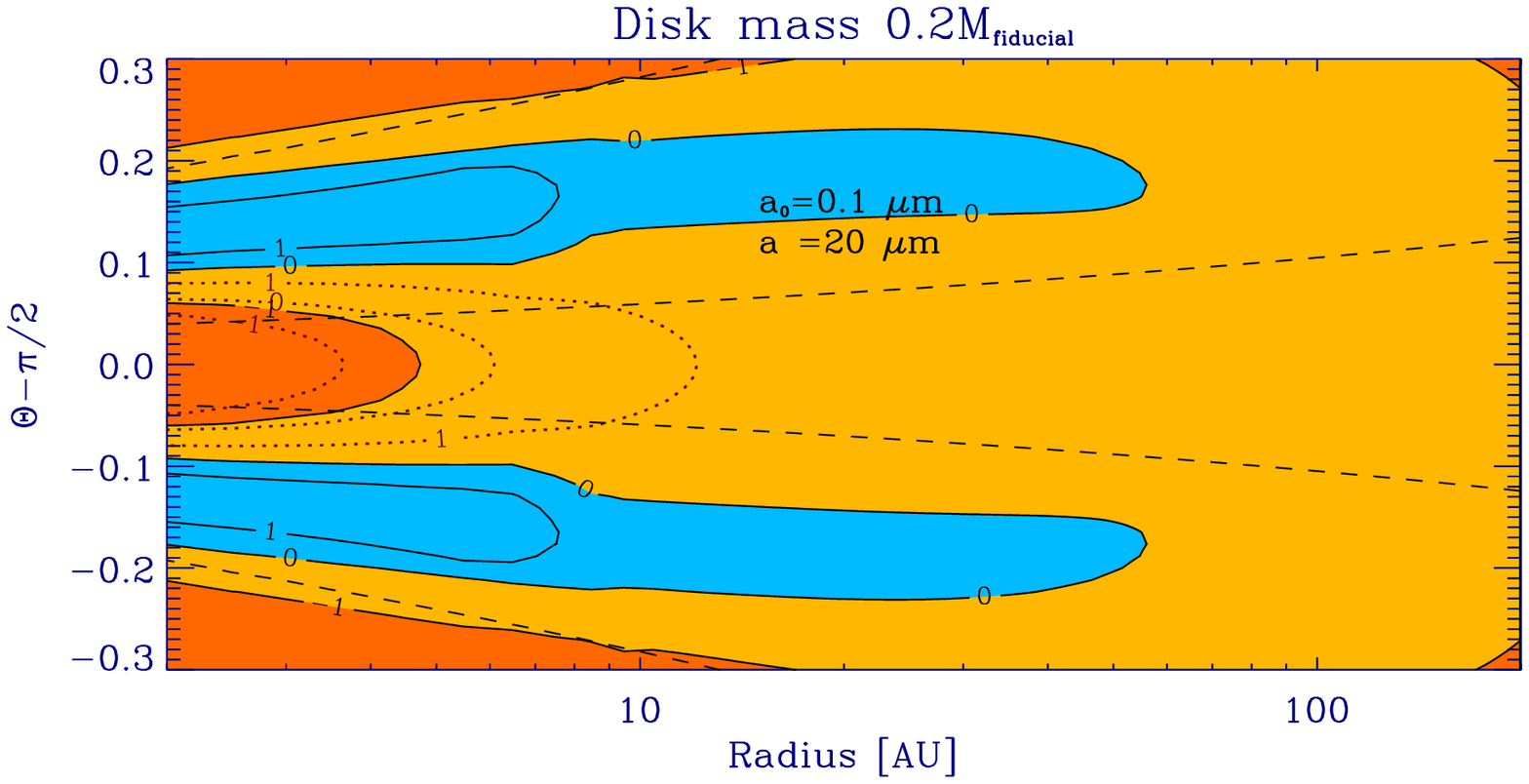}
\includegraphics[width=3.5in]{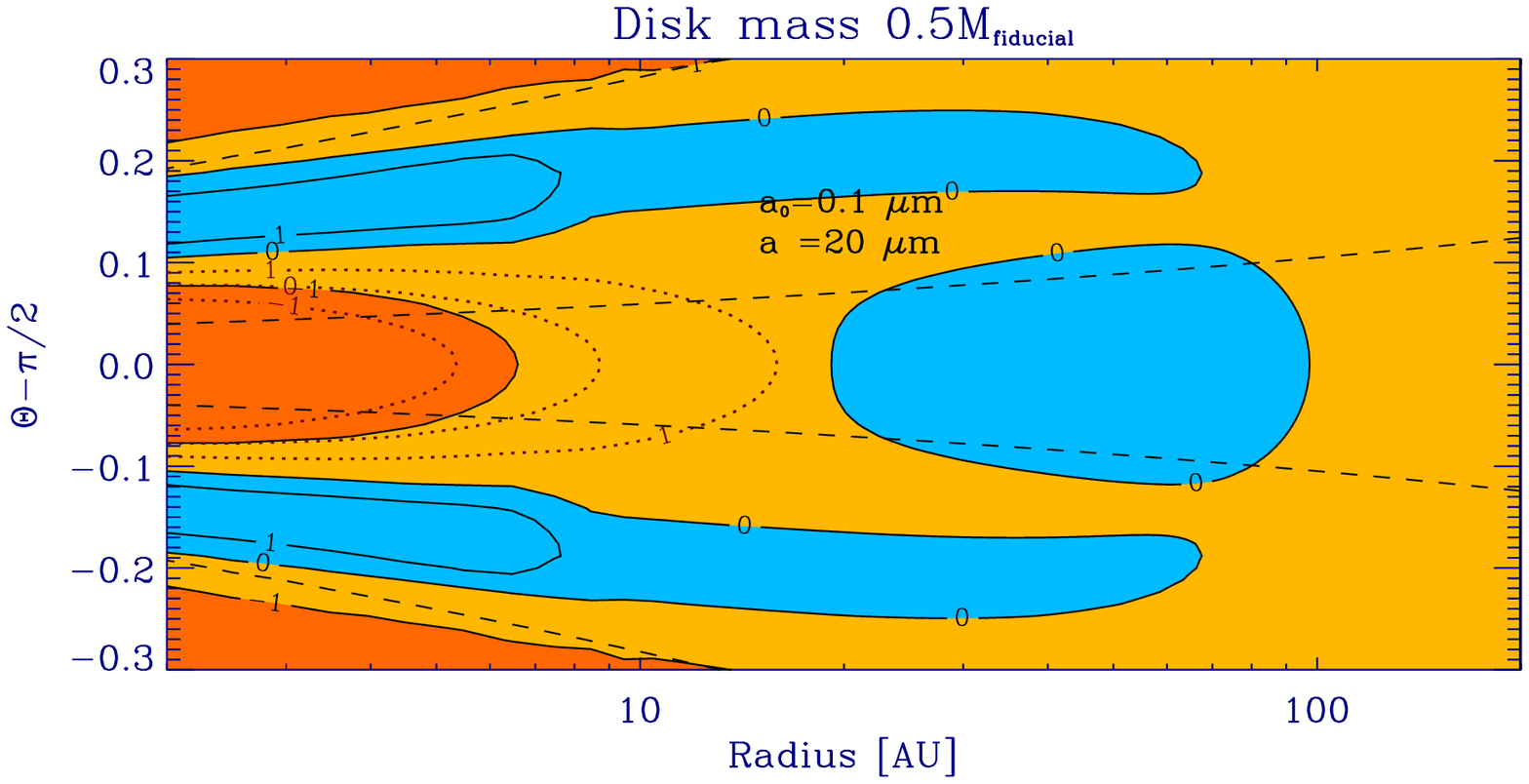}
}
\hbox{
\includegraphics[width=3.5in]{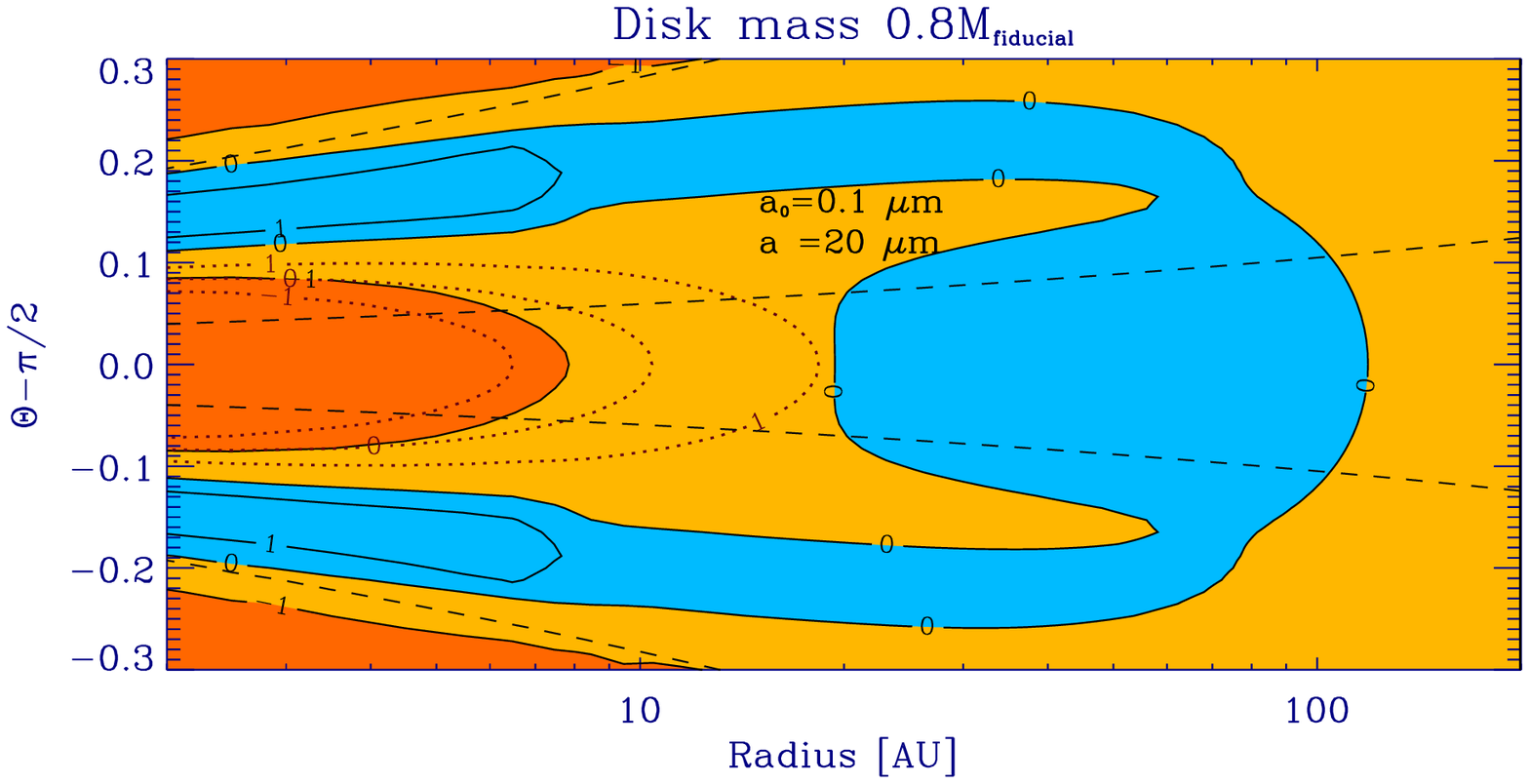}
\includegraphics[width=3.5in]{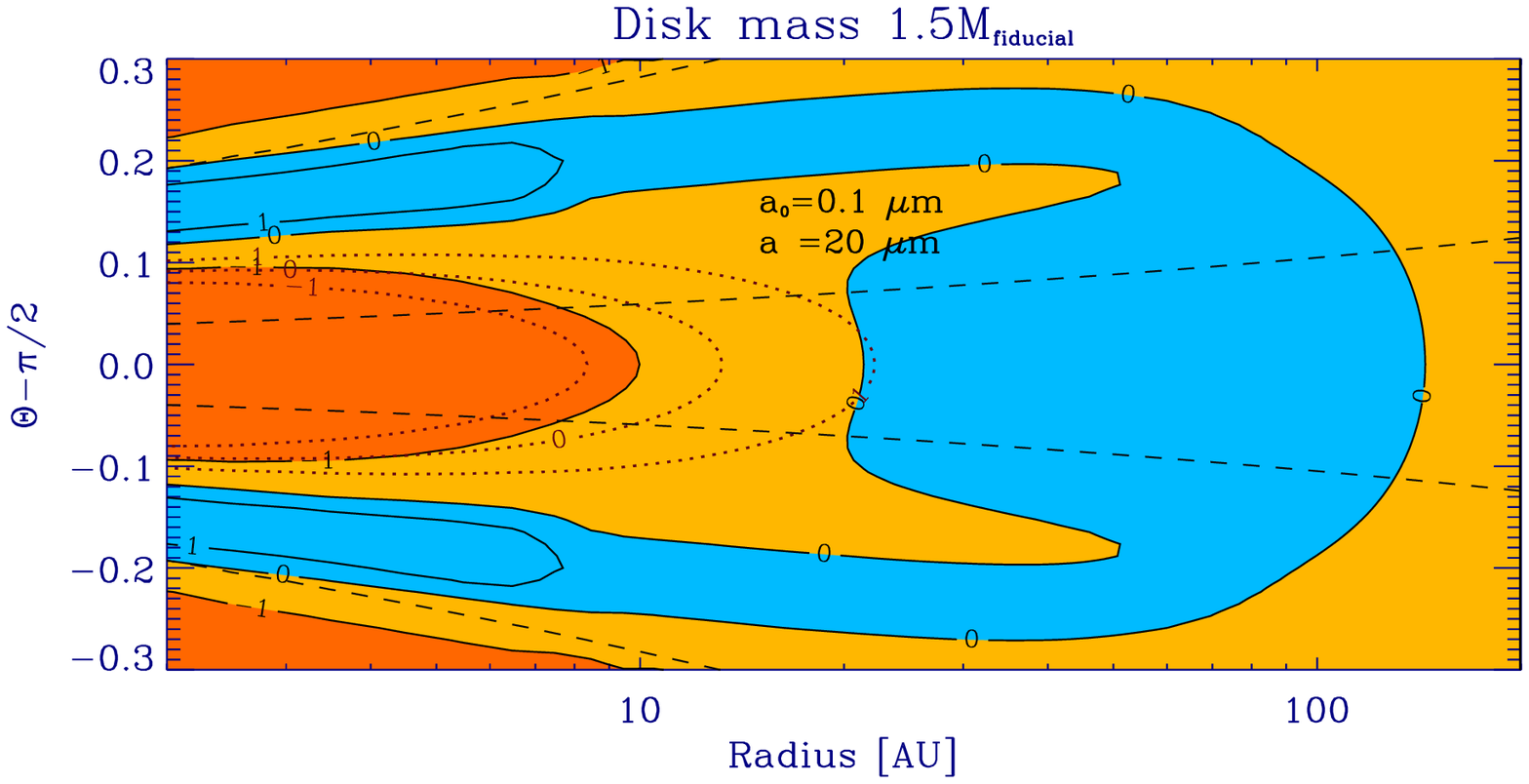}
}
\caption{ MRI-active regions' appearance for disk mass 0.2, 0.5, 0.8
  and 1.5 times the fiducial model.  Colors and symbols are as in
  figure~\ref{fi-els}.  }
\label{disk1}
\end{center}
\end{figure}

In set~3 from table~\ref{sets}, we vary the disk mass by scaling the
fiducial model's densities from eq.~3 (section~2.1).  Increasing the
mass makes more of the outer disk active, if we hold fixed the ratio
of gas to magnetic pressure (Fig.~\ref{disk1}).  The reason is that
higher densities make the ambipolar diffusion term less important
relative to the induction term.  The outermost border of the active
zone shifts from 100~AU for the disk with $M_{\rm disk}=0.5 M_{\rm
  fiducial}$ to 140~AU for $M_{\rm disk}=1.5 M_{\rm fiducial}$,
following the gas isodensity contour $\rho=10^{-15}\rm\ g\,cm^{-3}$.
At the same time, increasing the disk mass has little effect on the
location and thickness of the active layers at higher latitudes.  The
active layer base simply moves slightly further from the midplane, as
does the layer absorbing the ionizing cosmic rays and X-rays.  The
active layer becomes a little thinner because its top is still
constrained by the requirement that magnetic pressure is not too much
greater than gas pressure.

For the smallest disk mass in set~3 ($0.2 M_{\rm fiducial}$)
MRI-active layers are found only at the higher latitudes, and the
midplane is either dead or transitional.  For larger disk masses, the
midplane active island appears (Fig.~\ref{disk1}, top right).  In the
two most massive disks from set~3, the active surface layers and
midplane island merge.  Over the same mass range, the interior dead
zone grows due to the reduced ionization at the higher mass columns.

The transitional region between active and dead zones generally
narrows with increasing disk mass, as the expanding inner dead zone
approaches the expanding outer active island.  The transition region
is 13~AU, 12~AU and 10~AU wide for $M_{\rm disk}=0.5, 0.8$ and $1.5
M_{\rm fiducial}$.  At the lowest disk mass, the transitional region
stretches in the midplane from 4 to at least 200~AU.  A substantial
volume is thus capable of MRI turbulence at reduced levels.

\begin{figure}
\begin{center}
\includegraphics[width=4.5in]{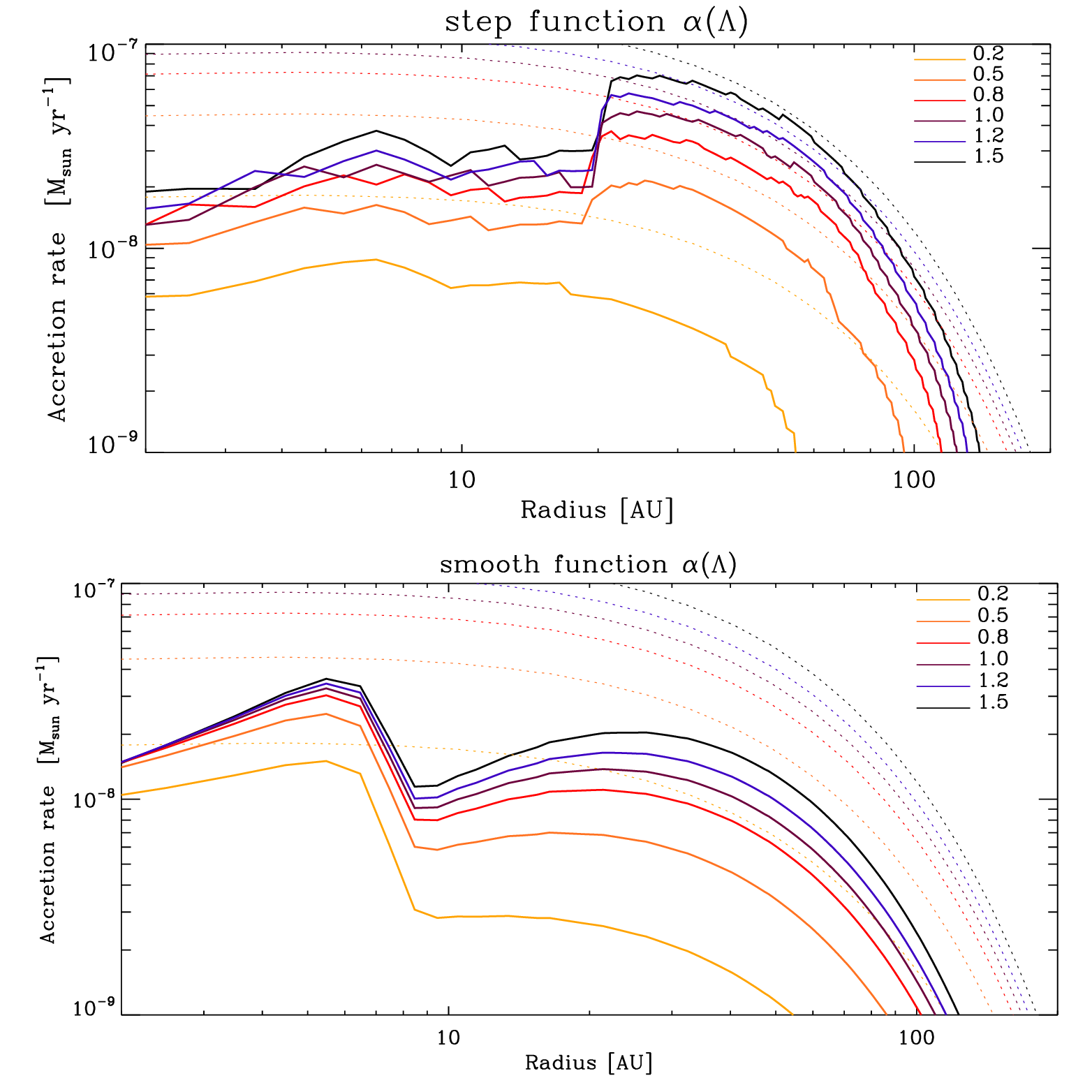}
  \caption{ Accretion rate vs.\ radius in set~3 where the disks have
    masses 0.2, 0.5, 0.8, 1, 1.2 and 1.5 times the fiducial model.
    Dotted lines show the accretion rates if the disks were active
    throughout.  \label{fig:set3mdot}}
\label{disk2}
\end{center}
\end{figure}

The radial profiles of the accretion rate are shown in
Fig.~\ref{disk2}.  Note that we keep the dust-to-gas ratio and monomer
size same as in the fiducial model.  The accretion rate is about
$10^{-8}$~M$_\odot$~yr$^{-1}$ at our inner boundary.  For the smooth
$\alpha(\Lambda)$ function the accretion rate dips at 9-10~AU near the
metal line.  In the more massive disks, the accretion profile 
becomes less  dependent of  mass.

The radial profiles of the accretion rate are shown in
Fig.~\ref{disk2}.  Note that we keep the dust-to-gas ratio and monomer
size the same as in the fiducial model.  The accretion rate is about
$10^{-8}$~M$_\odot$~yr$^{-1}$ at our inner boundary.  For the smooth
$\alpha(\Lambda)$ function the accretion rate dips at 9~AU near the
metal line.  In the more massive disks, the accretion profile is
almost independent of the mass.

\subsection{Surface Density Slope}

In set~4 (Table~\ref{sets}) we vary the power-law index $p$ in the
surface density profile $\Sigma\propto r^{-p}$ from 0.5 to 1.5,
keeping the disk mass and other parameters as in the fiducial model.
The slope $p=0.5$ is frequently adopted for numerical simulations
\citep{fro06,dzy10,flo11}.  Millimeter interferometer measurements of
the dust continuum emission yield a median index $p=0.9$
\citep{and09}, while Hayashi's minimum-mass Solar nebula model has
$p=1.5$ to approximate the distribution of mass in the Solar system.

\begin{figure}
\begin{center}
\hbox{
\includegraphics[width=3.5in]{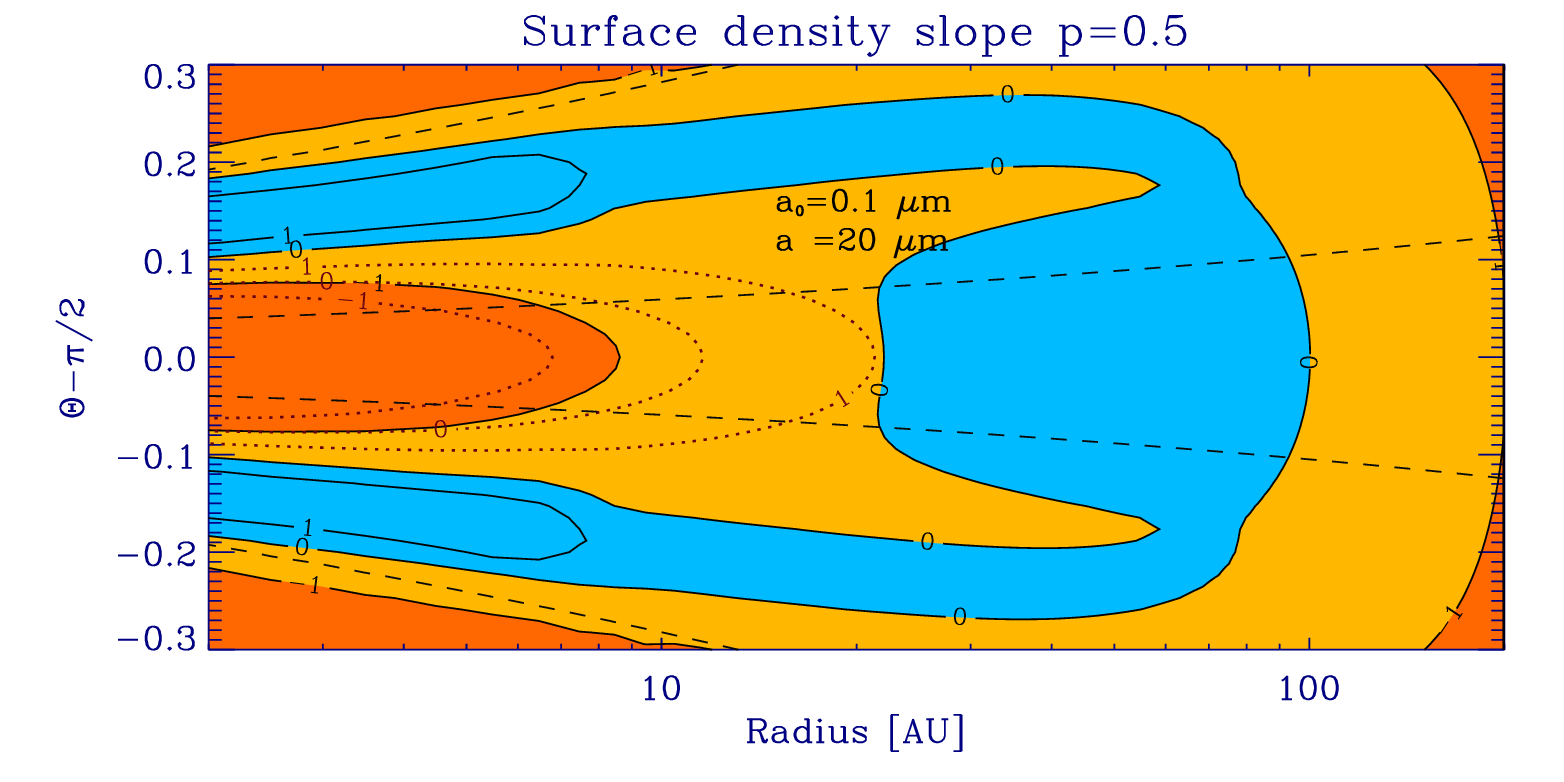}
\includegraphics[width=3.5in]{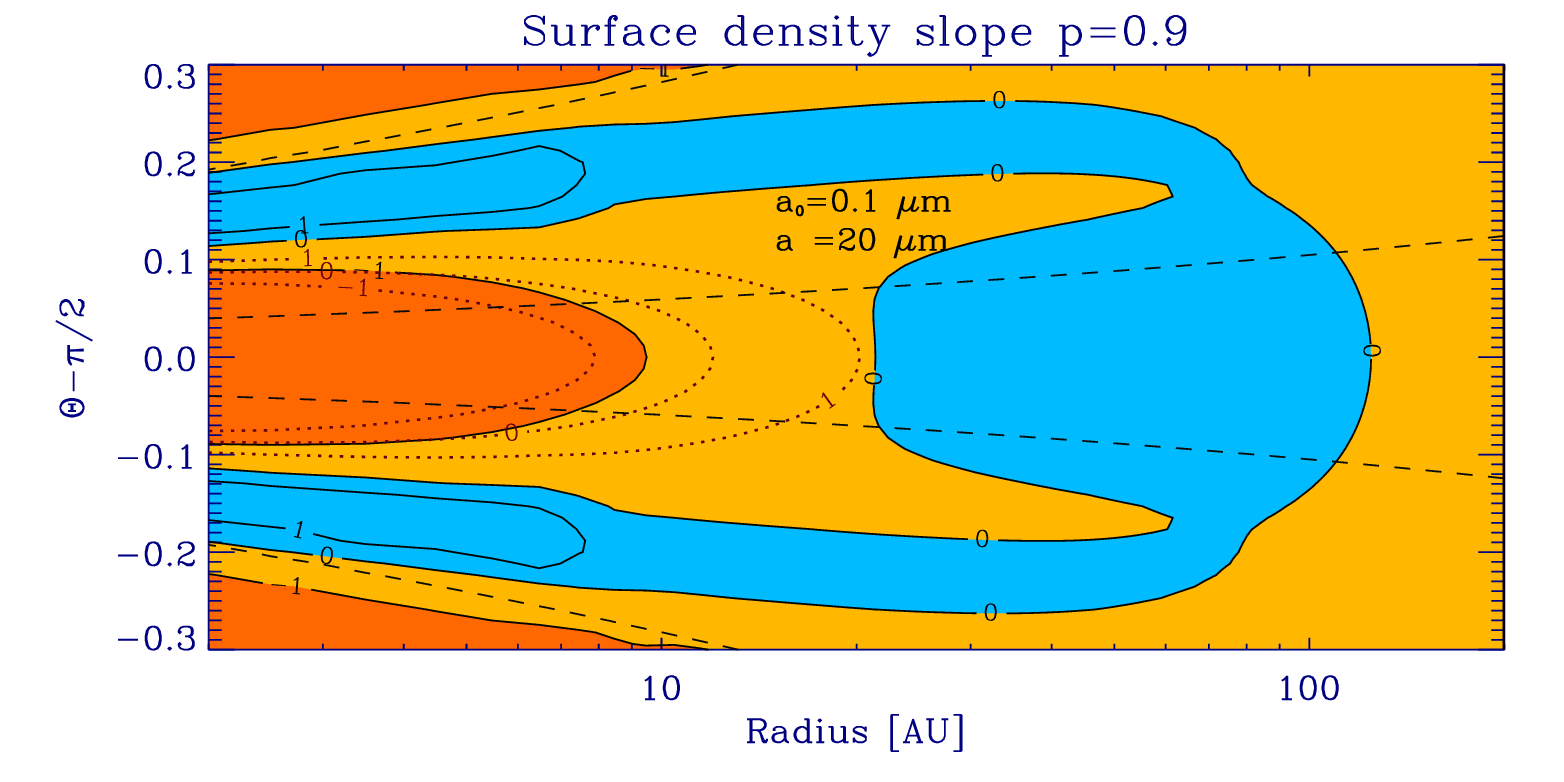}
}
\hbox{
\includegraphics[width=3.5in]{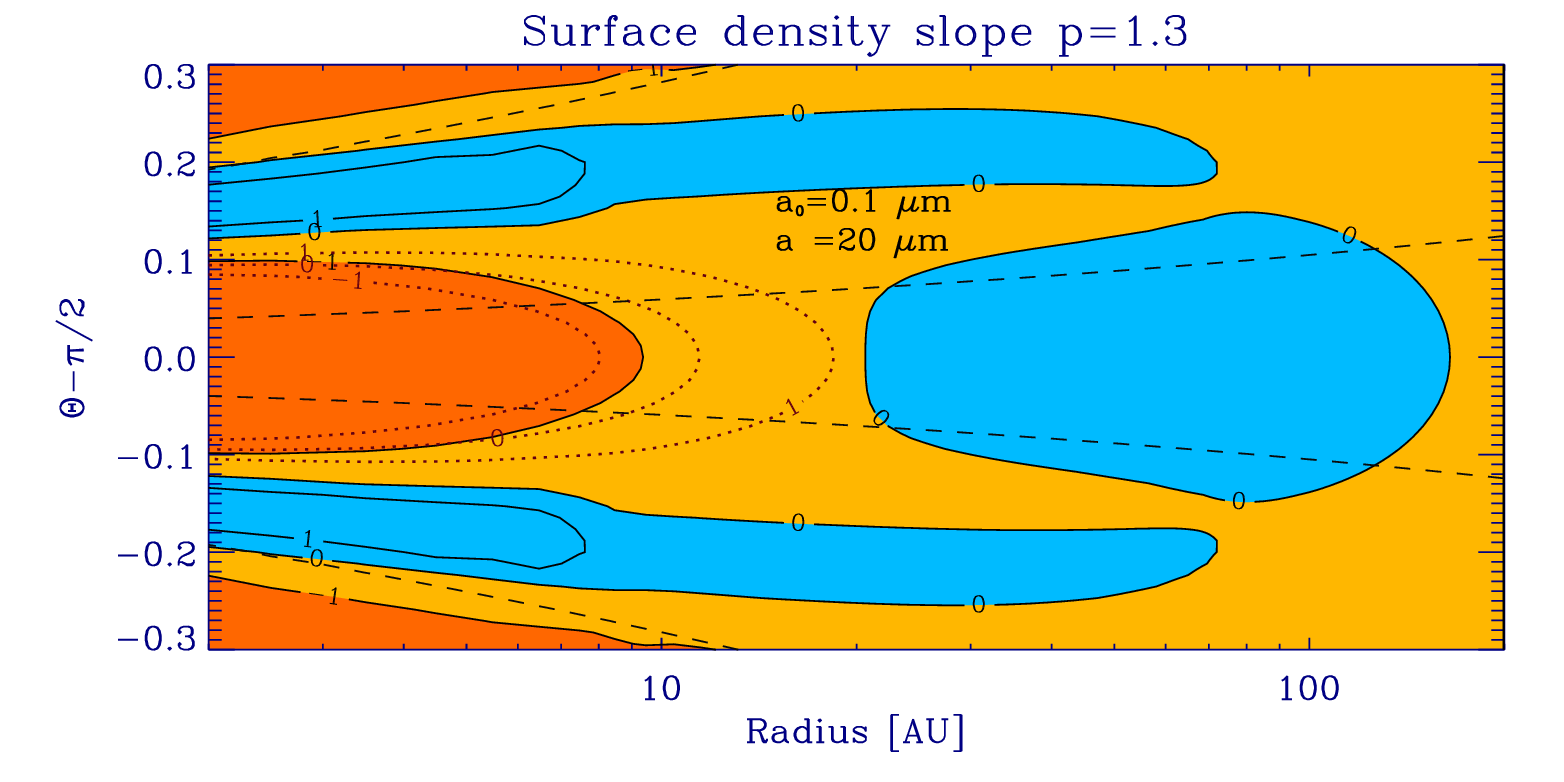}
\includegraphics[width=3.5in]{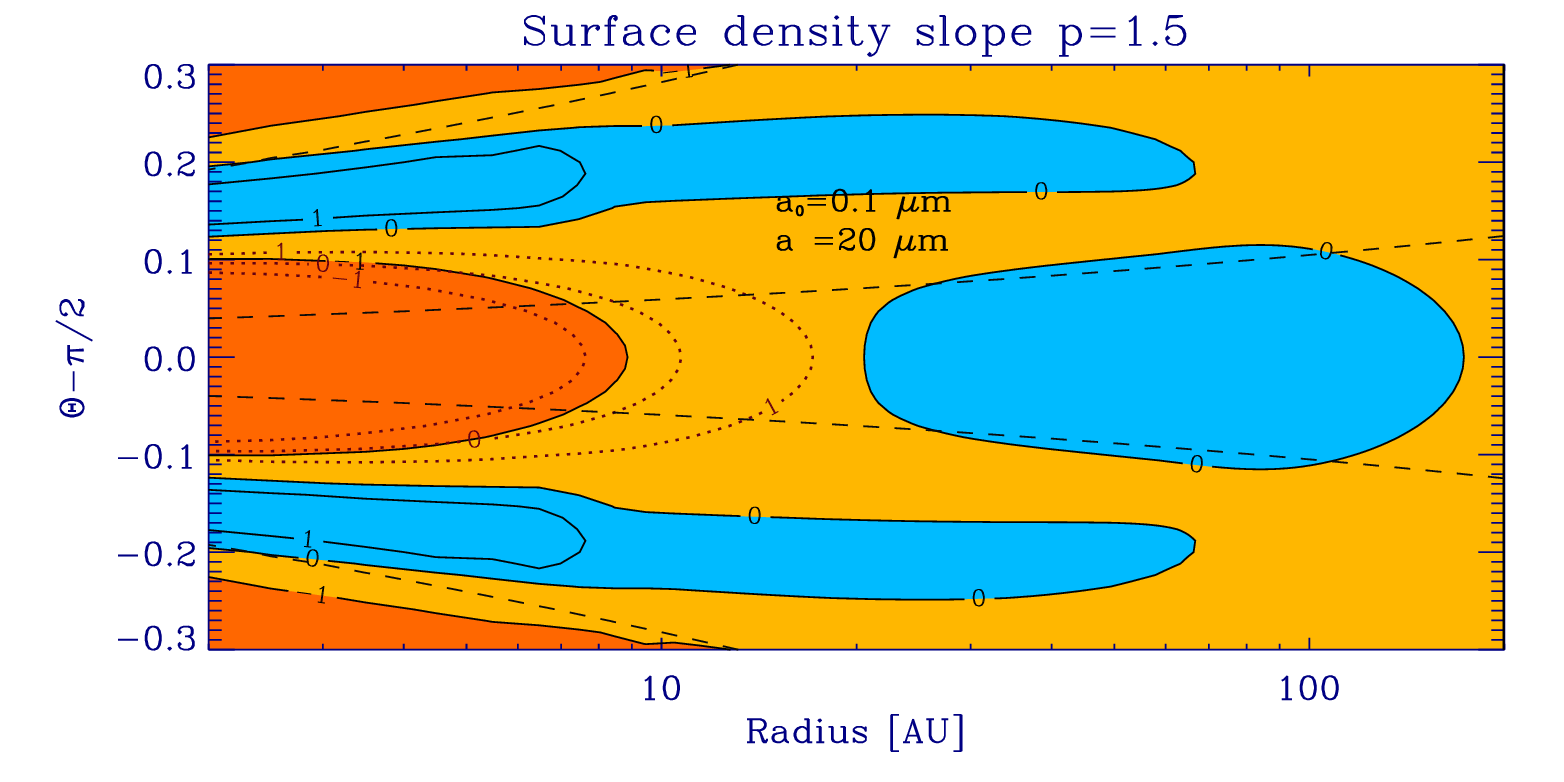}
}
  \caption{ MRI-active regions appearance for surface density slope
    0.5, 0.9, 1.3, 1.5, assuming a constant disk mass.  Colors and
    symbols are as in figure~\ref{fi-els}.  }
\label{slope1}
\end{center}
\end{figure}

With increasing surface density slope, the active zone expands outward
from 100~AU to 195~AU (Fig.~\ref{slope1}).  At the same time the
midplane dead zone thickens near 1~AU.  For slopes $p$ between 0.5 and
1.3 we obtain a single MRI-active region with a fish-tail shaped
transitional layer.  For $p=1.5$ the outer disk's low densities make
the ambipolar diffusion stronger still.  The transitional region is
larger and isolates the magnetic activity in the two surface layers
from the midplane island of activity.  The transitional layer between
the island and active layers at 50~AU (Fig.~\ref{slope1}, bottom
right), is only about one scale height thick.  As MRI turbulence can
overshoot the dead zone boundary by a scale height \citep{tur10}, the
transitional layers could be turbulent thanks to the activity above
and below.

\begin{figure}
\begin{center}
\includegraphics[width=4.5in]{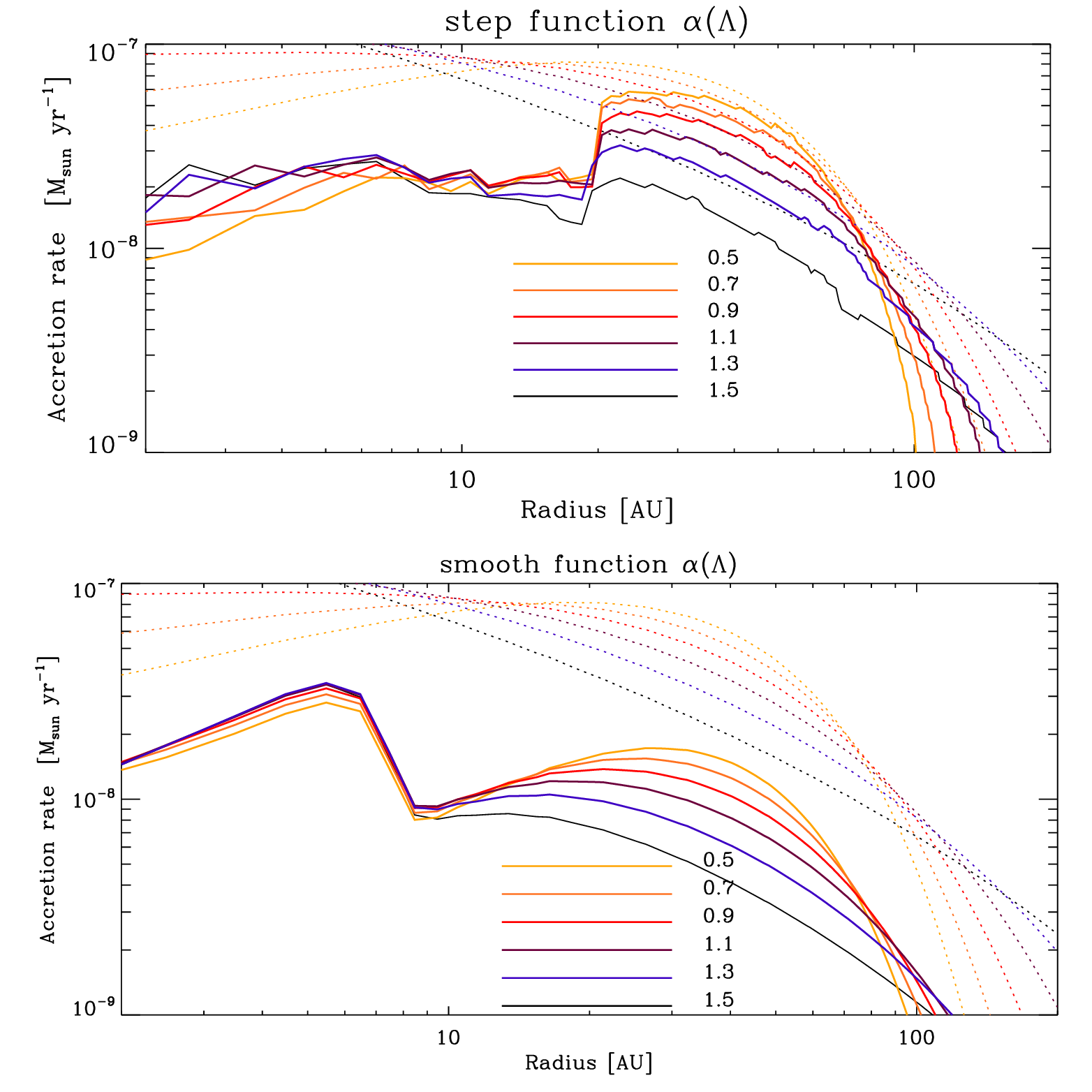}
  \caption{ Accretion rate vs.\ radius in the disks with surface
    density slopes $p=0.5, 0.9, 1.3$ and $1.5$.  Dotted lines show the
    accretion rates if the disks were active
    throughout.  \label{fig:set4mdot}}
\label{slope2}
\end{center}
\end{figure}

In Fig.~\ref{slope2} we plot the accretion rates for the models in
set~4.  The inner border of the MRI-active region moves inward very
little as the slope is increased.  Similar to the results in the
previous section, the accretion flow shows no gaps under the
step-function $\alpha(\Lambda)$.  The accretion rate is about
$10^{-8}$~M$_\odot$~yr$^{-1}$ at our 1~AU inner boundary.  For the
smooth $\alpha(\Lambda)$ function, $\dot{M}$ peaks near the metal
line.  The accretion rate dip is deeper for shallower surface density
profiles.

\subsection{Temperature}

Here we explore Set~5 (Table~\ref{sets}) in which we vary the gas
temperatures.  The domain we consider remains too cool for significant
thermal ionization.  Nevertheless the temperature is important for the
conductivity because it regulates the disk thickness and hence the gas
density, determines the gas-phase metal abundance, and sets the
particles' thermal speeds.  We consider a range of temperatures from
those observed at the starlight-heated surfaces of disks in nearby
star-forming regions \citep{bouw01,fur09} down to those inferred for
the disk midplanes through radiative transfer modeling
\citep{chi97,dul07}.  We consider temperatures at 1~AU, $T_0$, between
150 and 600~K, keeping the temperature's radial power-law slope
constant at $-0.5$.

\begin{figure}
\begin{center}
\hbox{
\includegraphics[width=3.5in]{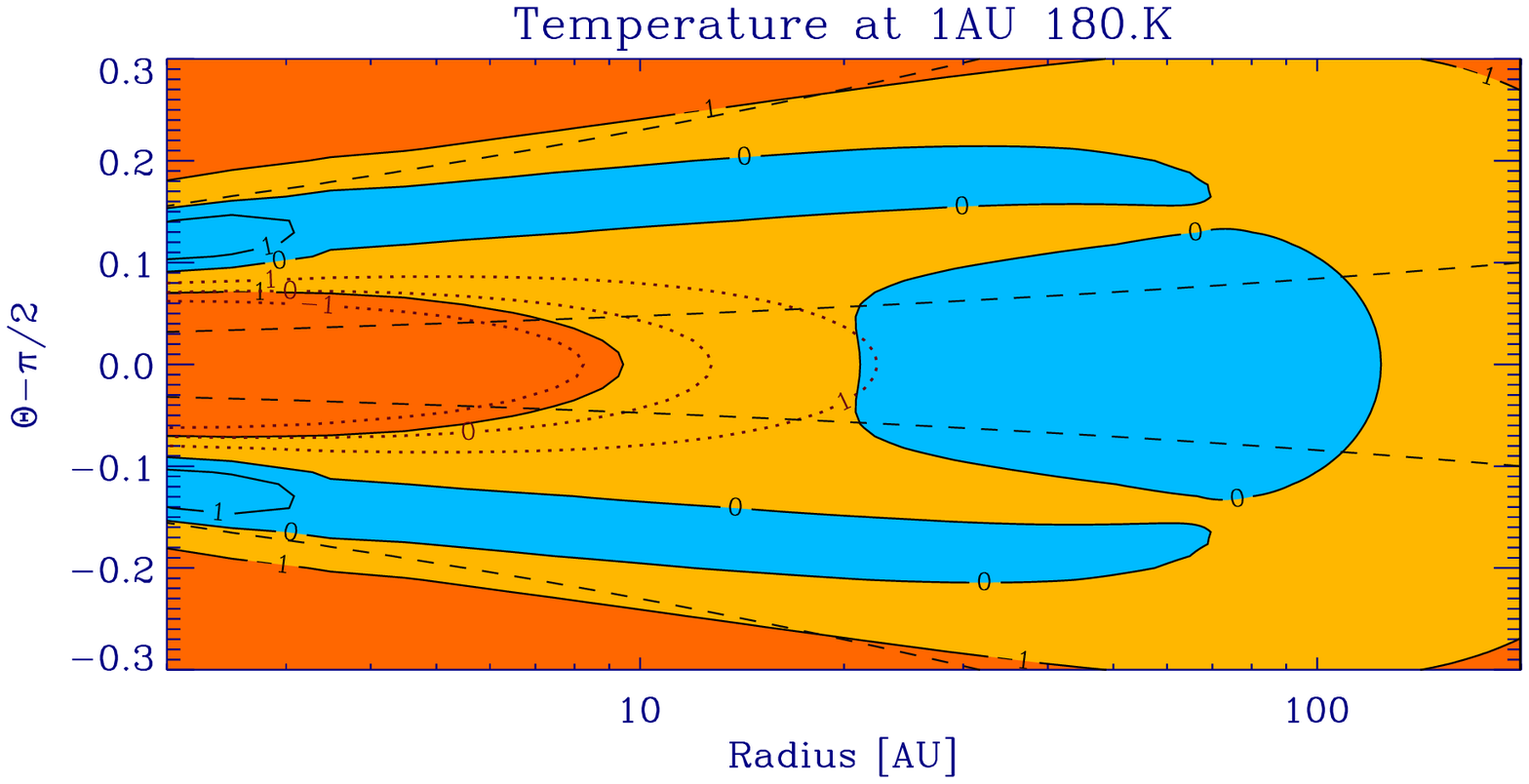}
\includegraphics[width=3.5in]{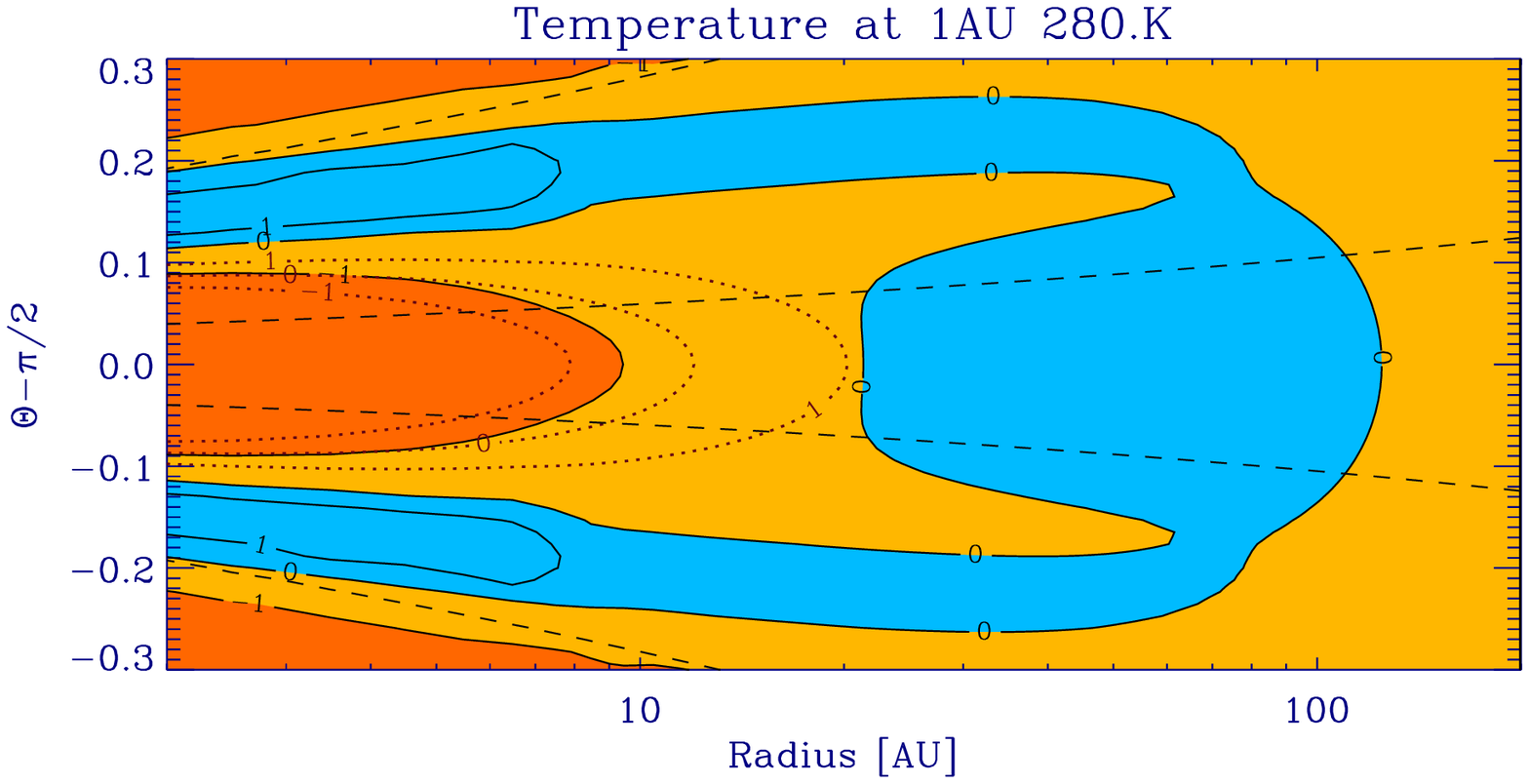}
}
\hbox{
\includegraphics[width=3.5in]{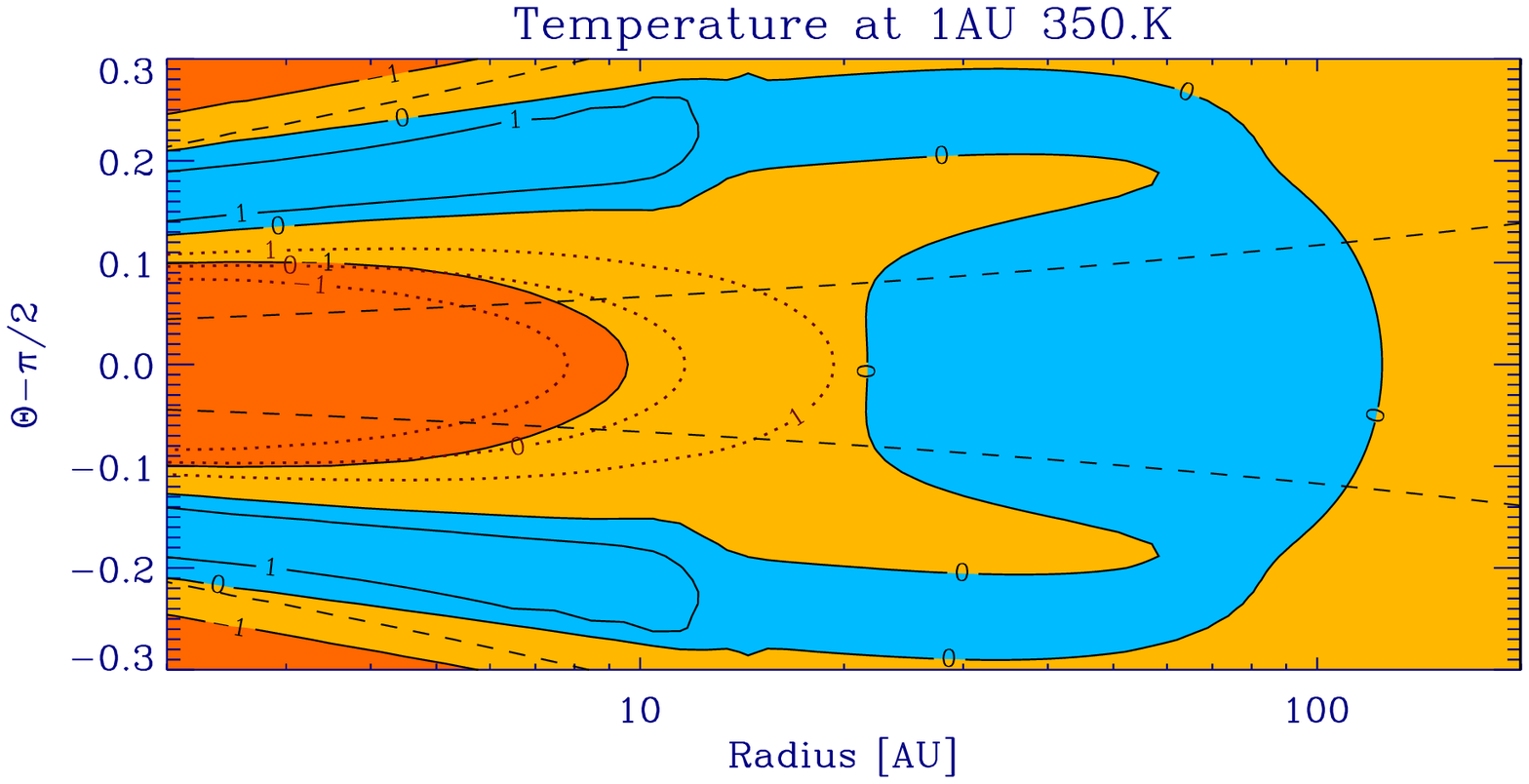}
\includegraphics[width=3.5in]{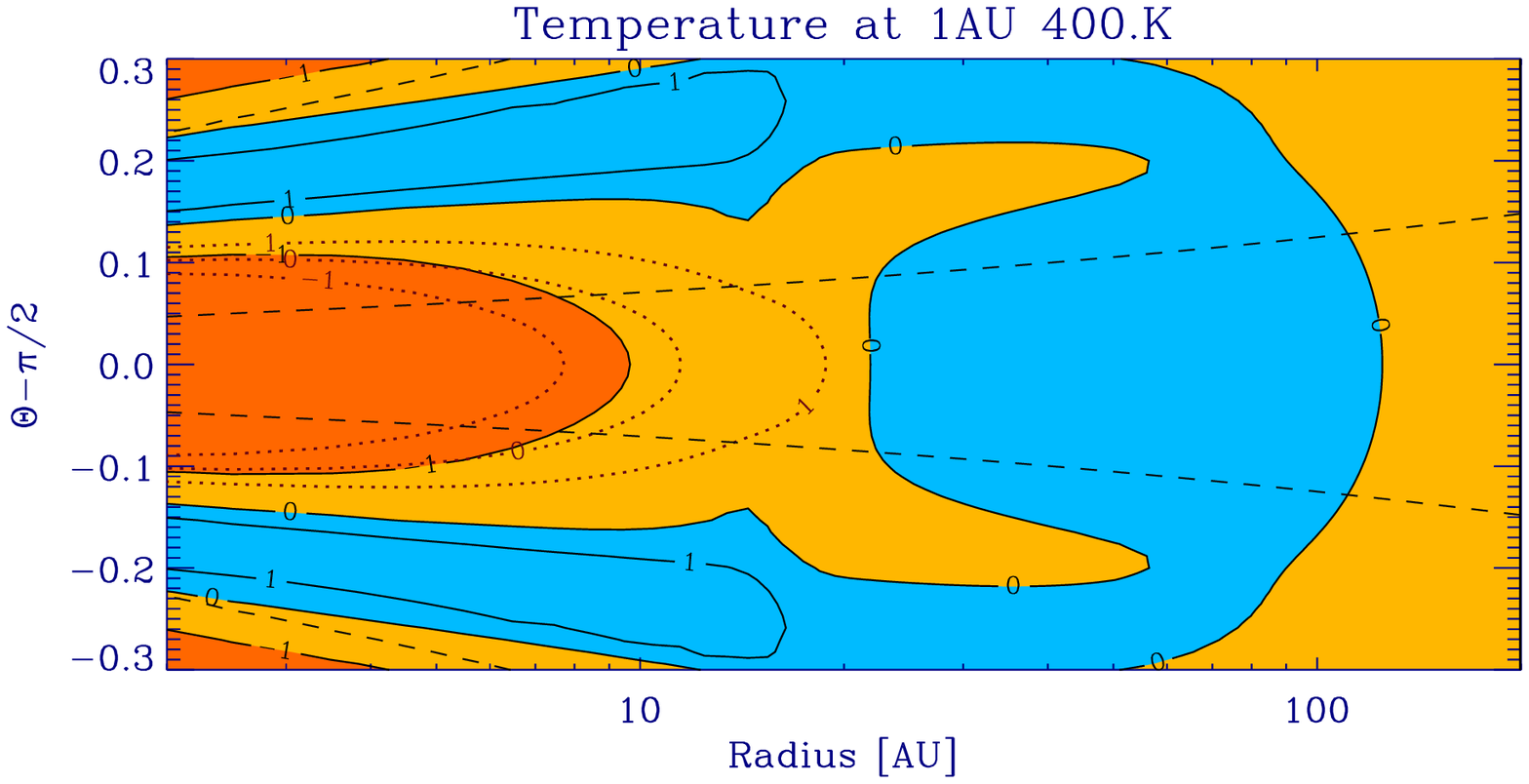}
}
\hbox{
\includegraphics[width=3.5in]{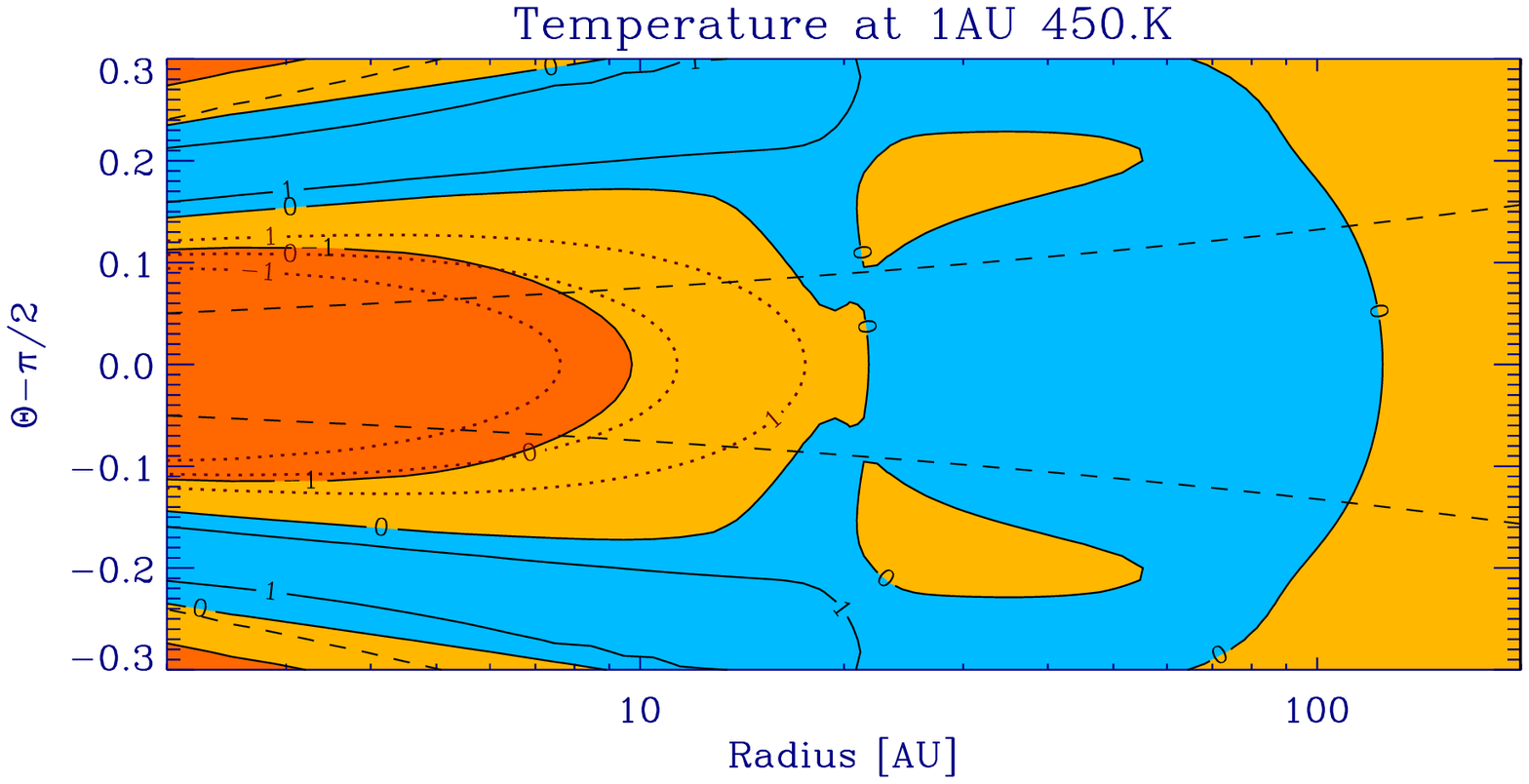}
\includegraphics[width=3.5in]{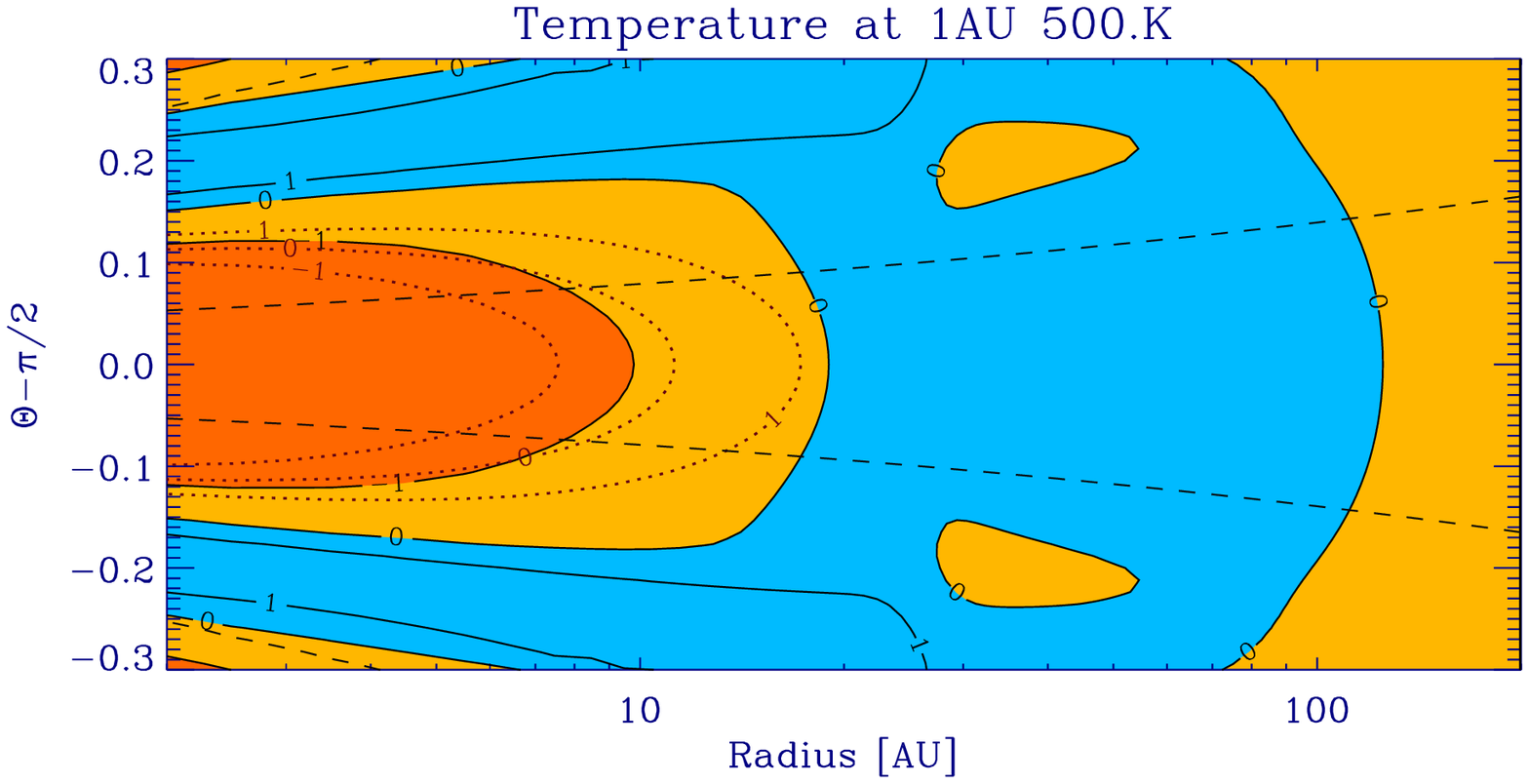}
}
  \caption{ Magnetic coupling maps for disks with a range of
    temperature profiles $T=T_0{(r/1AU)}^{-0.5}$, with $T_0=180,
    \ldots, 500$~K.  Colors and symbols are as in figure~\ref{fi-els}.
  }
\label{temp1}
\end{center}
\end{figure}

The resulting dead zones are shown in Fig.~\ref{temp1}.  In each case
the MRI-active layers contract across the radius where the metal ions
freeze out.  The magnesium in our models freezes out at 2.5~AU if
$T_0=180$~K, 4.5~AU if 280~K and 11~AU if 350~K.  At $T_0=450$~K the
MRI-active layers reach deep towards the midplane.  We observe that
the transitional layer's fish-tail shape is present only for
$T_0<450$~K.  For still higher temperatures, simple almond-shaped dead
and surrounding transitional zones appear.  The remnants of the
fish-tail appear as islands in yellow roughly $2H$ above and below the
midplane (Fig.~\ref{temp1}, bottom).  Although the islands stretch
over several tens of AU in radius, their vertical thickness is less
than one scale height, and they could conceivably be activated by the
neighboring better-coupled regions.

\begin{figure}
\begin{center}
\includegraphics[width=4.5in]{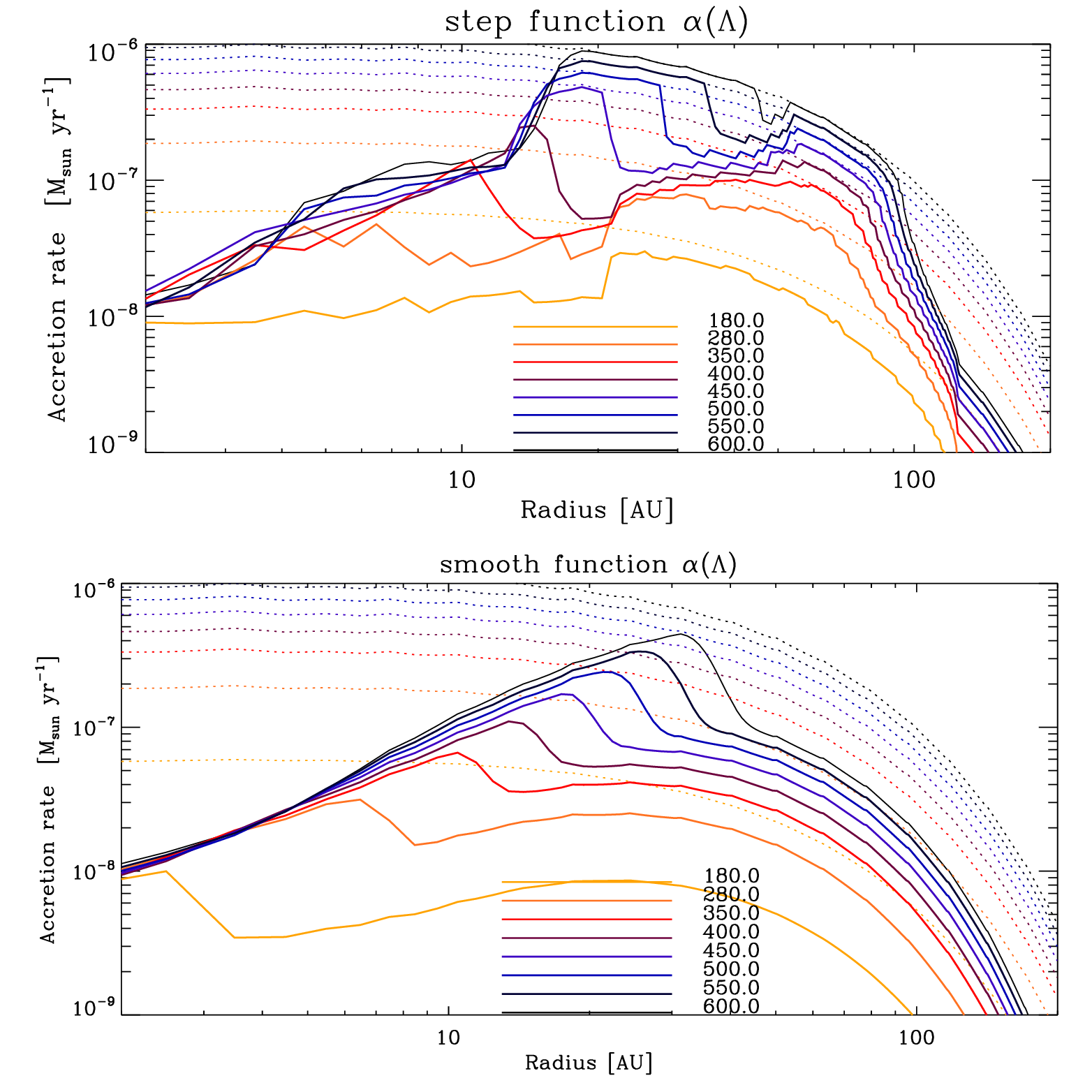}
  \caption{ Accretion rate vs.\ radius in disks with various
    temperature profiles $T=T_0 ({r}/1AU)^{-0.5}$ where $T_0=180,
    \ldots, 600$~K.  Dotted lines indicate the accretion rates when
    the whole disk is active.  \label{fig:set5mdot} }
\label{temp2}
\end{center}
\end{figure}

The accretion rates for the models in set~5 are shown in
Fig.~\ref{temp2}.  The islands of transitional coupling for
temperatures $T_0$ from 450 to 600~K, seen in yellow in the bottom
panels of Fig.~\ref{temp1}, lead to small reductions in the accretion
rate if we take $\alpha(\Lambda)$ to be a step function.  If we
instead use the smooth $\alpha(\Lambda)$ function, the accretion rate
peak shifts outward following the isotherm $T\approx 100$~K, reaching
50~AU for the hottest disk (Fig.~\ref{temp2}, bottom).

\subsection{Stellar Mass}

In set~6 (Table~\ref{sets}) we vary the stellar mass, which determines
the orbital frequency, and through the vertical component of gravity
affects the disk's gas density.  Furthermore the stellar luminosity
grows with the stellar mass, affecting disk temperatures.  We let the
disk mass be proportional to the stellar mass, $M_{\rm disk}=0.064
M_*$.  Luminosities for stars between 0.3 and 2~$M_\odot$ are taken
from evolutionary tracks computed with the STELLAR code \citep{bod07}.
For simplicity we neglect the influence of the gas accretion on the
evolutionary tracks.

\begin{figure}
\begin{center}
\includegraphics[width=5.5in]{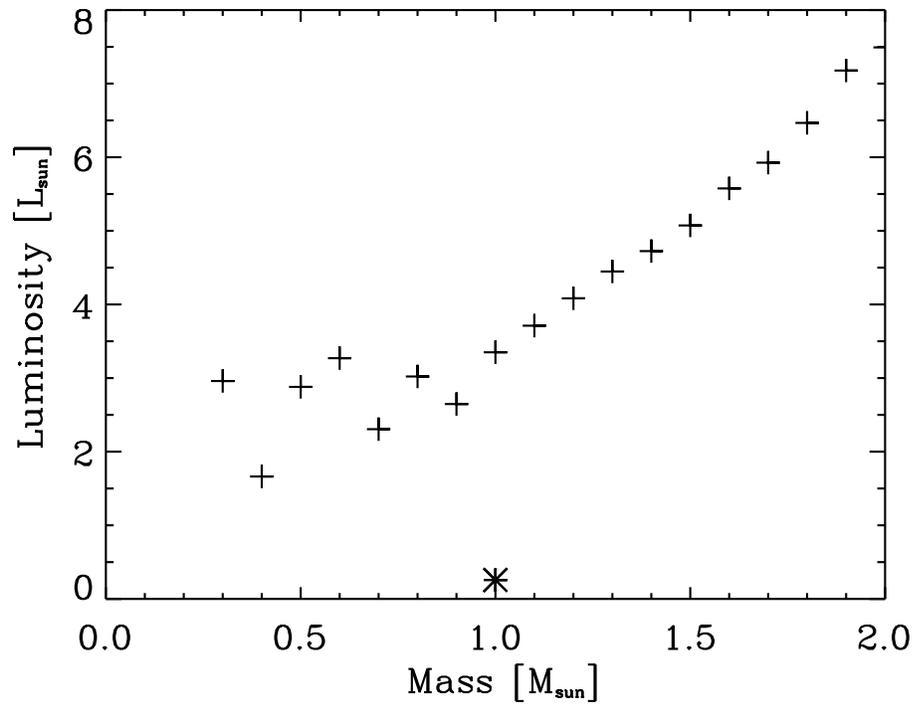}
  \caption{ Stellar luminosity vs.\ stellar mass from evolutionary
    tracks at 1~Myr.  A star symbol indicates the stellar luminosity
    in our fiducial model.  }
\label{star1}
\end{center}
\end{figure}

The stellar luminosity is plotted vs.\ the stellar mass in
Fig.~\ref{star1}.  We choose the time-step closest to 1~Myr as a
typical age for disk-bearing stars.  From the luminosity we
reconstruct temperatures in the circumstellar disk as follows.  The
location of the dust sublimation front, where the temperature is about
$T_s=1500$~K, can be calculated as follows:
\begin{equation}
  R_{in}
  =R_* \left(\frac{T_*}{T_s}\right)^2
  =\left(\frac{L_*}{4\pi\sigma T_s^4}\right)^{1/2}.
\end{equation}
From this point we assume the temperature falls off as $T = T_s
\sqrt(R_{in}/R)$ corresponding to the optically-thin approximation
used in the minimum-mass Solar nebula model on which our fiducial
model is based.  This overestimates the temperature since the disk is
optically-thick, but our point here is to see whether the dead zone
depends on the temperature.

For masses 0.4, 0.8, 1.2, 1.6 and 2~$M_\odot$, the temperatures at
1~AU are 446.8, 518.7, 559.3, 604.6 and 650.9~K, respectively.  The
Solar-mass star aged 1~Myr has a higher luminosity than our fiducial
model, yielding a temperature at 1~AU of 532.3~K instead of 280~K.

\begin{figure}
\begin{center}
\hbox{
\includegraphics[width=3.5in]{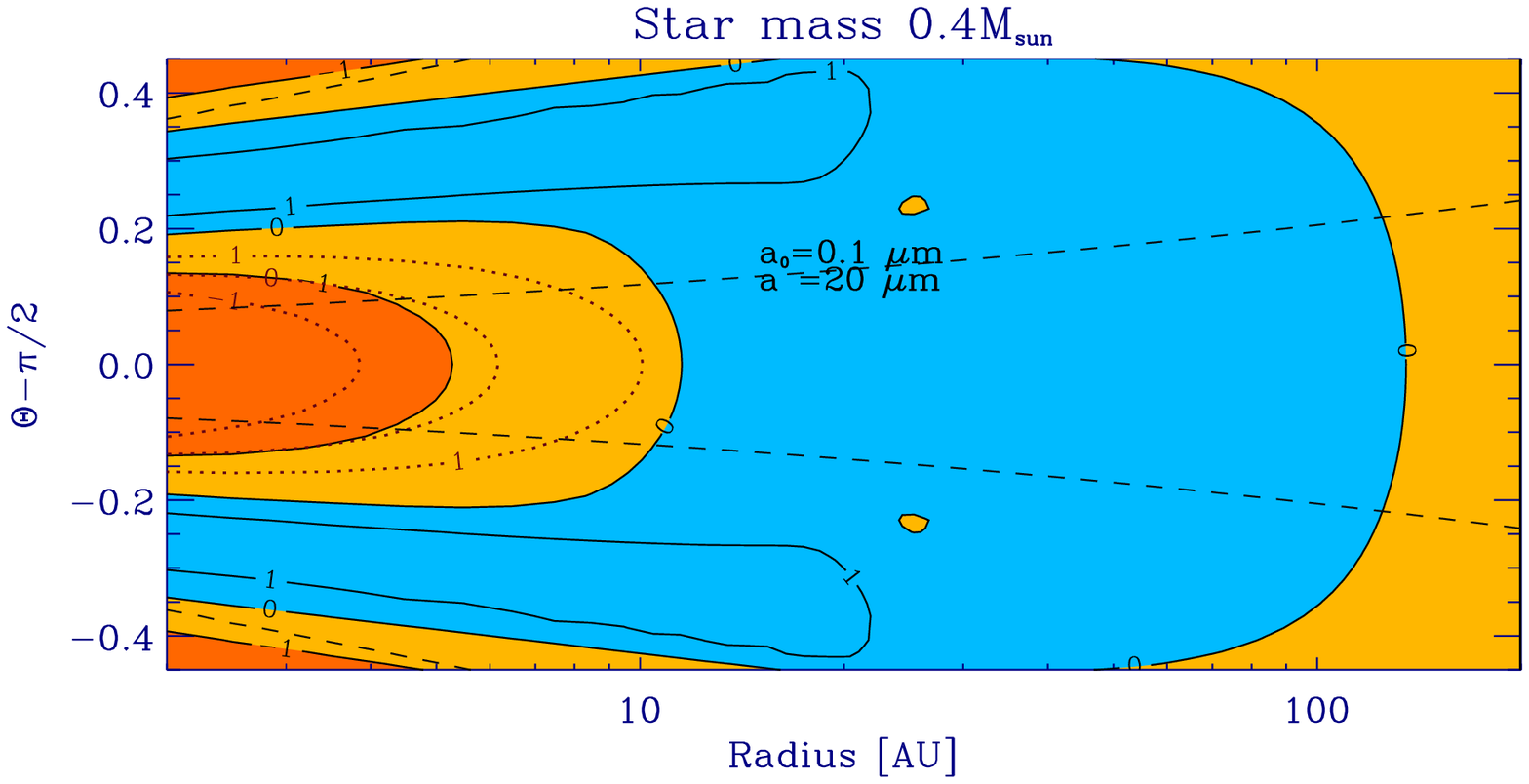}
\includegraphics[width=3.5in]{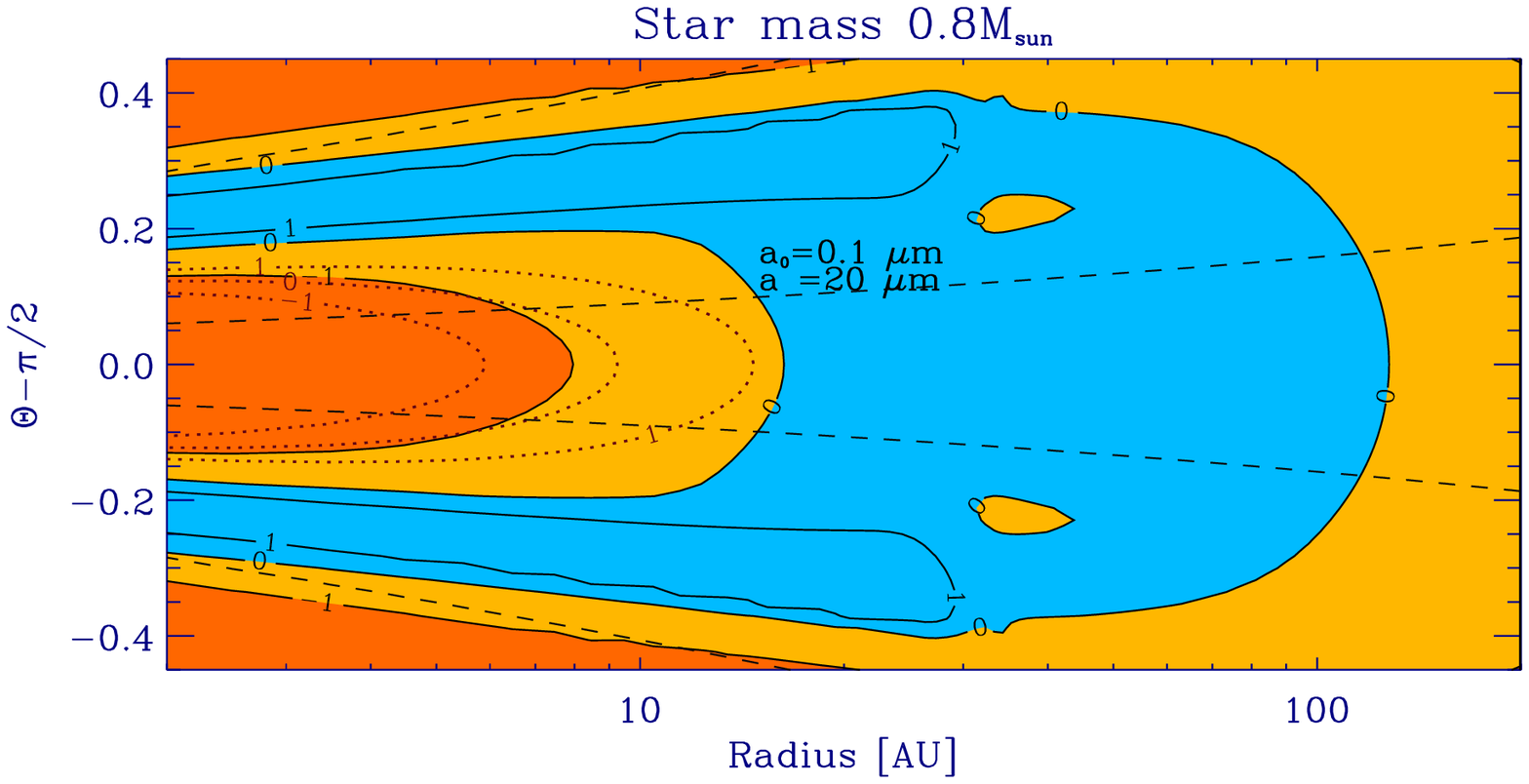}
}
\hbox{
\includegraphics[width=3.5in]{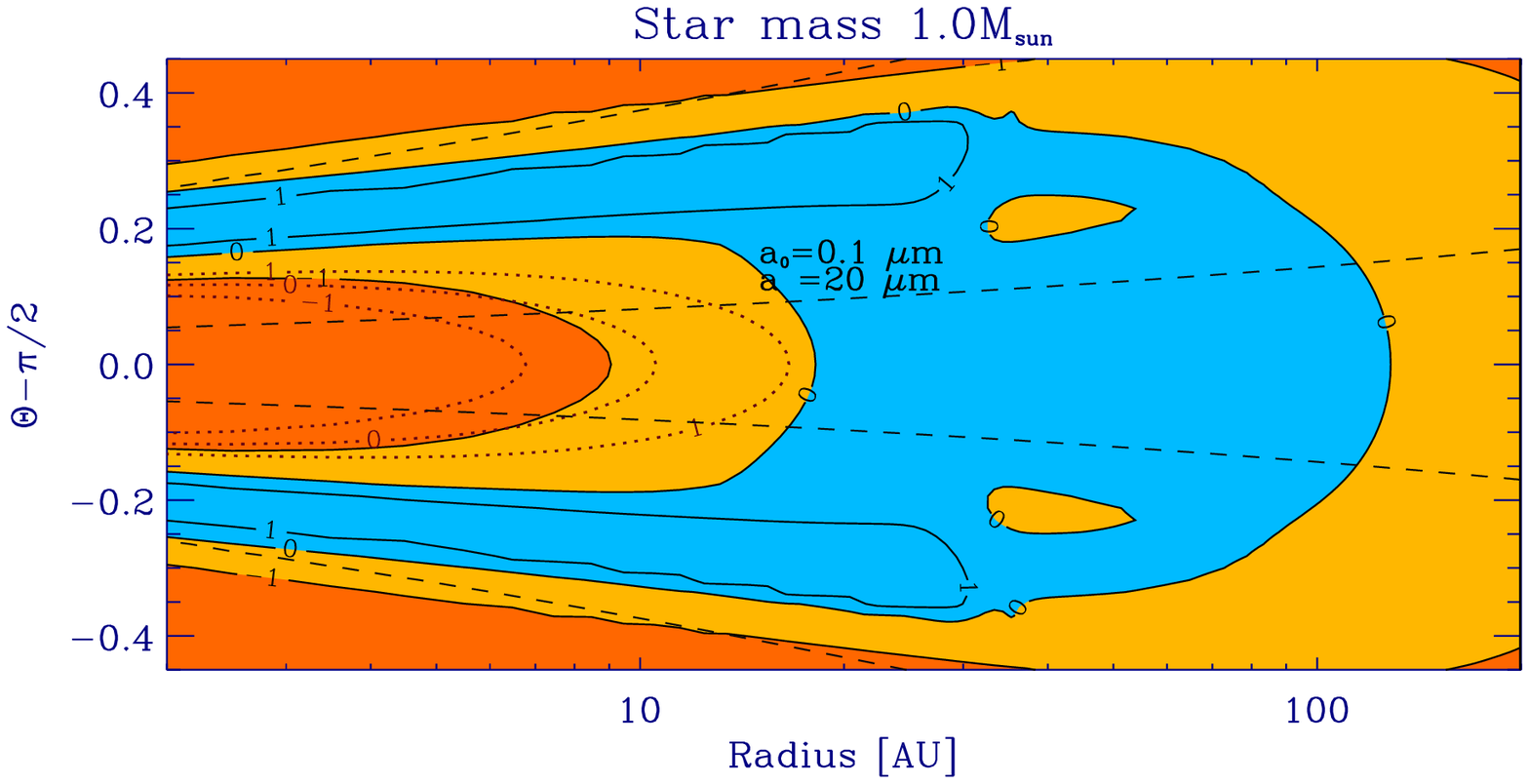}
\includegraphics[width=3.5in]{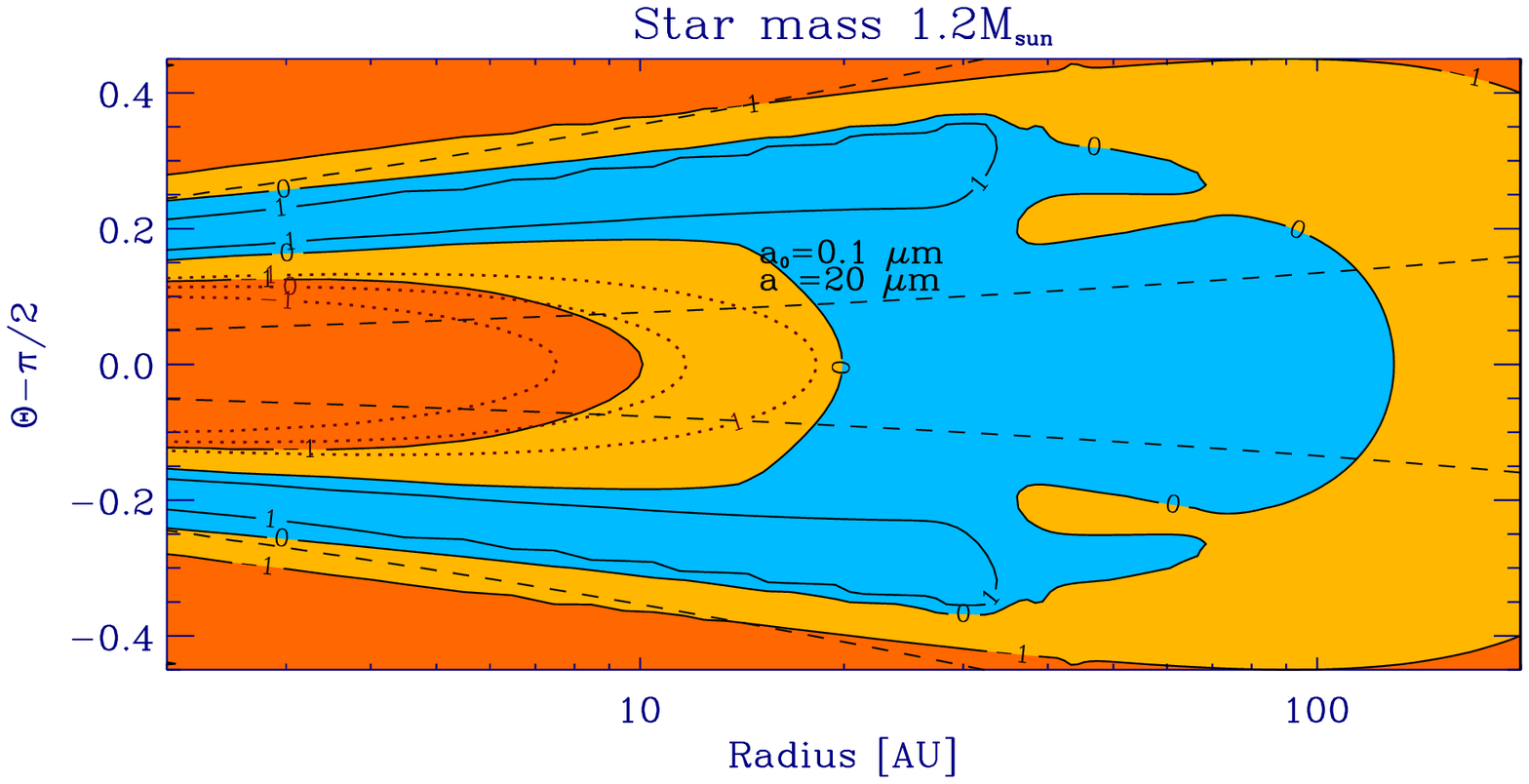}
}
\hbox{
\includegraphics[width=3.5in]{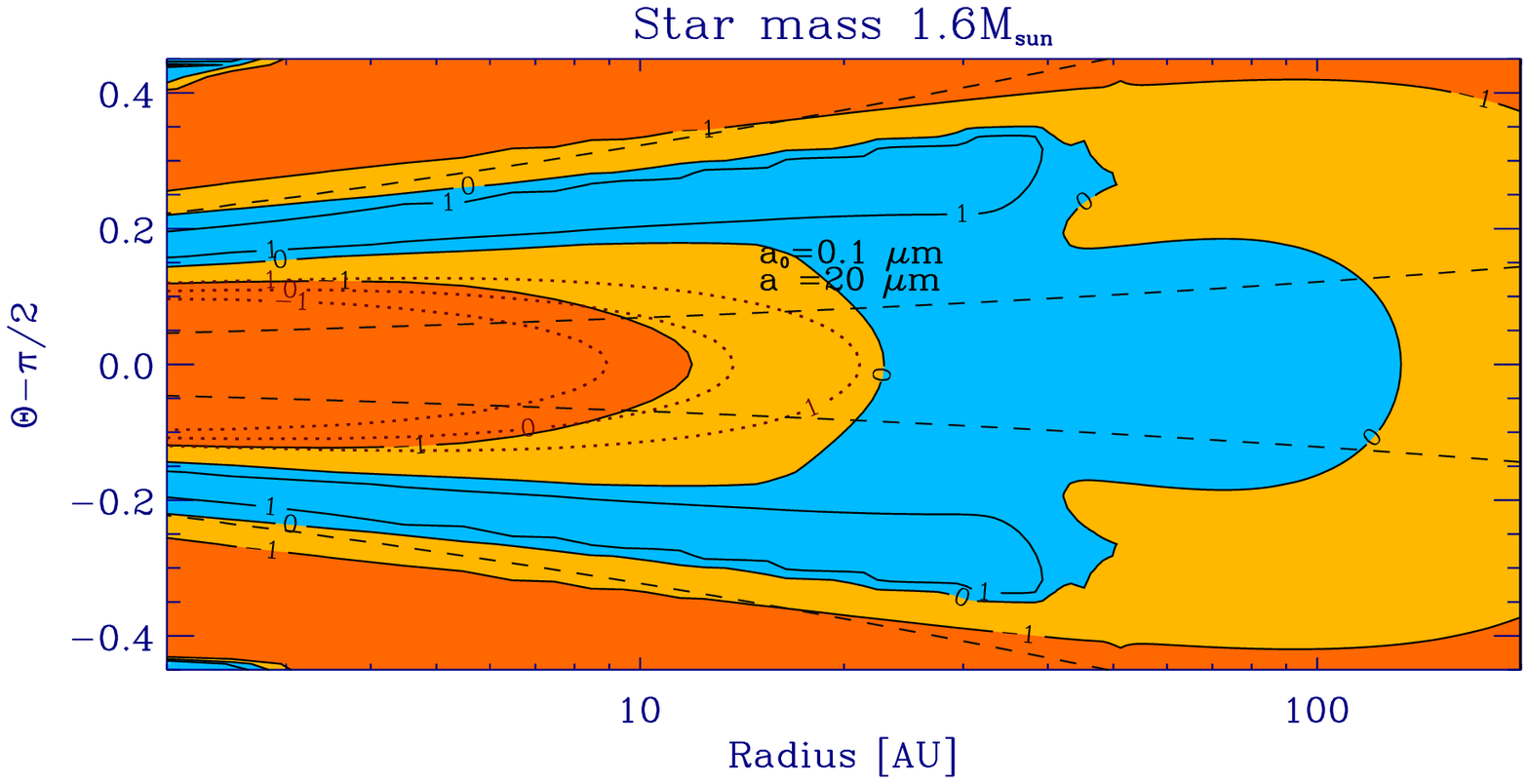}
\includegraphics[width=3.5in]{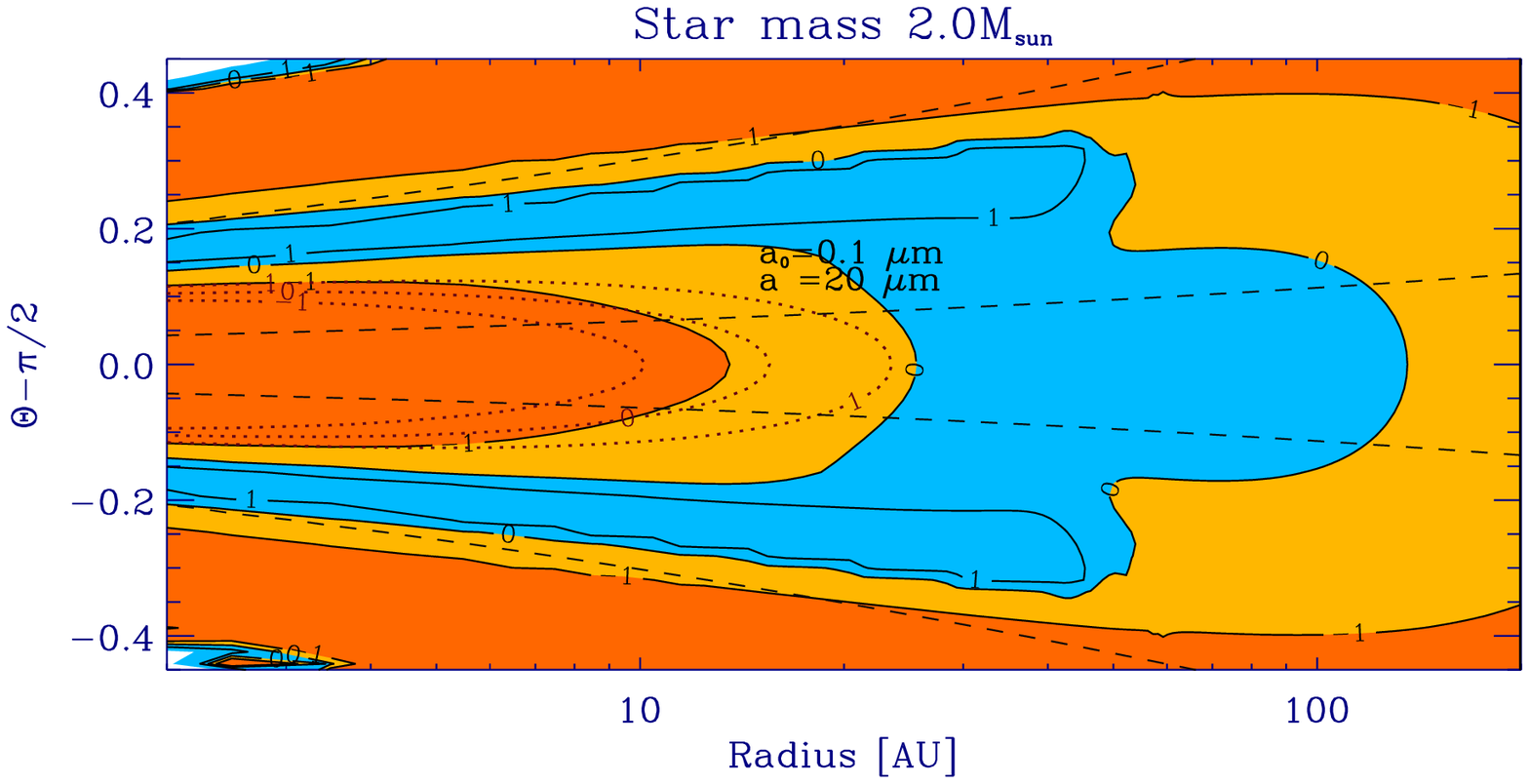}
}
\caption{ MRI-active region appearance for stars of various masses,
  with luminosities corresponding to age 1~Myr.  Colors and symbols
  are as in figure~\ref{fi-els}.  }
\label{star2}
\end{center}
\end{figure}

Higher luminosities stretch the metal-dominated region further from
the star than in the fiducial model (Fig.~\ref{star2}).  The Elsasser
number $\Lambda=10$ contour stretches along with the Mg$^+$-dominated
region, reaching 20~AU for the 0.4-$M_\odot$ star and 45~AU for the
2-$M_\odot$ star.  When the metal-dominated region reaches much beyond
10~AU, the fish-tail shape no longer appears in the transitional
layer.  Disks around stars with masses 0.4 to 1~$M_\odot$ (top panels
in Fig.~\ref{star2}) have once again islands of transitional coupling,
located about $2H$ above and below the midplane (see section~4.7).
The threshold between Mg$^+$ and HCO$^+$ makes itself visible in the
shape of the MRI-active region's outer border too.  For the disks
around stars from 1.2 to 2~$M_\odot$, one can clearly see the
MRI-active Mg$^+$ region overlapping with the HCO$^+$ region (bottom
panels in Fig.~\ref{star2}).  Interestingly, while the
Mg$^+$-dominated layers extend further out around heavier stars (see
for example the $\Lambda=10$ contours), the ambipolar term still beats
out the induction term at $r>40 $AU and $3>(\Theta - \pi/2)/c_0>2$.
As this effect occurs outside the exponential surface density cut-off,
where the density falls rapidly with radius, we expect this would have
little effect on the accretion rate.

\begin{figure}
\begin{center}
\includegraphics[width=4.5in]{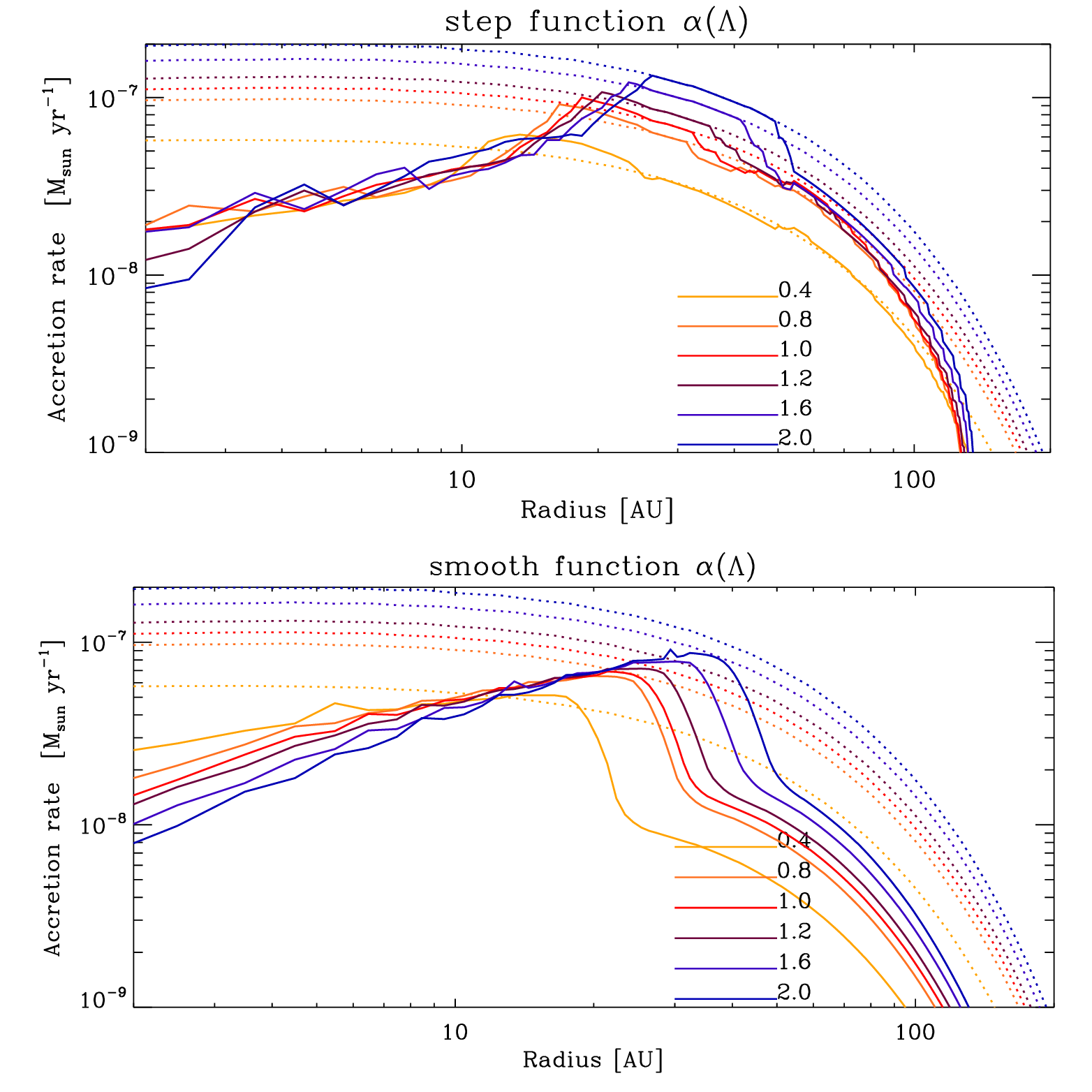}
  \caption{ Accretion rate radial profiles in disks around stars with
    masses from 0.4 to 2~$M_\odot$.  Dotted lines indicate the
    profiles when the whole disk is active.  \label{fig:set6mdot} }
\label{star3}
\end{center}
\end{figure}

The disks around the lower-mass stars have larger aspect ratios $H/R$
or $c_s/u_{\rm Kep}$.  The Keplerian velocity $u_{\rm Kep}$ increases
faster with the stellar mass than does the sound speed.  Since we fix
the ratio of disk to star mass, the surface density increases with the
mass of the central object.  This leads to larger dead zones in the
more massive disks.  We observe in Fig.~\ref{star2} that indeed the
disks around the higher-mass stars have smaller MRI-active regions.
The separation between the inner and outer radial boundaries of the
transitional layer on the midplane changes slowly with the stellar
mass, being $[6, 9, 10, 11,12, 13]$~AU around stars of $[0.4, 0.8, 1,
1.2, 1.6, 2] M_\odot$.  Thus over the stellar mass range, the
transitional layer widens from 6 to 10~AU, while the transitional
layer's inner edge ($log(\Lambda)=-1$ contour, or yellow-to-orange
transition in Fig.~\ref{star2}) moves outward by 4~AU.  For each
stellar mass the accretion rate peaks near the Mg$^+$ freeze-out
radius, beyond which the outer disk is less strongly accreting,
considering the smooth dependence of the stress parameter on the
Elsasser number (Fig.~\ref{star3}, bottom).

\subsection{Cosmic Rays}

The final parameter we vary is the cosmic ray ionization rate.  We
consider values 10, 100 and 1000~times greater than in the fiducial
model (set~7, Table~\ref{sets}).  A wide range of cosmic ray
ionization rates is inferred from models of interstellar chemistry,
and observations of HD, OH, D and especially H$_3^+$ and He$^+$ ions
indicate rates up to $\zeta_{CR0}=1.2\times 10^{-15}\rm\ s^{-1}$ in the
diffuse interstellar medium
\citep{spitz68,webb98,vandish86,lepeti04,feder96,indri07,mcal03,ind09}.
The distance to the galactic ridge is also important \citep{mel11}.

\begin{figure}
\begin{center}
\hbox{
\includegraphics[width=3.5in]{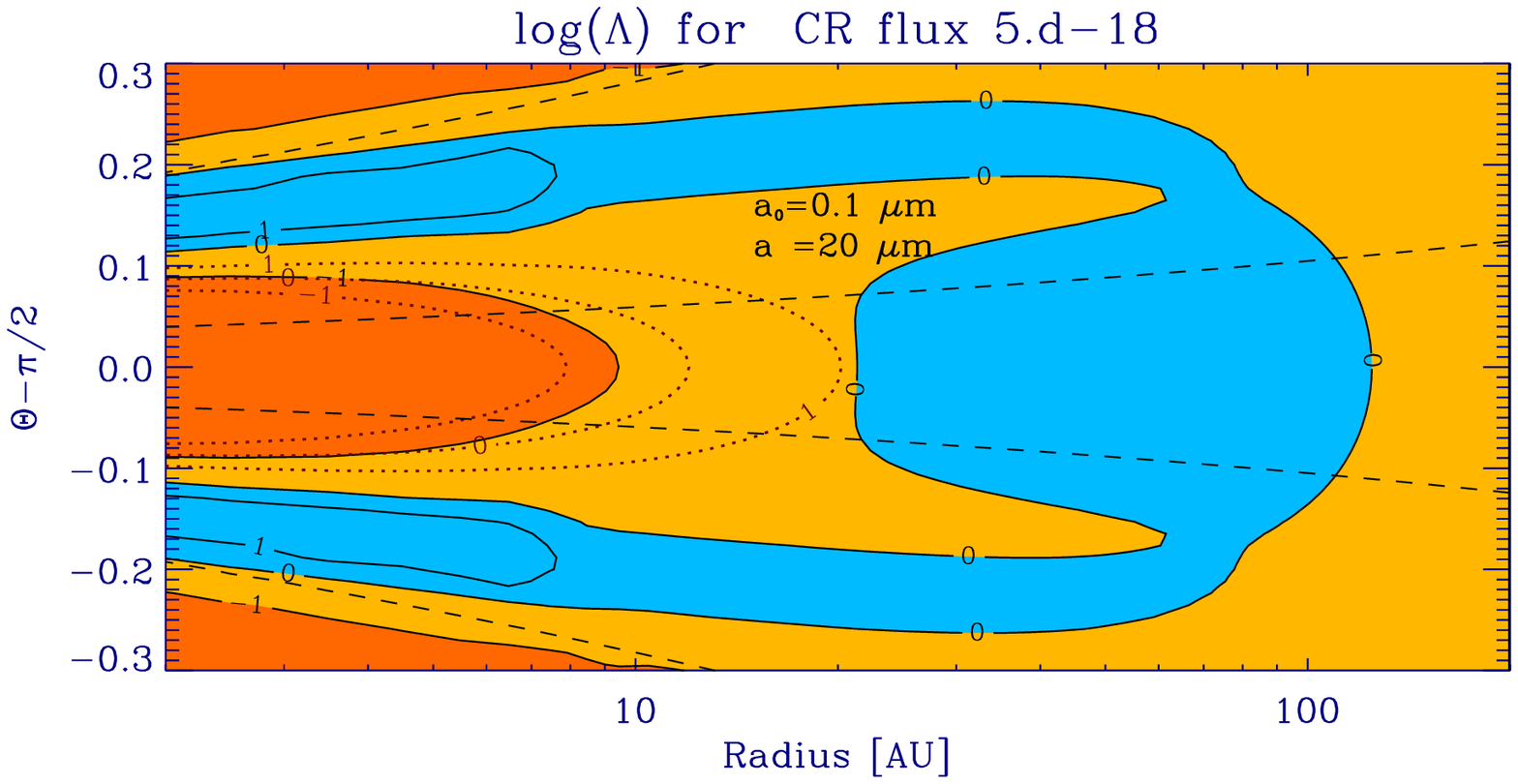}
\includegraphics[width=3.5in]{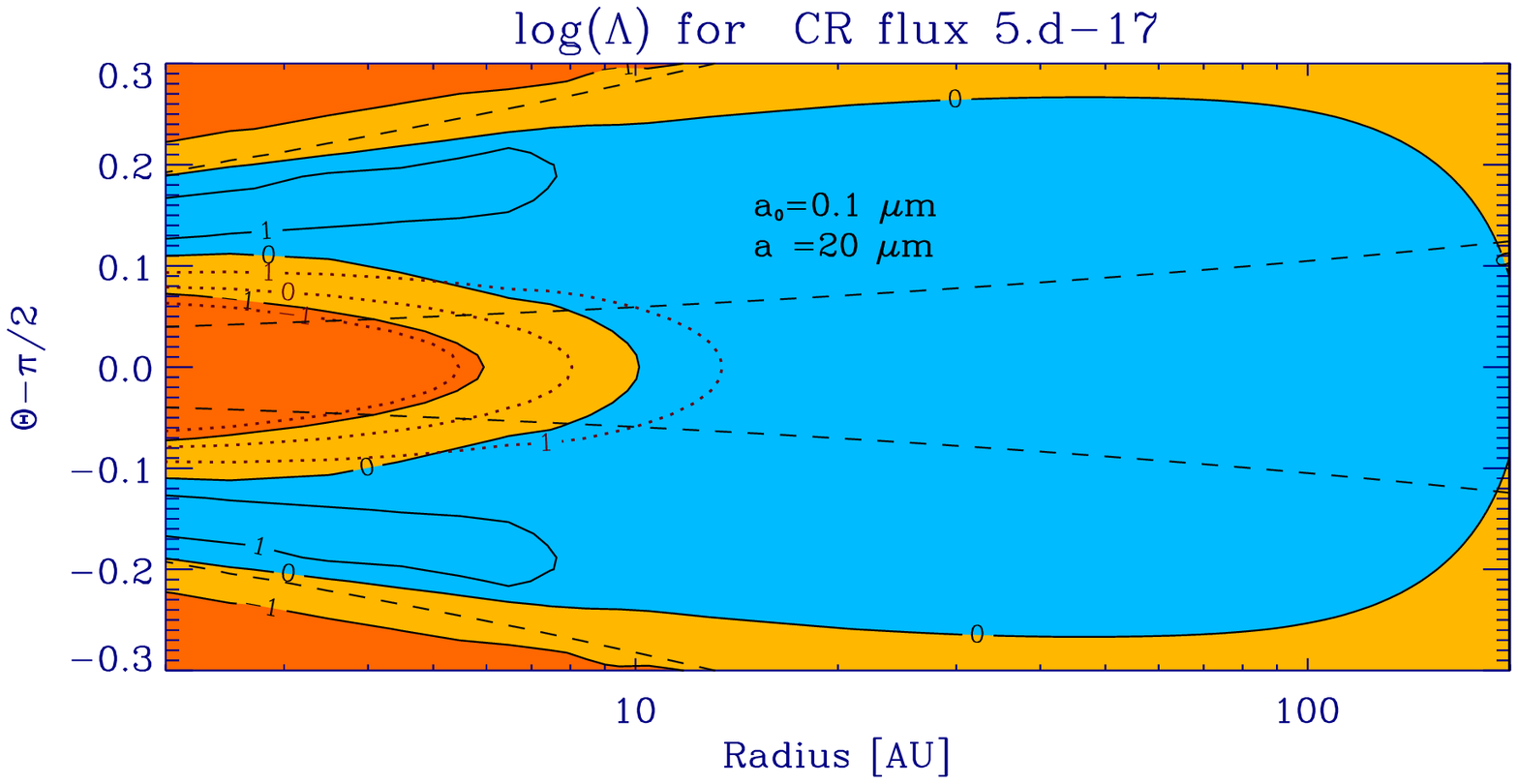}
}
\hbox{
\includegraphics[width=3.5in]{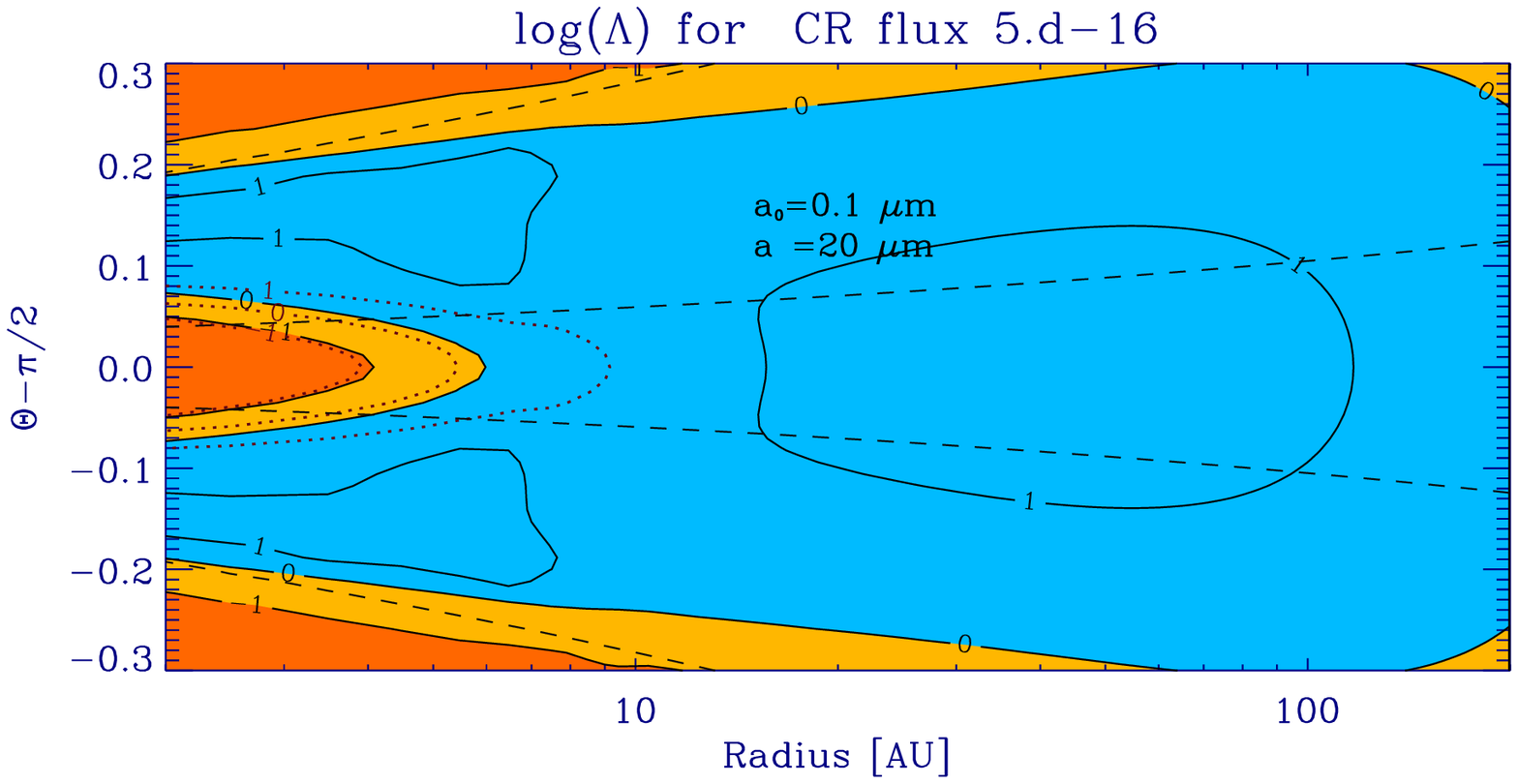}
\includegraphics[width=3.5in]{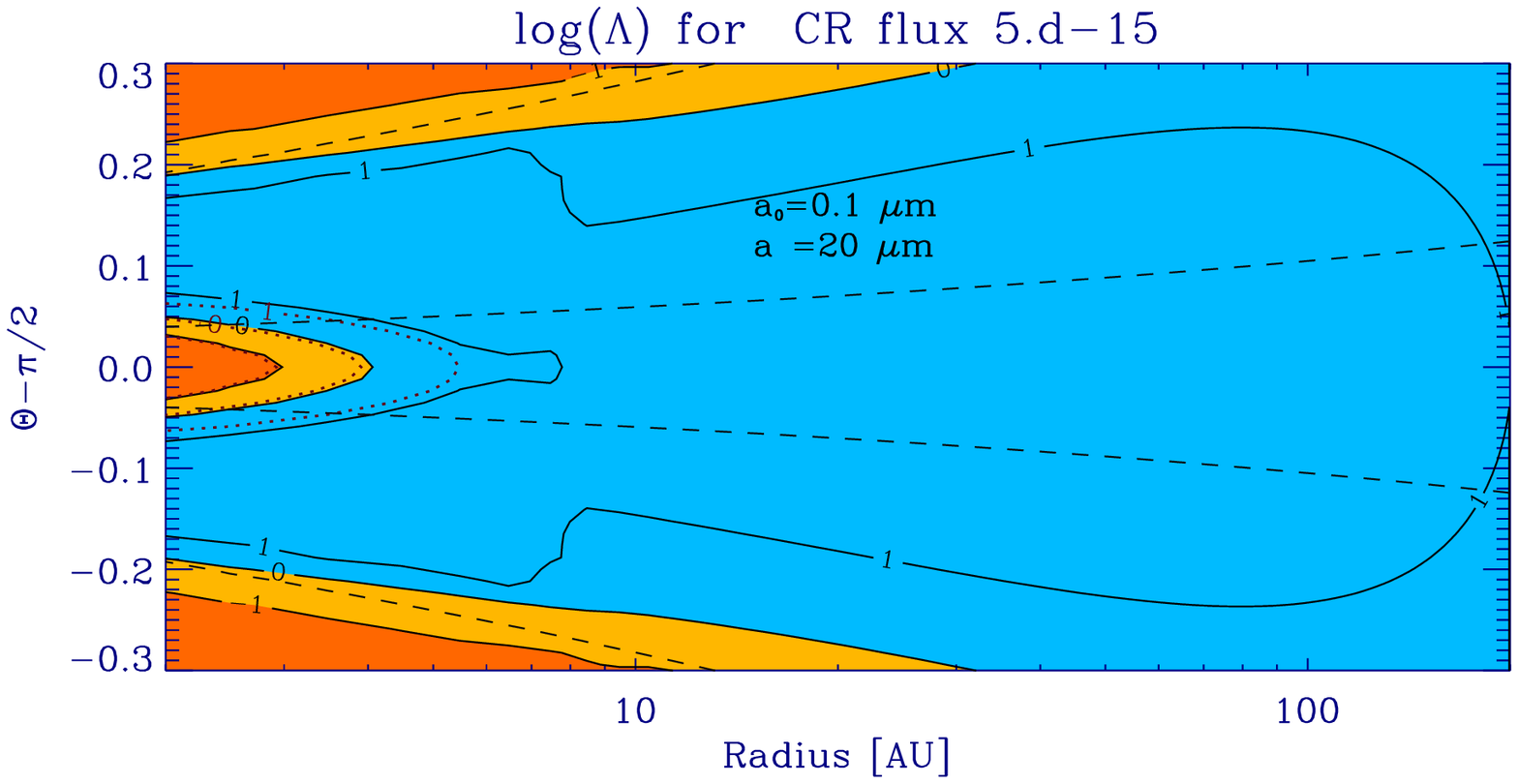}
}
  \caption{ Dead, transitional and active zones with the cosmic ray
    ionization rate varying from $5\times 10^{-18}\rm\ s^{-1}$ to
    $5\times 10^{-17}\rm\ s^{-1}$, $5\times 10^{-16}\rm\ s^{-1}$ and
    $5\times 10^{-15}\rm\ s^{-1}$.  Colors and symbols are as in
    figure~\ref{fi-els}.  }
\label{cosm1}
\end{center}
\end{figure}

\begin{figure}
\begin{center}
\includegraphics[width=4.5in]{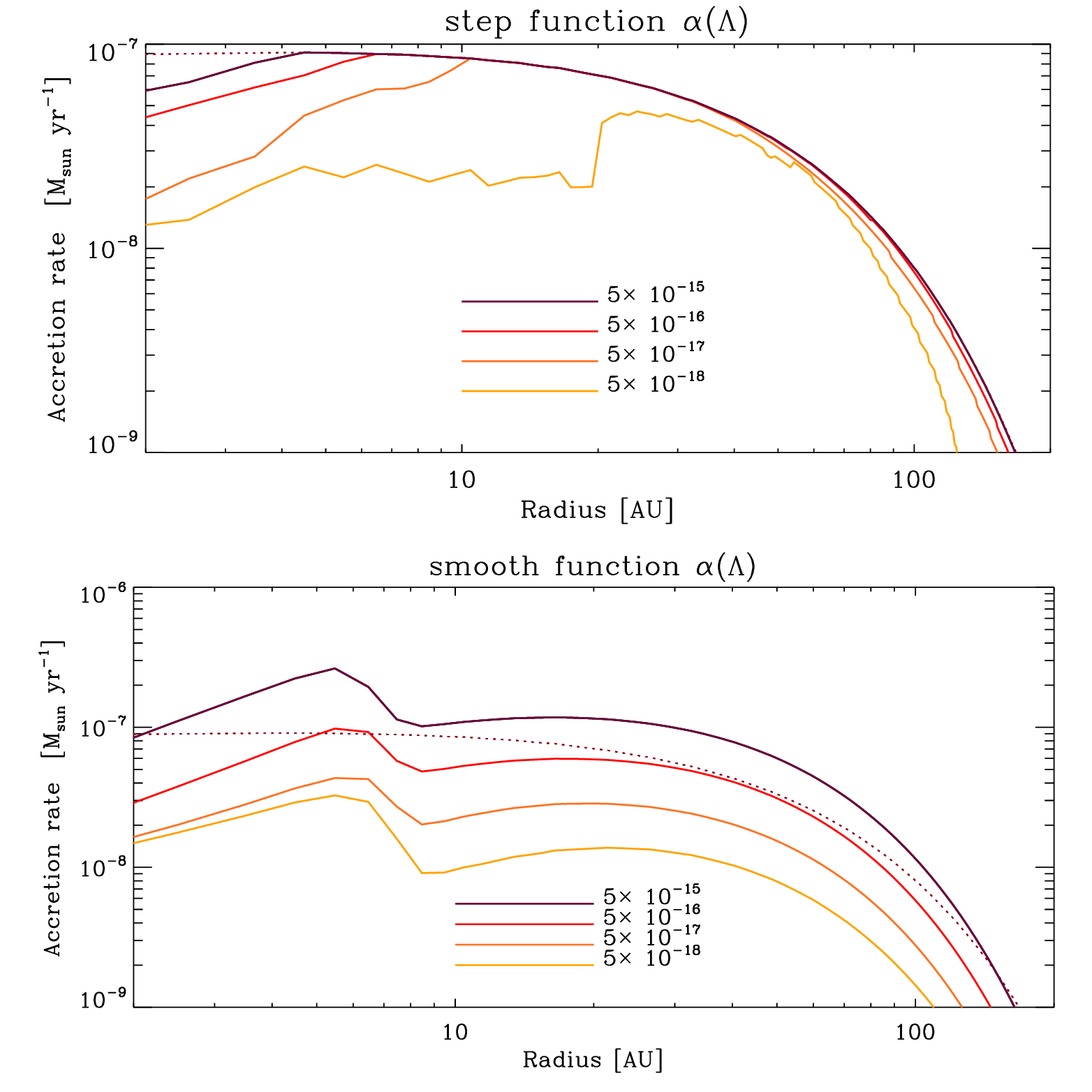}
  \caption{ Accretion rates vs.\ radius in set~7 with cosmic ray
    fluxes at the disk surface from $5\times10^{-18}$ to
    $5\times10^{-15}$.  The dotted line indicates the profile when the
    whole disk is active.  Note that the smooth dependence of
    turbulent stress on Elsasser number leads to higher accretion
    rates in the $\Lambda>10$ regions (see section~3.2).  }
\label{cosm2}
\end{center}
\end{figure}

From Fig.~\ref{cosm1} we see that ambipolar diffusion loses much of
its impact at a cosmic ray ionization rate of $10^{-16}-10^{-15}\rm\ 
s^{-1}$.  In the most extreme case, the Ohmic and ambipolar dead zones
coincide.  We also varied the dust-to-gas ratio, monomer size and
surface density power-law index for a cosmic ray ionization rate of
$5\times 10^{-17} \rm\ s^{-1}$ (not shown here).  The ambipolar
diffusion still determines the dead zone outer edge location and the
width of the transitional region.  A sharp outer edge was proposed as
a potential pressure trap for solid material in
\citet{lyr09,hase10a,hase10b}.  For high cosmic ray ionization rates
(two bottom panels in Fig.~\ref{cosm1}), a large region with Elsasser
number greater than $10$ appears.  We have a large volume with nearly
ideal MHD conditions.  However, the width of the transitional region
is still 1-2~AU, seen from the radial separation of the Elsasser
number contour lines.  The corresponding accretion rates for
step-function and for smooth function in $\alpha(\Lambda)$ are shown
in Fig.~\ref{cosm2}.  In the case of smooth $\alpha(\Lambda)$, there
is a peak in accretion rate at 8~AU visible for all models in set~7.
At the inner disk boundary, $\dot M$ increases from $10^{-8}$ to
$10^{-7} \rm M_\odot yr^{-1}$.

\section{Discussion}

\begin{figure}
\begin{center}
\includegraphics[width=6.8in]{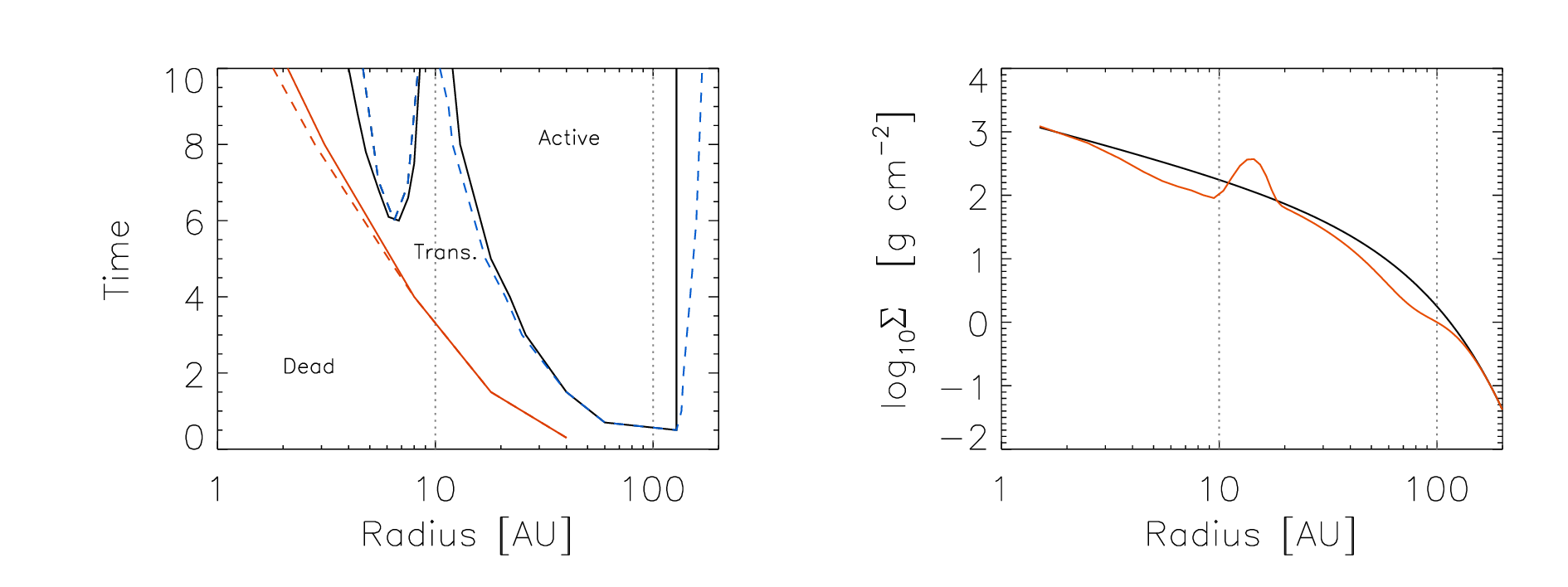}
  \caption{Left: Schematic picture of the time evolution of the dead,
    transitional and active regions. Solid lines show the regions'
    borders when only the dust-to-gas ratio is allowed to decrease
    with time.  Dashed lines include corrections for the gas surface
    density and disk mass evolving simultaneously.  Right: The black
    line shows the surface density in our fiducial model.  The red
    line shows the expected evolution to form a local gas pressure
    maximum capable of trapping solid bodies near the metal line
    (8~AU), surrounded by valleys in the surface density which will be
    better-ionized and potentially MRI-active.  }
\label{draw1}
\end{center}
\end{figure}

\paragraph{\bf Time Evolution of the Dead Zone:}
Over the lifetime of a protostellar disk, changes are likely in the
dust-to-gas ratio $f_{dg}$ as solid material is incorporated into
planetesimals and planets.  Let us consider the possibility that the
dust evolves down two paths simultaneously.  Most of the aggregates
remain in place, while a fraction grows in mass through mutual
collisions that are ``lucky'', fast enough to overcome electrostatic
forces and be compacted, but slow enough to avoid disruption.  If the
monomers' size distribution remains unchanged, the fluffy aggregates'
mass fraction relative to the gas will decline with time, causing the
dead zone to shrink and change shape.  In Fig.~\ref{draw1} left panel
we show schematically how the dead, transitional and active regions
change their midplane radial extent with time, and so with decreasing
dust-to-gas ratio.  Time increases upward and is labeled in arbitrary
units.  The picture is made more complicated by the shorter timescales
for evolution near the star.

In our first set of models we studied the impact of the monomer size
$a_0$ on the ionization state.  Real aggregates surely contain
monomers with a distribution of sizes, possibly resembling the
interstellar grain size distribution $n(a)\propto a^{-3.5}$
\citep{mat77}.  Accounting for the size distribution would likely make
the Hall effect stronger at intermediate gas densities \citep{war99}.
Future work should clarify how the dispersion in $a_0$ in combination
with the radius-dependent dust-to-gas ratio changes the dead zone's
shape.

The models in sets~3 and~4 show how changing disk mass and surface
density shift the active zone inward or outward.  In Fig.~\ref{draw1},
dashed lines represent the correction to the dead and active zone
borders when the density slope (or disk mass) changes at the same time
the dust-to-gas ratio is decreasing.  As the outer disk loses mass,
the active zone expands inward (compare blue solid and dashed lines).
The dead zone is situated in the inner part of the disk, where mass is
gained slowly over time due to the weakness of the MRI-activity in the
outer disk.  For low dust-to-gas ratio, a second MRI-active zone
appears inside 10~AU.  We expect the accretion flow to drain mass from
this new zone faster than mass is re-supplied from the outer parts.
The surface density may be reduced locally, leading to a valley in the
surface density.  Over the long term the valley in surface density
near the metal line will grow, becoming better and better ionized and
leading to still higher accretion rates locally.  If this process runs
away, it could clear the inner disk of material.  The right panel in
Fig.~\ref{draw1} shows the expected time evolution of the surface
density.  Please note that Fig.~\ref{draw1} is a schematic
representation.  A more precise picture requires quantitative disk
evolution models.

\paragraph{\bf Dependence of the Stress on Elsasser Number:}
Gaps appear in the accretion rate's radial profile in many cases when
we assume the stress parameter $\alpha$ is a step function across
Elsasser number unity (figures~\ref{fig:set1mdot}, \ref{fig:set3mdot},
\ref{fig:set5mdot}).  A step function adequately approximates the
turbulent stress' variation with height in the inner disk, where
conductivity gradients are steep.  The step function is harder to
justify when modeling the radial variation in the outer disk, however.
We demonstrated here that the outer disk can have a quite different
accretion rate profile if $\alpha(\Lambda)$ is a smooth function.  The
reason is that the Elsasser number is near unity over a large range of
radii $r>10$~AU.  Thus much of the outer disk is marginally
MRI-active.

We assume a field strength such that the magnetic pressure is a fixed
fraction of the midplane gas pressure at each radius.  A more complete
approach would involve making the stress proportional to the magnetic
pressure in the saturated turbulence.  \citet{bai11c} provides a
fitting function for the minimum possible plasma beta of the saturated
fields in the presence of ambipolar diffusion.  We have constructed
the corresponding turbulent stress profile (not shown in this paper),
applying the floor for plasma beta mentioned above.  It is reassuring
that the radial profiles of the accretion rate are basically the same
as those we present here.  When compared to the accretion rates
derived from $\alpha\propto 1/\beta$, the accretion is faster near the
inner radial boundary using Bai's fit, because the weaker magnetic
fields leave more of the upper layers with $\beta<1$.  This deserves
to be investigated in more detail using 3-D MHD simulations of MRI
turbulence including both Ohmic and ambipolar diffusion.

\paragraph{\bf Accretion Variability:}
It is noteworthy that all models show an accretion rate peak just
inside the metal line, which lies at 8~AU in our fiducial model.  This
radial location is affected by the disk temperature.  The temperature
thus affects the location and shape of the MRI-active zone.

The luminosity produced by the typical T~Tauri accretion rate $10^{-8}
\rm M_\odot/yr$ on a star of Solar mass and 2~Solar radii is
$GMM_\odot/r \sim 0.1 L_\odot$, probably not enough to significantly
change temperatures inside the disk region we consider.  However
during FU~Orionis outbursts, the accretion luminosity rises
sufficiently to increase the overall system brightness a hundredfold
\citep{har96}, requiring disk temperatures to at least triple.  This
is likely to expand the Mg$^+$-dominated region.  The outbursts last
about 100~years, a few orbits at 10~AU in our fiducial model and long
enough for MRI turbulence to reach saturation in the newly-coupled
gas.  Thus we expect the peak in the accretion rate to move to larger
radii as in set~5.

\paragraph{\bf Limitations:}
The outer radial border of the MRI-active zone coincides with the gas
density contour $10^{-15}\rm\ g cm^{-3}$, lying at 128~AU in our
fiducial model (section~4.1).  Since gas-phase molecular reactions
dominate the recombination there and the assumption about fast chemical 
equilibrium may become inaccurate, it would be useful to
recalculate the ion and electron densities there using the extended
chemical network, in order to validate our findings on the ambipolar
diffusion.
Additionally, we use the results from non-stratified simulations 
of MRI with ambipolar diffusion \citep{bai11}, to construct
  our smooth turbulent stress function and the accretion rates (see Section 3.2).
In the follow-up work of \citet{bai13}, it is shown that the MRI-active
layers at 1AU can first become turbulent but quickly be stabilized by
the ambipolar diffusion. If this finding holds, 
 this would affect our predictions of the MRI-driven accretion rate. 

Our chemical network includes only one representative
metal, magnesium, but of course many metals can occur in the gas
phase, and each has its own thermal desorption temperature.  There
should then be multiple metal lines, one for the freeze-out of each
metal.

All the metal lines will shift when we consider a more detailed
temperature distribution.  In our simple model the temperature is
constant on cylinders, an approximation which is appropriate only for
the disk's optically thick interior.  The surface layers are typically
hotter because they are directly illuminated by the starlight
\citep{dul07}, so the metal lines would bend away from the star with
increasing distance from the equatorial plane.

We may also ask whether a steeper radial variation in the Elsasser
number can be obtained with a radial gradient in, for example,
dust-to-gas ratio or grain charging properties.  A steeper
$\Lambda(r)$ would yield bigger differences in the stresses between
the transitional, dead and MRI-active zones.  Such variations also
open the possibility of planetesimal traps at multiple radii.

\section{Conclusions}

We determine dead zone sizes and shapes in disks where the ionization
state is controlled by recombination on fluffy dust aggregates.  We
study the effects of parameters including the constituent monomer
size, the aggregates' mass fraction, the magnetic field strength, the
masses of the disk and star, the slope of the surface density profile,
the disk temperature and the cosmic ray ionization rate.  The three
parameters most strongly affecting the dead zone are the monomer size,
the dust mass fraction and the cosmic ray rate.

We find that ambipolar diffusion generally defines the dead zone's
outer radial boundary, as well as the thickness of the transitional
layer just inside the boundary.  The ambipolar diffusion is overridden
by Ohmic dissipation near the dead zone edge only if the magnetic
fields are weak enough that the midplane plasma beta is greater than
4000, corresponding to field strengths $B<$0.15~mGauss at 100~AU.

The fractal aggregates are efficient in sweeping up free electrons and
ions, providing a major recombination pathway at gas densities greater
than $10^{-14}\rm\ g cm^{-3}$.  Nevertheless the disk surface layers
couple to the magnetic fields sufficiently for MRI turbulence to drive
accretion.  Accurately estimating the resulting mass flow rates
requires carefully treating the stresses' dependence on the Elsasser
number.  A step function is an inadequate description because so much
of the disk has Elsasser numbers near the critical value.  The mass
flow is $10^{-9}$--$10^{-8} M_\odot yr^{-1}$ at 1~AU for our fiducial cosmic
ray ionization rate of $5\times 10^{-18}\rm\ s^{-1}$.  This is similar
to the findings of \citet{bai11c}, who considered compact grains
rather than fractal aggregates.

The dead zone's outer edge, which we call the transitional region and
define by Elsasser numbers from 0.1 to 1, is generally very broad at
the midplane.  Its width is 12~AU in our fiducial model and ranges
from 2~AU in our most strongly-ionized model from set~7, to the entire
disk at the maximum dust-to-gas ratio from set~1.  The mass flow rates
indicate that the transitional region will accumulate no sharp density
bump as the disk evolves.  On the other hand we find that the magnetic
coupling, and therefore the flow rate, declines steeply where the
temperature falls below a threshold allowing metal atoms to thermally
adsorb on the grains.  Such a metal freeze-out can cause a local
maximum in the flow rate, leading to a gap in the surface density with
a long-lived pileup of material just outside.  This kind of accretion
peak occurs in all models across the parameter study, shifting in
radius according to the disk temperature distribution.

By varying each of the parameters in turn, we find:
\begin{itemize}
\item At dust aggregate mass fractions near $10^{-2}$, thin MRI-active
  surface layers yield mass flow rates $10^{-9}\rm\ M_\odot yr^{-1}$.
  At mass fractions below $10^{-4}$ the dead zone takes a fish-tail
  shape in meridional cross-section, and the mass flow rate is around
  $10^{-8}\rm\ M_\odot yr^{-1}$.  Even at very low dust
  concentrations, there is a patch at $r\gtrsim 10$~AU where the
  magnesium atoms freeze out and ambipolar diffusion pushes the
  Elsasser number into the transitional regime $0.1<\Lambda<1$.
\item The monomer size also affects the dead zone shape.  When the
  monomers are 10~nm or smaller, giving the aggregates a large
  cross-section for a given solid mass, the activity is confined to
  narrow isolated surface layers along with a blob of marginally
  active gas near 100~AU.  When the monomers are 100~nm or larger, the
  dead zone takes the fishtail shape.
\item Varying the surface density slope or the disk mass reveals the
  dead zone's origins.  With increasing disk mass we observe larger
  MRI-active zones in the outer parts, as higher densities lead to
  more frequent collisions with neutrals, thwarting ambipolar
  diffusion.  A similar effect comes from reducing the surface density
  slope so that more of the mass resides far from the star.  The dead
  zone's outer edge lies at 20~AU for disk masses between 0.5 and 1.5
  times the fiducial model.  At surface density slopes 1.3 and
  steeper, ambipolar diffusion isolates a midplane marginally-active
  region outside 20~AU.
\item Making the disk hotter or colder also changes the dead zone's
  shape.  The fishtail appears strongest when the magnesium freezeout line lies
  near the inner edge of the annulus where ambipolar diffusion at
  intermediate heights separates magnetically-active layers in the
  better-ionized surface and denser interior. The fins of the fishtail-shaped
  transitional zone appear to be sensitive to the local ion density 
  and therefor to the disk parameters.
\item The dead zone's dependence on stellar mass combines the impacts
  of disk mass, gravity and temperature.  For our chosen stellar
  masses and ages (0.4 to 2 $M_\odot$, all at 1~Myr), the disks are so
  hot that the Mg$^+$-dominated region extends far from the star. 
   Both the magnesium freezeout line and
  the corresponding accretion rate peak are shifted to larger radii
  with increasing stellar mass.  The magnesium line ranges from 10 to
  25~AU over the stellar mass range we consider.
\end{itemize}

We identify a new way to create a local pressure maximum capable of
trapping solid particles.  The steep gradient in mass flow rate across
the metal freezeout line means the magnetic stresses will evacuate the
side of the line near the star, while material piles up beyond.  Gas
drag forces from headwinds outside the maximum, and tail winds inside,
will bring solid particles toward the peak.  The radial gradients
could grow strong enough to trigger the formation of vortices,
contributing to angular momentum transport in the weakly magnetically
coupled transitional region \citep{flo12,rae12} and further
concentrating solid material.  Testing the effectiveness of such traps
is a challenging task.  Probing the distribution of the field strength
in the MRI turbulence requires global 3-D MHD simulations with
ambipolar diffusion.  Ionization chemistry and the metal freezeout
must be treated.  The pressure profile's evolution near the metal line
over the longer accretion timescales might be explored using
azimuthally-averaged mass transport calculations with turbulent
stresses obtained from MHD simulations.  Since the MRI turbulence and
dust are mutually coupled and feedback is possible, an important part
of the story will be the evolving dust abundance and the corresponding
changes in the shape of the dead zone.

\acknowledgments
 
N.~Dzyurkevich was supported by the Deutsche
For\-schungs\-ge\-mein\-schaft (DFG) through For\-scher\-gruppe 759,
``The Formation of Planets: The Critical First Growth Phase''. 
We acknowledge support from the Deutsches Zentrum f\"ur Luft- und Raumfahrt (DLR),
 support code 50 OR 0401. 
 N.~J.\
Turner's participation was supported by the NASA Solar Systems Origins
program under grant 07-SSO07-0044, and by the Alexander von Humboldt
Foundation through a Fellowship for Experienced Researchers.  The work
was carried out in part at the Jet Propulsion Laboratory, California
Institute of Technology.
We thank Prof. W. Brandner and M. Gennaro for the tutoring on the 
STELLAR code. We thank S. Okuzumi for a very thorough referee
report that greatly improved the quality of this paper. 

\bibliographystyle{apj}

\bibliography{nat5apj}

\begin{appendix}

  \section{Determining Gas-Phase Metal
    Abundance \label{gasphasemetal}}

  Most of the metal atoms are incorporated into grains.  The remainder
  can occur either free in the gas, where they are available to
  participate in the ionization balance, or adsorbed on grain
  surfaces.  We determine here the number density of the metal atoms
  in the gas phase, which depends on temperature.  We consider only
  the atoms' thermal adsorption on and desorption from the grains.
  Gas-phase species $X$ when captured on the grains is labeled $gX$.
  The equilibrium number density on the grains is given by
  \begin{equation}
    n[gX]=\frac{k_1S_x}{k_2}n[X]n_{\rm gr}=a n[X],
  \end{equation}
  where the coefficients $k_1=\sigma_{\rm gr}v_{\rm th}$ and $k_2$ are
  adsorption and desorption rates.  The number density in the gas
  \begin{equation}
    n[X]=n[X]^{\rm total}-n[gX].
  \end{equation}
  We determine the amount of magnesium on the grain surfaces using the
  expression
  \begin{equation}
    n[g{\rm Mg}]=a\frac{n[{\rm Mg}]^{total}}{1+a},
  \end{equation}
  and in the gas phase
  \begin{equation}
    n[{\rm Mg}]=\frac{n[{\rm Mg}]^{total}}{1+a}.
  \end{equation}
  The initial abundances are 1 for H$_2$, 9.75$\times10^{-2}$ for He,
  3.62 and 8.53 $\times10^{-4}$ for C and O, and $1\times10^{-11}$ for
  Mg \citep{ilg06}.  Not every collision with a grain leads to
  sticking, and the sticking probability is
  $S_X=\exp(-E_{kin}/\sqrt{(2E_D\Delta E_{trans})})$, with $\Delta
  E_{trans}\simeq 2\times 10^{-3}$~eV for the energy transfer to the
  grain via lattice vibrations.  $E_D^{\rm H_2}=450$, $E_D^{\rm
    HCO^+}=1510$ and $E_D^{\rm Mg}=5300 k_B$~K.  The desorption
  coefficient $k_2=\nu_0 \exp(-E_D/kT)$ with $\nu_0=\sqrt{(3\times
    10^{15} {\rm cm}^{-2} E_D/(\pi^2 m_x))}$ \citep{hase92}.  After
  finding the gas-phase metal abundance, we compute the total
  recombination rate $c_t$ for the dominant ion by applying an
  iterative approach (section~\ref{sec:gasphaserecomb}).

\begin{figure}
\begin{center}
\hbox{
\includegraphics[width=2.5in]{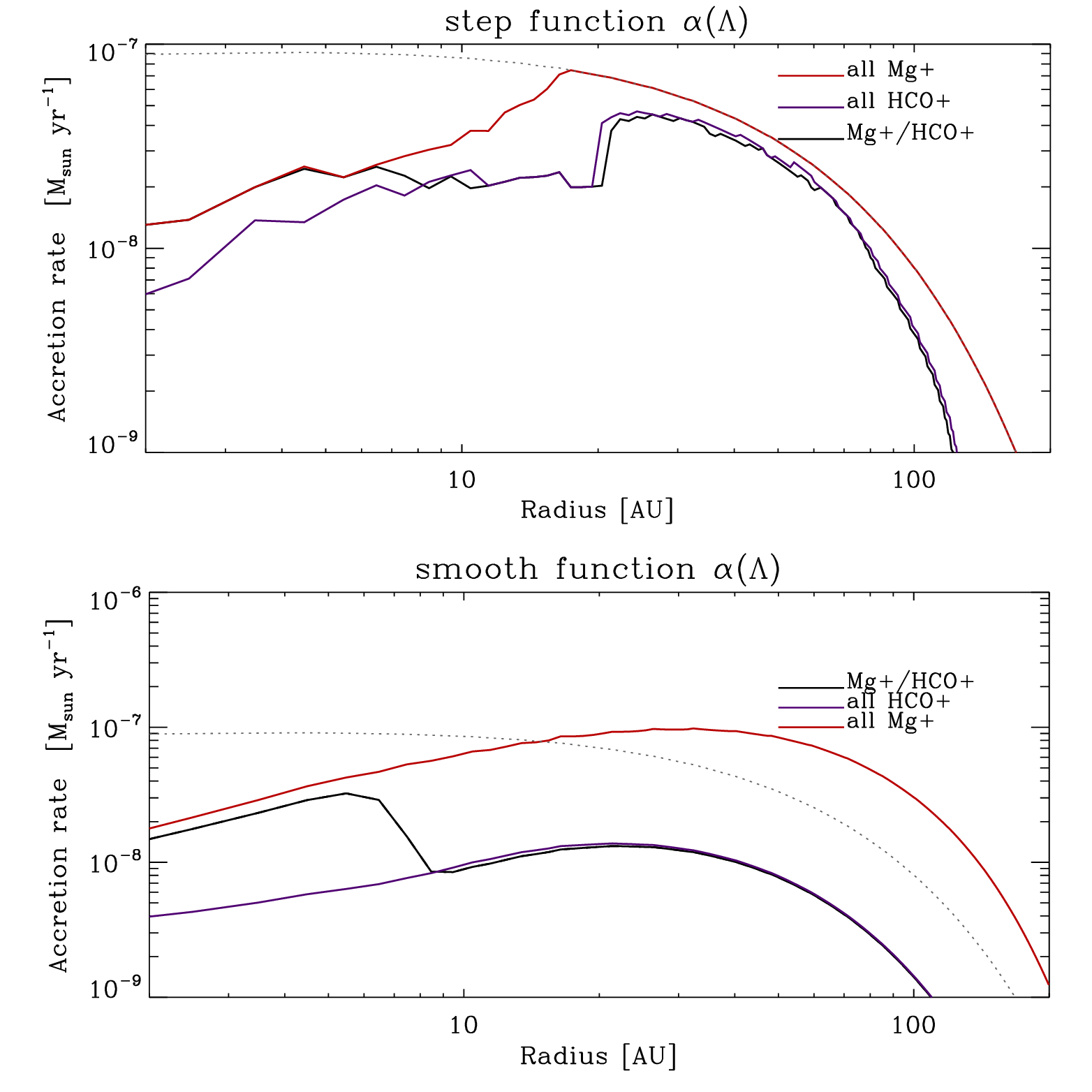}
}
  \caption{Accretion rates in solar masses per year for models with
    the representative ion Mg$^+$ or HCO$^+$ (red and purple lines). Black
    line shows the fiducial model with the transition between the two ions, 
    which occures between 5 and 9 AU.  The dotted line shows the accretion rate for the case when
    the whole disk is active with the turbulent stress from eq.~29.
    Note that the smooth-function turbulent stress can lead to higher
    accretion rates in the $\Lambda>10$ regions (see section
    3.2). \label{gasphasem}}
\end{center}
\end{figure}

 The transition between two representative ions leads to the jump in 
 Elsasser number and in the corresponding accretion rates (Fig.~\ref{gasphasem}).
 Here we show how the accretion rates would look like, if the representative ion
 is chosen to be Mg$^+$ everywhere, or HCO$^+$ everywhere. The accretion rates
for the fiducial model follow the case of pure Mg$^+$  inside of $r<5$AU, and
the case of  pure HCO$^+$  outside of $r>9$AU.

  \section{Gaussian Charge State Distribution -- Limit of
    Validity\label{chargeneutrality}}

\begin{figure}
\begin{center}
\hbox{
\includegraphics[width=2.5in]{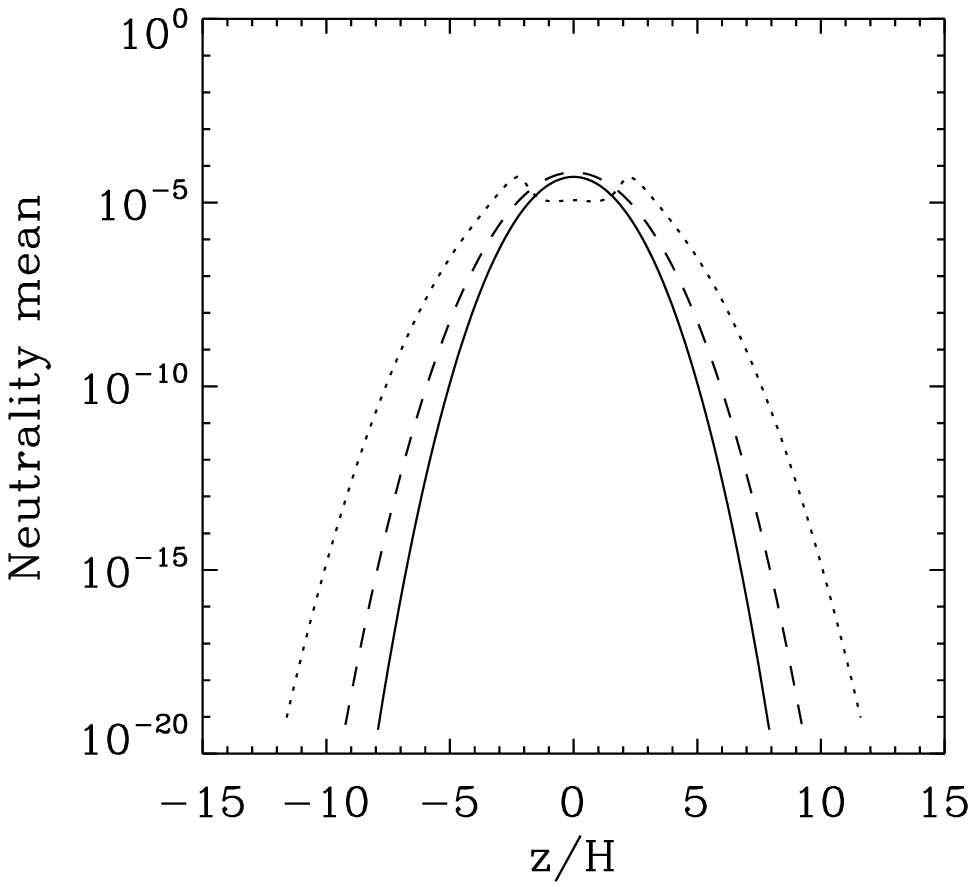}
\includegraphics[width=2.5in]{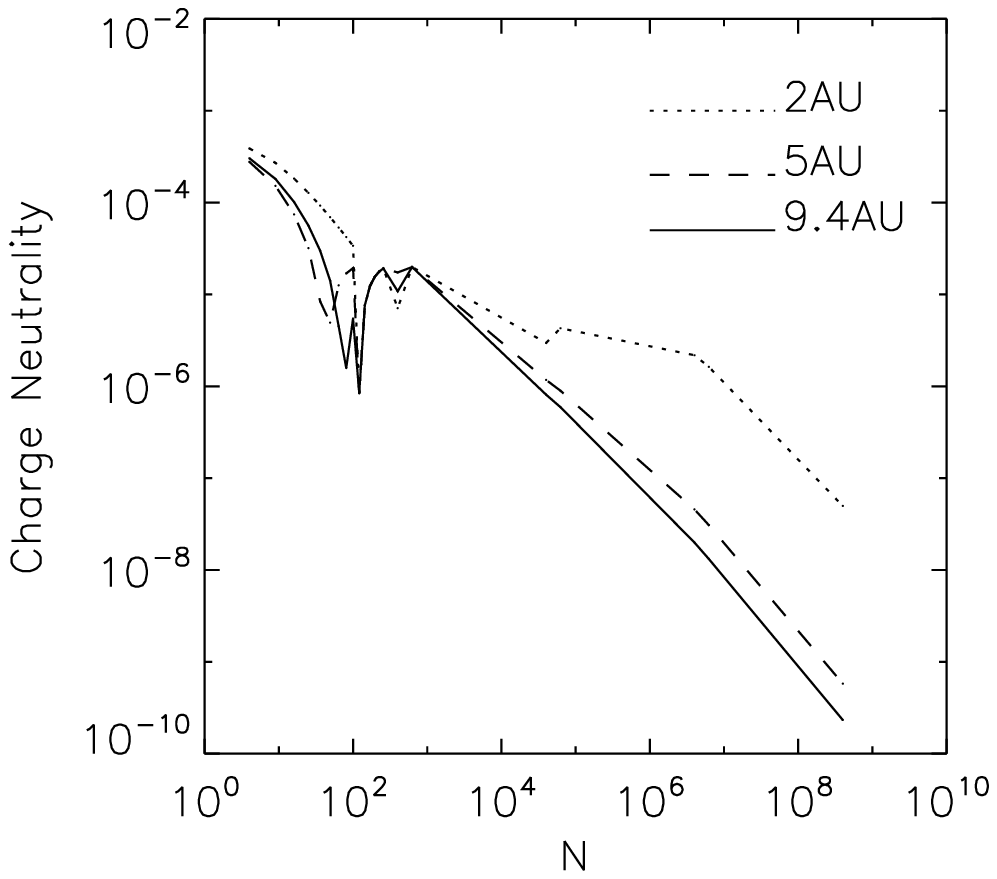}
}
  \caption{Left: Charge neutrality for $R=2$, $5$ and $9.4$~AU, for
    $N=400$.  Right: same labeling for charge neutrality at the
    midplane, for variable number of monomers $N$.  }
\label{neut}
\end{center}
\end{figure}

  We find the abundances of the ions, electrons and charged grains
  from \citet{oku09a} eqs.~32, 33 and 22, letting the grains charge by
  up to 30~electrons either side of their mean charge.  The grains'
  charge state distribution is very well-approximated by the Gaussian
  when the grains carry many charges each.  Small aggregates however
  charge weakly enough that their few discrete charge states can
  differ noticeably from a Gaussian.  As a result the plasma is not
  quite charge-neutral overall.  Fig.~\ref{neut} demonstrates the
  accuracy of the charge neutrality for three different radii in the
  fiducial model.  The largest errors occur near the midplane where
  most of the charge resides on the grains.  Accuracy is good when the
  fluffy grains are made of $N>400$ monomers, while relatively large
  departures from neutrality occur when $N<10$.  Several of the
  assumptions used by \citet{oku09a} break down when the number of
  monomers is too small, including the assumption of negligible
  electric polarization.  But most importantly, charge neutrality is
  violated when the grains' charge spread is too small to be
  represented with a Gaussian.

 \section{MRI Criterion When Ambipolar Diffusion
    Dominates\label{mricriterionA}}

  Whether MRI turbulence can operate under an ambipolar-dominated
  magnetic diffusivity is sometimes judged using the ion magnetic
  Reynolds number.  In this Appendix we examine the general conditions
  for this number to be equivalent to the ambipolar Elsasser number.
  That is, we seek the requirements for $Re_i \approx \Lambda_A$ or
  \begin{equation}
    \frac{\gamma_i\rho_i}{\Omega} \approx \frac{v_A^2}{\eta_A\Omega}
  \end{equation}
  or
  \begin{equation}\label{eq:problem}
    \eta_A \approx \frac{B^2}{4\pi\rho\gamma_i\rho_i}.
  \end{equation}

  Recall
  \begin{equation}\label{eq:etaAcond}
    \eta_A
    = \frac{c^2}{4\pi}
    \left( \frac{\sigma_P}{\sigma_\perp^2} - \frac{1}{\sigma_O} \right)
  \end{equation}
  from eq.~\ref{eq:eta}.  Let each of the three conductivities be
  dominated by a single species.  Label the main Ohmic species' mass
  $m_O$, charge $q_O$, number density $n_O$, collision coefficient
  $\gamma_O$ and coupling parameter $b_O$, and similarly for the Hall
  and Pedersen conductivities:
  \begin{eqnarray}
    \sigma_O &=& \frac{ec}{B} n_Oq_O
      \left(\frac{eBq_O}{m_Oc\gamma_O\rho}\right)\\
    \sigma_H &=& \frac{ec}{B} \frac{n_Hq_H}
      {1+\left(\frac{eBq_H}{m_Hc\gamma_H\rho}\right)^2}\label{eq:sigmaHalt}\\
    \sigma_P &=& \frac{ec}{B}
      \frac{n_Pq_P\left(\frac{eBq_p}{m_Pc\gamma_P\rho}\right)}
      {1+\left(\frac{eBq_P}{m_Pc\gamma_P\rho}\right)^2}
  \end{eqnarray}
  We have used charge neutrality to derive eq.~\ref{eq:sigmaHalt} from
  eq.~\ref{eq:sigmaH}.  Notice that $\sigma_O$ is independent of the
  field strength, so we can end up with $\eta_A\propto B^2$ as in
  eq.~\ref{eq:problem} only if the final term in eq.~\ref{eq:etaAcond}
  is negligible.

  Now the perpendicular conductivity $\sigma_\perp =
  \sqrt{\sigma_H^2+\sigma_P^2}$ can be dominated by either $\sigma_H$
  or $\sigma_P$.  We consider the two possibilities in turn.  If
  $\sigma_H\gg\sigma_P$, the remaining piece of the right-hand side of
  eq.~\ref{eq:etaAcond} is
  \begin{equation}\label{eq:case1}
    \frac{\sigma_P}{\sigma_H^2}
    = \frac{n_Pq_P^2}{m_P\gamma_P\rho}
      \left(\frac{B}{cn_Hq_H}\right)^2
      \left(1+\left[\frac{eBq_P}{m_Pc\gamma_P\rho}\right]^2\right)^{-1}
      \left(1+\left[\frac{eBq_H}{m_Hc\gamma_H\rho}\right]^2\right)^2.
  \end{equation}
  This reduces to eq.~\ref{eq:problem} if three conditions are met:
  \begin{enumerate}
  \item The ions dominate the Pedersen conductivity.
  \item The ions have the same number density and charge magnitude as
    the Hall species, i.e.\ $n_H=n_P=n_i$ and $|q_H|=|q_P|=|q_i|$.
    For example, the ions could also dominate the Hall conductivity.
  \item Both the ions and the Hall species are poorly coupled, so
    $b_i=b_P\ll 1$ and $b_H\ll 1$ and the terms not
    quadratic in $B$ can be neglected.
  \end{enumerate}

  In the other case, where $\sigma_H\ll\sigma_P$, the right-hand side
  of eq.~\ref{eq:etaAcond} becomes
  \begin{equation}\label{eq:case2}
    \frac{1}{\sigma_P}
    = {\left(n_Pq_P^2\frac{e^2}{m_P\gamma_P\rho}\right)}^{-1}
      \left(1+\left[\frac{eBq_P}{m_Pc\gamma_P\rho}\right]^2\right).
  \end{equation}
  This reduces to eq.~\ref{eq:problem} if the ions again determine the
  Pedersen conductivity, and in addition the term proportional to
  $B^2$ dominates, i.e.\ the ions couple to the fields and $b_i=b_P\gg
  1$.

  We next look at how the Ohmic term becomes negligible in the
  ambipolar diffusivity when the other requirements for
  eq.~\ref{eq:problem} are met.  In the first case, where
  $\sigma_H\gg\sigma_P$, the Ohmic term is small if $\sigma_O \gg
  \sigma_H^2/\sigma_P$ or
  \begin{equation}
    n_Oq_Ob_O \gg \left(\frac{n_H^2q_H^2}{n_Pq_P}\right)
      \frac{1+b_P^2}{b_P(1+b_H^2)^2} \label{eq:case1noO}.
  \end{equation}
  Now in this first case, eq.~\ref{eq:problem} works only when
  conditions~1-3 are satisfied.  Under the conditions,
  eq.~\ref{eq:case1noO} simplifies to
  \begin{equation}\label{eq:case1noOsimplified}
    b_O \gg \frac{n_iq_i}{n_Oq_O} b_i^{-1}.
  \end{equation}
  That is, the Ohmic species must not decouple too much from the
  magnetic fields.  Throughout our fiducial model disk, and over a
  wide range of other models, the Ohmic conductivity comes from the
  electrons.  Since these have the same charge magnitude as the ions,
  and a coupling parameter $b_e\approx 1000b_i$ due to their low mass,
  eq.~\ref{eq:case1noOsimplified} means the ion coupling must satisfy
  $1000b_i^2\gg n_i/n_e$.  Also the electrons have the same number
  density as the ions in the ion-electron plasma limit where
  grain-surface recombination is unimportant, while in the opposite
  limit of an ion-dust plasma the electrons' depletion onto grains
  yields $n_e\approx 10^{-2}n_i$ \citep[eq.~38 of][]{oku09a}.
  Eq.~\ref{eq:case1noOsimplified} then requires $b_i\gg 0.03$ and
  $b_i\gg 0.3$, respectively.  These are incompatible or just barely
  compatible with $b_i\ll 1$ needed by condition~3.
  Eq.~\ref{eq:problem} is therefore valid at most in a narrow range of
  ion coupling parameters when $\sigma_H\gg\sigma_P$.

  In the second case, where $\sigma_H\ll\sigma_P$, the Ohmic term can
  be dropped when $\sigma_O\gg\sigma_P$, that is when
  \begin{equation}
    n_Oq_Ob_O \gg \frac{n_Pq_Pb_P}{1+b_P^2}.
  \end{equation}
  Since eq.~\ref{eq:problem} works in this case only if $b_P\gg 1$, it
  is enough to say the Ohmic term is small when
  \begin{equation}
    b_O \gg \frac{n_iq_i}{n_Oq_O}b_i^{-1},
  \end{equation}
  the same form as eq.~\ref{eq:case1noOsimplified}.  Even the stronger
  of the two constraints derived above for
  eq.~\ref{eq:case1noOsimplified}, $b_i\gg 0.3$, is already met in
  this case by our requirement that $b_i=b_P\gg 1$.

  Considering the two cases together, we see that the ion Reynolds
  number is equivalent to the ambipolar Elsasser number if the ions
  couple to the field and dominate the Pedersen conductivity, which in
  turn is larger than the Hall conductivity.

\end{appendix}

\end{document}